\documentclass[10pt]{iopart}
\usepackage{iopams} 
\usepackage[english]{babel}
\usepackage{wrapfig} 
\usepackage{tikz}
\usepackage{ifsym}
\usepackage{subcaption}
\usepackage{bm}

\usepackage{longtable}

\usepackage[toc,title]{appendix}
\usepackage{url}
\usepackage{booktabs}
\usepackage{xltabular}
\usepackage{array}
\usepackage[normalem]{ulem} 

\expandafter\let\csname equation*\endcsname\relax

\expandafter\let\csname endequation*\endcsname\relax

\usepackage{amsmath}
\usepackage{amsthm} 
\usepackage{amssymb}
\usepackage{amsfonts}
\usepackage{mathrsfs}
\usepackage{mathtools}
\usepackage{dsfont} 
\usepackage{textcomp} 
\usepackage{physics} 

\usepackage{hyperref}
\usepackage{cleveref}
\usepackage[acronym, sort=standard, nopostdot ]{glossaries}

\makeglossaries

\newglossarystyle{longleft}{%
  \setglossarystyle{long}%
    {%
      \renewcommand{\arraystretch}{1.2}%
     
      \begin{longtable}[l]{l @{\hspace{10em}} p{0.65\textwidth}}%
    }%
    {\end{longtable}}%
}

\newacronym{cptp}{CPTP}{Completely Positive Trace-Preserving}
\newacronym{ucp}{UCP}{Unital Completely Positive}
\newacronym{gkls}{GKLS}{Gorini--Kossakowski--Lindblad--Sudarshan}
\newacronym{qmc}{QMC}{Quantum Markov Chain}
\newacronym{ks}{KS}{Kraus--Sudarshan}
\newacronym{qds}{QDS}{Quantum Dynamical Semigroup}
\newacronym{dfs}{DFS}{Decoherence-Free Subspace}
\newacronym{dqls}{DQLS}{Dissipatively Quasi-Locally Stabilizable}
\newacronym{ckks}{CKKS}{Chru{\'s}ci{\'n}ski-Kimura-Kossakowski-Shishido}
\newacronym{dfa}{DFA}{Decoherence-Free Algebra}

\newtheorem{theorem}{Theorem}[section]

\newtheorem{corollary}[theorem]{Corollary}
\newtheorem{definition}[theorem]{Definition}
\newtheorem{example}[theorem]{Example}

\newtheorem{proposition}[theorem]{Proposition}

\newtheorem{remark}{Remark}

\definecolor{greenlink}{rgb}{0.3, 0.6, 0.3}
\definecolor{greenclear}{rgb}{0.7,0.95,0.7}
\definecolor{grey}{rgb}{0.8,0.8,0.8}
\definecolor{greenother}{rgb}{0.75,0.9,0.75}

\DeclareMathOperator{\Ker}{Ker}
\DeclareMathOperator{\Attr}{Attr}

\DeclareMathOperator\Fix{Fix}
\DeclareMathOperator{\Ran}{Ran}
\DeclareMathOperator{\spec}{spec}
\DeclareMathOperator{\Sem}{\mathbf{\Phi}}

\newcommand{\II}{\mathbb{I}}

\newcommand{\C}{\mathbb{C}}
\newcommand{\Bo}{\mathcal{B}}
\newcommand{\St}{\mathcal{S}}
\newcommand{\Hi}{\mathcal{H}}

\newcommand{\DoA}{\mathrm{DoA}}

\usepackage[utf8]{inputenc}
\usepackage[english]{babel}

\usepackage[
    backend=biber,
    biblabel=brackets,
    articletitle=true,
    style=phys, 
    sorting=nyt,
    maxnames=10,
    minnames=1,
]{biblatex}
\usepackage{hyperref}

\hypersetup{
    colorlinks=true,
    linkcolor=[rgb]{0.2, 0.4, 0.4},
    filecolor=[rgb]{0.2, 0.4, 0.4},
    citecolor=[rgb]{0.2, 0.4, 0.4},
    urlcolor=[rgb]{0.2, 0.4, 0.4},
}
\begin{document}
\title[Long-time dynamics]{Long-time dynamics of open quantum systems: mathematical and physical aspects}
\author{Daniele Amato$^{1,2}$, Paolo Facchi$^{1,2}$, Giovanni Gramegna$^{1,2}$ and Arturo Konderak$^3$}
\address{$^1$Dipartimento di Fisica, Universit\`a di Bari, I-70126 Bari, Italy}
\address{$^2$INFN, Sezione di Bari, I-70126 Bari, Italy}
\address{$^3$Center for Quantum-Enabled Computing, Center for Theoretical Physics, Polish Academy of Sciences, Aleja Lotnik\'{o}w 32/46, 02-688 Warsaw, Poland}
\ead{daniele.amato@uniba.it}
\begin{abstract}
Quantum channels provide the mathematical framework for describing the evolution of open quantum systems, i.e., quantum systems interacting with an environment. This review focuses on the long-time or asymptotic dynamics of open quantum systems, a topic of growing interest due to its relevance for quantum information theory and noisy intermediate-scale quantum (NISQ) devices. We present a comprehensive analysis of homogeneous evolutions in both discrete and continuous time, treating the dynamics in full generality within the Schr{\"o}dinger and Heisenberg pictures. By systematically studying the spectral properties of quantum channels, we present general structure theorems that characterize asymptotic behaviour, illustrated through several paradigmatic examples. We further discuss how the asymptotic dynamics can be exploited for dissipative state engineering, the identification of decoherence-free subspaces, the design of quantum error-correction schemes, and the evaluation of the asymptotic capacities of quantum channels.
\end{abstract}

\tableofcontents

\markboth{}{}

\section{Introduction}
\label{histo_sec}
\subsection{Historical overview and motivation}

The unitary evolution of a quantum system, governed by the Schr{\"o}dinger equation, describes the dynamics of a \emph{closed} or \emph{isolated} quantum system. Many physical phenomena, however, require a description that accounts for interactions with external degrees of freedom. Examples include the decay of an excited atom to its ground state and the decoherence effects affecting the position of a particle coupled to a thermal environment. Such processes are naturally described within the framework of \emph{open quantum systems}, in which the system of interest interacts with its surroundings, usually called the \emph{environment},  \emph{reservoir}, or  \emph{bath}. 
The  interaction between the system and its environment can lead to various phenomena, such as decoherence, dissipation, as well as a possible backflow of information, giving rise to memory (or non-Markovian) effects.

An axiomatic description of the evolution of an open quantum system was first presented by Sudarshan \emph{et al}~\cite{SMR_61,Jordan_Sudarshan_61,Jordan_Sudarshan_62} and Kossakowski~\cite{kossa_1972,AK-1,AK-2}.
The missing ingredient in these early works, which laid the foundations of open-quantum-system dynamics, was the crucial notion of complete positivity, first introduced in the physics literature by Ludwig~\cite{Ludwig_68} and further developed in subsequent works~\cite{Hellwig_Kraus_68,Hellwig_Kraus_70,Doplicher_Haag_Roberts_71,Kraus_71}. In particular, Markovian evolutions received special attention at the dawn of the theory, due to their relevance in quantum optics and nuclear magnetic resonance~\cite{alickilendi}. This class of evolutions was characterized in the two seminal papers~\cite{GKS76,Lindblad76} and was shown to be governed by the Gorini--Kossakowski--Lindblad--Sudarshan \emph{(GKLS) master equation},  
which can be thought of as a generalization of the Schr{\"o}dinger equation to open quantum systems. 
From a probabilistic perspective, open-quantum-system theory can be regarded as the generalization of classical probability theory to the noncommutative setting, initiated by Accardi \emph{et al}~\cite{Accardi_Lewis_82,Accardi_Cecchini_82} 
and further developed by Hudson and Parthasarathy~\cite{Hudson_Parthasarathy_84} with the quantum stochastic calculus; see e.g.~\cite{Accardi_Alicki,Accardi_Lu_2000,Barchielli_91}. The reader is referred to the reviews~\cite{fagnola_flows,Saverio_Darek_review,Chr_review_2022,benatti2026fifty} for more detailed historical accounts. 

\begin{figure}
    \begin{center}
    \includegraphics[width=\textwidth]{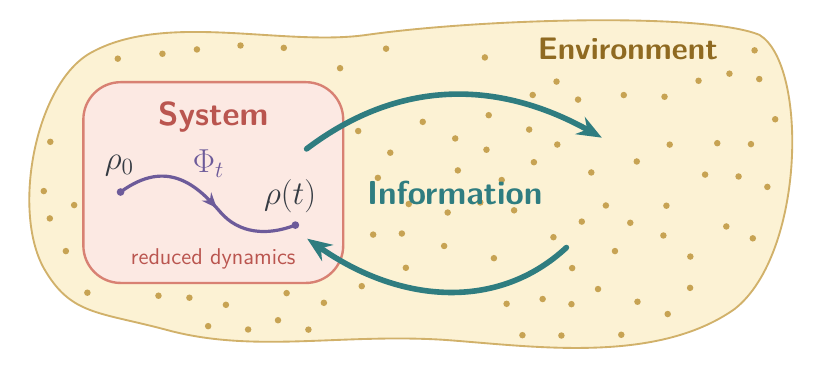}
    \end{center}
    \caption{Schematic representation of an open quantum system: a system (light red box) interacts with the environment (light yellow patch), and its evolution over the time interval $[0,t]$ describes a trajectory in the state space (solid purple line). The teal arrows represent the two-way flow of information between system and environment, possibly leading to memory effects.} 
\label{fig:representation_open_system}
\end{figure}

This review article is dedicated to the analysis of the \emph{long-time evolution} or \emph{asymptotics} of open quantum systems. Historically, this topic has drawn the attention of the physics community since the birth of open-quantum-system theory. Indeed, one of the main motivations behind the study of this field is the notion of irreversibility and approach to equilibrium~\cite{kossa_inga_1975}, which lies at the heart of thermodynamics and statistical mechanics~\cite{huang}. 
In the quantum realm, as shown by~\cite{davies1973harmonic,davies1974markovian}, open-quantum-system theory has been addressed even before the characterization of quantum dynamical semigroups in~\cite{GKS76,Lindblad76}, which, however, paved the way for the formulation of general sufficient conditions for irreducibility and the existence of a unique equilibrium state, called the relaxing property~\cite{spohn1976approach,evans1977_irred,Spohn_77,Frigerio_78,Frigerio_Verri_82}; see~\cite{spohn1980kinetic,nigro2019uniqueness} for historical accounts. Moreover, irreducible and primitive (completely) positive maps, see~\cite{wolf2012quantum}, received much interest since the seventies, as they provide the natural quantum generalization of irreducible and primitive stochastic matrices~\cite{evanskrohn,albeverio,groh_irre,groh_1982,groh_1984,farenick}. In particular, one of the main interests was the noncommutative extension of the Frobenius-Perron theory~\cite{perron,frobenius}, providing information on the peripheral eigensystem of stochastic matrices. 

These initial efforts may be regarded as the first step towards the study of the asymptotics in the discrete-time dynamical setting. The interest in open-quantum-system asymptotics is also related to the quantum-to-classical transition~\cite{schlosshauer2007decoherence}, namely how classical features emerge from quantum systems. Besides the approach to the problem based on the semiclassical limit $\hbar \rightarrow 0$, classicality may also arise as a result of decoherence of quantum systems in the long-time limit~\cite{blanchard1993interaction,blanchard2003decoherence,chruscinski2012observables,alipour2017dynamical}. 
In this context, from an algebraic viewpoint, it is then natural to introduce, in some appropriate way, the contracted algebra of classical observables arising from the original quantum algebra in the asymptotic regime of the evolution.

The fundamental interest in the asymptotics of open quantum systems has then been supplemented by the relation of this topic to various applications in quantum information processing~\cite{zanardi1997noiseless,lidar_dec_1998,Lidar2003,choi2006method}, quantum state engineering~\cite{Zoller_res_eng,Verstraete2009,lin2013dissipative}, quantum error correction~\cite{choi2006method,ng2010simple,rahaman2017multiplicative}, quantum data classification~\cite{marshall2019classifying}, dissipative quantum computation~\cite{Verstraete2009}, quantum thermodynamics~\cite{brask2015autonomous,tacchino2018steady}, and quantum state synchronization~\cite{bucaalgebraic2022,vstverba2023asymptotic}. 
Thus, open-quantum-system asymptotics has continued to receive a lot of attention both from the quantum physics~\cite{albert2014symmetries,albert2016geometry,albert2019asymptotics,AFK_asympt_1,AFK_asympt_2,AFK_asympt3,AFK_asympt4,AF_bounds1,wolf2010inverse,jex_st_2012,jex_st_2018} and quantum probability communities~\cite{carbone2013decoherence,deschamps2016structure,fagnola2017,fagnola2019role,carbone2020period,fagnola2024basins,carbone2025fixed,carbone2021absorption,Sasso2023} over the last years, proving that this is a highly active field.

Physically, the asymptotic subspace of an open-system dynamics encompasses both stationary (or steady) states, which remain invariant under the evolution, and oscillatory states, characterized by persistent oscillatory behaviour determined by the peripheral eigenvalues of the dynamics. On the one hand, the stability of stationary states is the basis of quantum reservoir engineering or data classification, aimed at creating a desired quantum state or categorizing data by relaxation, respectively, as well as of noiseless codes at the heart of passive quantum error correction. On the other hand, the absence of noise and decoherence characterizing the asymptotic evolution unearthed the possibility of performing unitary gates within 
decoherence-free subspaces. In particular, active quantum error correction may be performed by applying recovery channels in order to restore a corrupted state.
Many of the  applications  mentioned above will be discussed in some depth from a theoretical perspective. 
\subsection{Outline}
The present work provides an updated overview of the long-time dynamics of open quantum systems, bringing together perspectives from quantum physics, quantum probability, mathematical physics, and quantum information theory. By connecting mathematical results with physical questions and applications, the review aims to serve as a resource for researchers interested in the foundations of open-system dynamics, as well as in the control and use of dissipation in quantum technologies.

First, we shall give a brief summary of the main results characterizing the general structure of the asymptotic evolution. The analysis will be carried out in parallel within two distinct, yet related, dynamical settings: discrete-time Quantum Markov Chains (QMCs) and continuous-time Quantum Dynamical Semigroups (QDSs). 
Second, we shall discuss the role of open-quantum-system asymptotics in various quantum information applications, with a primary focus on theoretical developments. Throughout this review, we will use the term ``asymptotic" only to refer to the long-time limit of the dynamics. Below is a detailed outline of the topics discussed throughout this work. 
\begin{itemize}
    \item In Section~\ref{qc_sec} we start by discussing the basics of the time-homogeneous dynamics of open quantum systems in the discrete-time setting, described by QMCs. In particular, we will introduce quantum channels and Unital Completely Positive (UCP) maps, which describe the unit-time 
    evolutions in the Schr{\"o}dinger and Heisenberg pictures, respectively. Then, in Section~\ref{repr_qc_sec} we shall recap the Stinespring and Kraus--Sudarshan (KS) representations as well as the Choi--Jamio{\l}kowski isomorphism of quantum channels and UCP maps. In Section~\ref{spec-qc-subsec} we  will recall the basic spectral properties of quantum channels and, similarly, of UCP maps. This analysis will allow us to introduce the attractor and steady subspaces of discrete-time Schr{\"o}dinger and Heisenberg evolutions, used to describe the asymptotic dynamics in Section~\ref{sec:asymptotics}. 
    \item In Section~\ref{QDS_sec} we will present the notion of a QDS, describing the continuous-time evolution of an open quantum system in the Markovian case. 
    In particular, we shall discuss the generator of the dynamics, which may be characterized through the GKLS representation in terms of Hamiltonian and dissipative parts. Then, we will overview the spectral properties of GKLS generators in Section~\ref{spec_lindbladian}, later employed in Section~\ref{asympt_QDS_sec} for the study of the asymptotic behaviour of QDSs.
    \item In Section~\ref{sec:asymptotics} we will focus on the asymptotic dynamics of QMCs. As discussed in Section~\ref{bas_notions_subsec},  the part of the spectrum relevant for our purpose corresponds to the unimodular eigenvalues of the dynamics, characterizing its attractor subspace. In Section~\ref{sec:examples}, the long-time dynamics of QMCs will be illustrated through several paradigmatic examples. In particular, these examples unveil the building blocks of the asymptotic dynamics of arbitrary QMCs, presented in Section~\ref{sec:structure_theorems} both in the Schr{\"o}dinger and Heisenberg pictures, see Theorems~\ref{Wolf_struc} and \ref{struc_Heis_th}, respectively. Finally, the last part of Section~\ref{sec:structure_theorems} is devoted to an alternative expression for the asymptotics in the Schr{\"o}dinger picture, also showing the relation between attractor and steady subspaces of the evolution, cf.\ Theorem~\ref{thm:cycles}.
    \item In Section~\ref{asympt_QDS_sec} we will discuss the asymptotic dynamics of quantum dynamical semigroups along the lines of the analysis done in Section~\ref{sec:asymptotics} for QMCs. In this case, attractor and steady subspaces are characterized by the peripheral spectrum of the GKLS generator of the dynamics. In particular, after the basic notions discussed in Section~\ref{subsec:asympt_QDS_1}, the structure of the asymptotic subspace will be presented in Section~\ref{subsec:asympt_QDS_2}.
       \item In Section~\ref{protecting_unitary_evolution} we explore how to employ the asymptotic dynamics for noise-free information processing and storage. In Section~\ref{unit_asympt_sec}, we identify the conditions under which the long-time dynamics becomes unitary. Section~\ref{sec:dfs} then turns to decoherence-free subspaces, where the evolution is unitary and therefore protected from decay and decoherence. Finally, Section~\ref{QEC_sec} recalls the closely related notions of noiseless and correctable subsystems.
    \item In Section~\ref{sec:asymptotic_bounds} we will detail several universal bounds on the dimensions of the attractor and steady subspaces, both for quantum channels (Section~\ref{qc_bounds_subsec}) and GKLS generators (Section~\ref{GKLS_bounds-subsec}), and we will also introduce the important classes of ergodic (mixing) QMCs and QDSs, exhibiting only one steady (asymptotic) state.
    \item In Section~\ref{conv_sec} we shall provide insights into the transient properties of QDSs, by looking at their convergence towards the asymptotic subspace from different viewpoints. Specifically, we will recall and characterize the domains of attraction of steady states in Section~\ref{dom_attr_subsec}, then we will address the related problem of engineering a desired state by relaxation in Section~\ref{QRE_subsec}, and we shall present several general results about the rates of convergence of QDSs in Section~\ref{sec:ckks_bound}.
    \item In Section~\ref{sec:symmetries}, we will explore the role of symmetries in the structure of the asymptotic evolution. Specifically, we relate the symmetries of the dynamics to decoherence-free subspaces and noiseless subsystems in Section~\ref{sec:sym_DFS}, then we connect symmetries and conserved quantities, yielding a generalization of Noether’s theorem to open quantum systems, and we show how to exploit conserved quantities to study long-time dynamics in Section~\ref{sec:Noether}.
    \item In Section~\ref{asympt_transm_sec} we will discuss how much classical and quantum information is protected by the action of a channel in the long-time limit, as quantified by its asymptotic classical and quantum capacities introduced and characterized in Section~\ref{asympt_cap_subsec}. In particular, quantum information is completely destroyed in the long-time limit by the class of asymptotically entanglement breaking channels, recalled together with similar concepts in Section~\ref{EB_subsec}.
        \item In Section~\ref{sec:alg_approach} we will address the asymptotic properties of QMCs and QDSs in the Heisenberg picture within the algebraic framework. The central object of the analysis, called the decoherence-free algebra, will be studied both for QMCs and QDSs in Section~\ref{sec:dfa}, while we will introduce the Choi--Effros product, see Equation~\eqref{CE_def}, in order to equip the attractor subspace of QMCs with an algebraic structure and, consequently, unveil the algebraic properties of the asymptotic evolution of QMCs in Section~\ref{alg_sec}.
    \item In Section~\ref{other_top_sec} we will briefly mention other topics related to open-quantum-system asymptotics not developed in the review: quantum associative memories in Section~\ref{sec:qam}, non-Markovianity in the asymptotic regime in Section~\ref{sec:asympt_non_Markov}, and a quick overview of numerical methods for steady states in Section~\ref{sec:numerical_methods}.
    \item Finally, the work includes four appendices organized as follows: in Appendix~\ref{classic_sec} we will recall the basics of Markov chains and semigroups, providing the classical counterparts of QMCs and QDSs studied in the core part of the review, in Appendix~\ref{Jordan_op_app} we will recap the Jordan decomposition of (linear) operators, which is then used to prove Proposition~\ref{trace_eigen_prop} about the properties of the generalized eigenoperators of quantum channels, in Appendix~\ref{intro_Cast} we will provide a concise overview on $C^\ast$-algebras and related notions, and in Appendix~\ref{sec:symbols} we will summarize the main symbols, the abbreviations, and the conventions used in this review. 
\end{itemize}

\section{Quantum channels and quantum Markov chains}
\label{qc_sec}
In this section, we shall overview the dynamics of \emph{open quantum systems} in the discrete-time setting~\cite{Heinosaari,watrous2018theory}; we shall defer the discussion of homogeneous continuous-time evolutions to Section~\ref{QDS_sec}.

In the standard description of quantum mechanics, a quantum system is described by a Hilbert space $\mathcal{H}$, which
 will be assumed to be finite-dimensional with dimension $d$, and a state of the system is represented by a \emph{density operator} $\rho$, which is a positive-semidefinite operator on $\mathcal{H}$ with unit trace. We will denote with $\mathcal{S}(\mathcal{H})$  the convex and compact set of density operators on $\mathcal{H}$ and with $\Bo(\Hi)$ the set of bounded linear operators on $\Hi$.

A density operator is said to be \emph{pure} if it is in the form $\rho = \ketbra{\psi}$, with $\ket{\psi} \in \mathcal{H}$ $\norm{\psi} = 1$ being a \emph{vector state}, and \emph{mixed} otherwise. 
Any density operator may be written as 
a convex combination of pure density operators 
\begin{equation}
\label{pure_state_dec}
\rho = \sum_{i\in[N]} p_i \ketbra{\psi_i},\quad\sum_{i\in[N]} p_i=1,\quad  p_i>0 \text{ for all }i\in[N], 
\end{equation}
where we use the  notation 
$[N]:=\{1,2,\dots,N\}$,   
with $N$ being a positive integer. The decomposition~\eqref{pure_state_dec}
can be interpreted as a statistical mixture: the state of the system is in the vector state  $\ket{\psi_i}$ with probability $p_i$. 
It is well known that the decomposition~\eqref{pure_state_dec} is not unique~\cite{Schrodinger_1935_ent}. However, there is a distinguished decomposition of $\rho$ which is obtained when $\{p_i\}$ and $\{\ket{\psi_i}\}$ are respectively the eigenvalues and the corresponding eigenvectors of $\rho$: in this case $N\leq d$ and \eqref{pure_state_dec} 
is the spectral decomposition of $\rho$.

On the other hand, an observable is represented by a Hermitian operator $O \in \mathcal{B}(\mathcal{H})$, and a duality between states and observables is encoded by the \emph{Hilbert--Schmidt scalar product} defined as
\begin{equation}
\label{HS_sp}
\braket{X}{Y}_{2} := \Tr(X^\ast Y), \quad X,Y \in \mathcal{B}(\mathcal{H}).
\end{equation}
Explicitly, the pairing $\braket{\rho}{O}_2=\Tr(\rho O)$ represents the expectation value of the observable $O=O^*$ over the state $\rho$.

The introduction of mixed states is particularly relevant for composite systems. Given two systems $A$ and $B$ described by the Hilbert spaces $\mathcal{H}_{A}$ and $\mathcal{H}_{B}$, the composite system $A+B$ is described by the tensor product Hilbert space $\mathcal{H}_{A} \otimes \mathcal{H}_{B}$. A state of the system is represented by a density operator $\rho_{AB}$ on $\mathcal{H}_{A}\otimes \mathcal{H}_{B}$. From the perspective of an observer who only has access to observables $X_A$ of system $A$ (and similarly for $B$), all expectation values can be obtained as
\begin{equation}\label{eq:PartialTrace}
    \Tr[(X_A\otimes \II_{B})\rho_{AB}]=\Tr[X_A\Tr_B(\rho_{AB})],
\end{equation}
where $\II_{B}$ denotes the identity on $\mathcal{H}_{B}$.
Thus, the state of the system is represented by the~\emph{reduced state} $\rho_A = \Tr_B(\rho_{AB})$, where $\Tr_B$ denotes the \emph{partial trace} over $\mathcal H_B$, which is the unique linear operator from $\Bo(\Hi_A\otimes\Hi_B)$ to $\Bo(\Hi_A)$ satisfying \eqref{eq:PartialTrace}. Equivalently, $\Tr_B$ is the unique linear extension of the map acting as $\Tr_B(X_A\otimes X_B)=X_A\Tr(X_B)$ to all the operators in $\Bo(\Hi_A\otimes\Hi_B)$~\cite{watrous2018theory}.
Interestingly, even if the global state $\rho_{AB}$ is pure, the reduced state $\rho_A$ is not generally so because of entanglement, which will be recalled later in this section. 

In the spirit of~\cite{SMR_61,Jordan_Sudarshan_61,Jordan_Sudarshan_62}, let us now describe the dynamics of an open quantum system in an axiomatic way, namely by a mapping between density operators $\rho \mapsto \Phi(\rho)$. In order to avoid paradoxical consequences, such as superluminal communication~\cite[Subsection 3.2.5]{preskillnotes}, we require $\Phi$ to preserve convex combinations, i.e.\ 
\begin{equation}
\label{pres_conv_comb_proper}
\Phi\Bigl(\sum_{i\in[N]} p_{i} \rho_i\Bigr) = \sum_{i\in[N]} p_i \Phi(\rho_i),\quad \rho_i \in \mathcal{S}(\mathcal{H}), \; p_i>0,\;\sum_{i\in[N]}p_i=1. 
\end{equation}
Therefore, the ensemble of $N$ states $\rho_i$ with probabilities $p_i$ is mapped into the ensemble of $N$ states $\Phi(\rho_i)$ again weighted with the initial probabilities $p_i$, implying the statistical equivalence of the two ensembles under the action of the dynamics $\Phi$. Importantly, any map on the state space $\mathcal{S}(\mathcal{H})$ satisfying~\eqref{pres_conv_comb_proper} may be uniquely extended to a linear map over the space $\mathcal{B}(\mathcal{H})$ of bounded operators on $\mathcal{H}$. From now on, we will only take into account linear maps on $\mathcal{B}(\mathcal{H})$. 

 Let us now recall that a map $\Phi$ on $\mathcal{B}(\mathcal{H})$ is called \emph{positive}(-preserving) if $\Phi(X)\geq 0$ whenever $X \geq 0$, and \emph{trace-preserving} if $\Tr(\Phi(X))=\Tr(X)$ for all $X\in\mathcal{B}(\mathcal{H})$. Since $\Phi$ must map states into states, we assume $\Phi$ to be a positive trace-preserving map on $\mathcal{B}(\mathcal{H})$. Note that any positive map is also \emph{Hermiticity-preserving}, namely $\Phi(X)^\ast = \Phi(X^\ast)$ for all $X \in \mathcal{B}(\mathcal{H})$, where $X^\ast$ denotes the adjoint of $X$.

A positive trace-preserving map on $\mathcal{B}(\mathcal{H})$ is not sufficient to properly describe the evolution of a quantum system. Indeed, if $\Phi_A$ and $\Phi_B$ are two positive trace-preserving maps on $\mathcal{B}(\mathcal{H}_{A})$ and $\mathcal{B}(\mathcal{H}_{B})$, respectively, then the tensor product map
\begin{equation}\label{eq:tensor_product_maps}
\Phi_A \otimes \Phi_B : \mathcal{B}(\mathcal{H}_{A} \otimes \mathcal{H}_{B}) \rightarrow \mathcal{B}(\mathcal{H}_{A} \otimes \mathcal{H}_{B})
\end{equation} 
is trace-preserving but is not positive in general. To be more precise, all \emph{separable} states, which are states that can be written as a probabilistic mixture of tensor products of states of A and B
\begin{equation}
\rho_{AB}=\sum_{i\in[N]} p_i \rho_{i}^{(A)} \otimes \rho_{i}^{(B)}, \quad \sum_{i\in[N]} p_i=1,\quad 
\rho_{i}^{(j)} \in \mathcal{S}(\mathcal{H}_{j}), \; p_i>0,\;
\; j\in\{A,B\}, \; i \in [N], 
\end{equation}
are mapped into (separable) states by the evolution~\eqref{eq:tensor_product_maps}, while this is not true in general for non-separable or \emph{entangled} states.
Entanglement represents purely quantum correlations between two (or more) systems and, according to Schr{\"o}dinger, is ``not one but rather the characteristic trait of quantum mechanics"~\cite{Schrodinger_1935_ent}. Therefore, we need to impose a stronger positivity condition on the dynamics $\Phi$. To express formally this property, let us introduce the $n$-th \emph{ampliation map} of $\Phi$:
\begin{equation}
\label{ampl_map}
\Phi^{(n)} := \Phi \otimes \mathsf{1}_n : \mathcal{B}(\mathcal{H}) \otimes  \mathcal{M}_{n}(\mathbb{C}) \rightarrow \mathcal{B}(\mathcal{H}) \otimes \mathcal{M}_{n}(\mathbb{C}),
\end{equation} 
where $\mathsf{1}_n$ is the identity operator on $\mathcal{M}_{n}(\mathbb{C}) = \mathcal{B}(\mathbb{C}^n)$, the space of complex matrices of order $n$. We can now recall the notion of \emph{complete positivity}.
\begin{definition}[Complete positivity]
\label{CP_def}
A positive map $\Phi$ on $\mathcal{B}(\mathcal{H})$ is said to be completely positive if $\Phi^{(n)}$ is positive for all $n \geq 1$.
\end{definition}
Importantly, a positive map $\Phi:\Bo(\Hi)\rightarrow\Bo(\Hi)$ is completely positive if and only if $\Phi^{(d)}$ is positive~\cite{Choi_1972}, with $d=\dim(\Hi)$.
If $\Phi_A$ and $\Phi_B$ are completely positive trace-preserving maps on $\mathcal{B}(\mathcal{H}_A)$ and $\mathcal{B}(\mathcal{H}_{B})$, respectively, then so is $\Phi_A\otimes \Phi_B$. 

In summary, the evolution of an open quantum system is described by a \emph{quantum channel}, as defined below.
\begin{definition}[Quantum channel]
\label{qc_def}
    A quantum channel is a \emph{\gls*{cptp}} map on $\mathcal B(\mathcal H)$.
\end{definition}

Note that the set of quantum channels on $\mathcal{B}(\mathcal{H})$ is closed under composition of maps, henceforth denoted with $\circ$. Furthermore, it is also convex and compact, as we have for the state space $\mathcal{S}(\mathcal{H})$. 

A quantum channel induces the evolution of an open quantum system in the \emph{Schrödinger picture}, as it describes how states change in time. Alternatively, it is possible to consider the evolution of observables, called \emph{Heisenberg dynamics}. The evolution $\Psi$ in the Heisenberg picture is obtained from the Schr{\"o}dinger dynamics $\Phi$ by requiring the equality of the expectation values
\begin{equation}
\Tr(\Phi(\rho)O) = \Tr(\rho \Psi(O)),\quad \forall \rho \in \mathcal{S}(\mathcal{H}), \,O=O^\ast \in \mathcal{B}(\mathcal{H}),
\end{equation}
and is therefore given by the Hilbert--Schmidt adjoint $\Phi^\dagger$  of the quantum channel $\Phi$. 
By definition, it directly follows that this is a \emph{\gls*{ucp}} map on $\mathcal{B}(\mathcal{H})$, where unitality refers to the property $\Phi^\dagger(\mathbb I)=\mathbb I$, with $\mathbb I$ being the identity over $\mathcal H$~\cite[Proposition 2]{Chr_review_2022}. 
\begin{remark}
Unless explicitly stated, we denote with $\Phi$ a quantum channel, and with $\Phi^\dagger$ a UCP map.  
\end{remark}

In the discrete-time setting~\cite{NielsenChuang} a quantum channel $\Phi:\mathcal B(\mathcal H)\rightarrow \mathcal B(\mathcal H)$ describes the unit-time evolution of the system in the Schr{\"o}dinger picture.
Under the 
homogeneity assumption, the evolution in the time interval $[0,n]$, with $n\geq 1$, is given by the $n$-fold composition 
\begin{equation}
    \Phi^n = \Phi \circ \Phi \circ \dots \circ \Phi
\end{equation}
 of $\Phi$. 
 \begin{figure}
    \centering
    \includegraphics[width=1\textwidth]{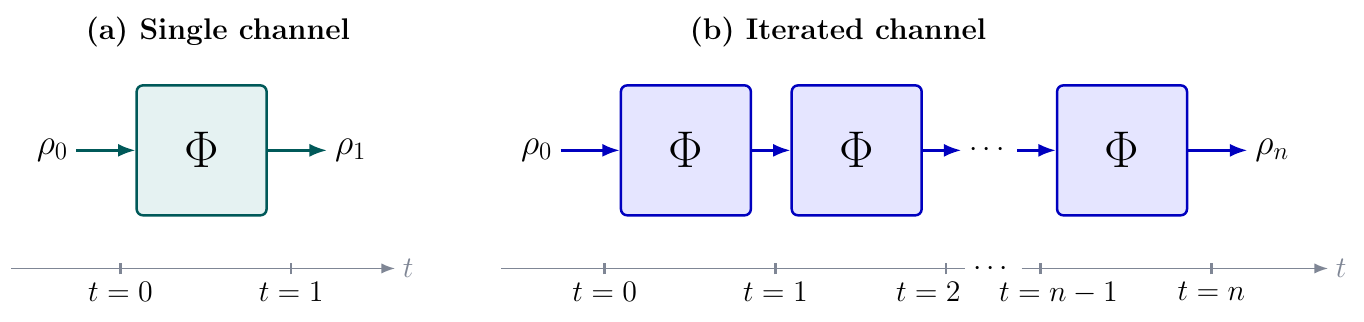}
    \caption{Representation of (a) a quantum channel $\Phi$ (Definition~\ref{qc_def}), and (b) the quantum Markov chain $\Sem = (\Phi^n)_{n \in \mathbb{N}}$ generated by $\Phi$. The initial state and the evolved one after one and $n$ time steps are denoted by $\rho_0$, $\rho_1$, and $\rho_n$, respectively.}
    \label{fig:repeated_channel}
\end{figure}
 Explicitly, if $\rho_0 = \rho(0)$ is the state of the system at the initial time $t=0$, then the state at time $t=n$ 
 is given by
 \begin{equation}
     \rho_0 \mapsto \rho_n=\Phi^n(\rho_0).
 \end{equation}
 In the jargon of  
 probability theory, this implements a (discrete-time) \emph{\gls*{qmc}} of quantum channels~\cite{jex_st_2012}, generating the transitions $\rho_0 \mapsto \rho_1 \mapsto \dots \mapsto \rho_n \mapsto \dots$, and it is also referred to as \emph{discrete-time quantum Markov semigroup}~\cite{singh2024zero,accardi1981topics,gudder2008quantum}. We will call $\Sem := (\Phi^n)_{n\in\mathbb N}$ the QMC generated by the quantum channel $\Phi$, see Figure~\ref{fig:repeated_channel} for its schematic visualization. In line with the adopted terminology, discrete-time Markov chains, briefly recalled in Appendix~\ref{classic_sec}, are the classical analogue of quantum ones and, in particular, a quantum channel is the quantum counterpart of a column-stochastic matrix, cf.\ Equations~\eqref{stoch-pro_1}--\eqref{stoch-pro_2}. Analogously, a UCP map $\Phi^\dagger$ induces a homogeneous Heisenberg dynamics of the system described by a discrete-time semigroup $\Sem^\dagger := ((\Phi^{\dagger})^n)_{n \in \mathbb{N}}$.  
As a final remark, note that the homogeneity assumption greatly simplifies the analysis of the long-time dynamics, as we shall see in Section~\ref{sec:asymptotics}; we refer the reader to~\cite{frigerio1993asymptotic,movassagh2021theory,movassagh2022ergodic,li_expo_2024,yoshida2026theory} for the asymptotics of non-homogeneous discrete-time evolutions.

\subsection{Representations of quantum channels}
\label{repr_qc_sec}
A quantum channel (Definition~\ref{qc_def}) can be explicitly represented in various ways, each of which sheds light on different properties of this class of maps. We will now recall three of these representations. The first one follows from a seminal paper by Stinespring~\cite{Stin_1955}, which provides an abstract structure theorem 
for completely positive maps~\cite[Subsection 3.1]{Chr_review_2022}.
\begin{theorem}[Stinespring representation theorem]\label{th:stinespring_dilatation}
Let $\Phi$ be a quantum channel on $\mathcal{B}(\mathcal{H})$. Then there exists a Hilbert space $\mathcal{H}_E$, a unitary operator $U$ on $\mathcal{H} \otimes \mathcal{H}_E$, and a vector state $\ket{\psi_E} \in \mathcal{H}_{E}$ such that
\begin{equation}
    \label{Stin_Schr}
    \Phi(X)=\Tr_E \bigl(U(X \otimes \ketbra{\psi_E} )U^\ast\bigr), \quad X \in \mathcal{B}(\mathcal{H}),
\end{equation} 
and
\begin{equation}
    \label{Stin_Heis}
    \Phi^\dagger(X)=\Tr_E\bigl((\mathbb{I}\otimes\ketbra{\psi_E})\,U^\ast(X \otimes \mathbb{I}_E )U\bigr),\quad X \in \mathcal{B}(\mathcal{H}),
\end{equation}
where $\mathbb{I}_E$ and $\Tr_{E}$ stand for the identity and the partial trace over $\mathcal{H}_E$. Equation~\eqref{Stin_Schr} is called \emph{Stinespring's representation} of $\Phi$.
\end{theorem}

The representation~\eqref{Stin_Schr} is also known as the Stinespring dilation of $\Phi$ because it shows that any quantum channel may be regarded as the result of the partial trace over the environment degrees of freedom of a unitary, i.e.\ closed-system, evolution, cf.\ Section~\ref{sec:examples}.  In fact, this representation~\eqref{Stin_Schr} is the starting point for the microscopic derivation of the GKLS master equation~\cite{davies1974markovian,spohn1980kinetic,TASAKI2007631,YUASA2007657,factormix}, which governs Markovian dynamics (see Section~\ref{QDS_sec}). We refer the reader to~\cite{BreuerPetruccione,alickilendi,RivasHuelga2012,Chr_review_2022} for modern introductions to the topic.
It is worth noticing that the Stinespring representation of a quantum channel $\Phi$ is not unique. In particular, the dimension of the ancillary Hilbert space $\mathcal{H}_E$ is not fixed, and a dilation is \emph{minimal} if $\mathcal{H}_E$ has the smallest possible dimension. Furthermore, once $\mathcal{H}_E$ is fixed, the unitary operator $U$ and the environment state $\ket{\psi_E}$ are not uniquely defined, since there is still a residual local unitary freedom on the environment~\cite[Subsection 4.2.2]{Heinosaari}.

A second representation of quantum channels, called the operator-sum or \gls*{ks} representation~\cite{SMR_61,Kraus_71}, allows one to write the action of a quantum channel in terms of operators acting on $\mathcal{H}$.

\begin{theorem}[Kraus--Sudarshan representation theorem]
\label{KS_theorem}
Let $\Phi$ be a quantum channel on $\mathcal B(\mathcal H)$. Then, $\Phi$ and its 
adjoint $\Phi^\dagger$ can be written as
\begin{equation}
    \label{op_sum_rep}
     \Phi(X)=\sum_{i\in[N]} A_i XA_i^*,\quad \Phi^\dagger(X)=\sum_{i\in[N]} A_i^\ast X A_i,\quad X\in\mathcal B(\mathcal H),
\end{equation}
where 
$\{ A_i \}_{i\in[N]}\subset \mathcal{B}(\mathcal{H})$
are called Kraus--Sudarshan (\emph{KS}) \emph{operators} of $\Phi$, and  satisfy
\begin{equation}
    \label{norm_con}
    \sum_{i\in[N]}A_i^\ast A_i = \mathbb{I}.    
\end{equation}
\end{theorem}
The representations~\eqref{op_sum_rep} of $\Phi$ and $\Phi^\dagger$ are valid for any completely positive map, while condition~\eqref{norm_con} is a consequence of the trace-preservation property of $\Phi$ or, equivalently, the unitality condition for $\Phi^\dagger$.
Moreover, notice that Theorem~\ref{KS_theorem} is the special case of a more general decomposition of any linear map $\Phi$ on $\mathcal{B}(\mathcal{H})$
\begin{equation}
    \label{oper_sum_rep_gen}
    \Phi(X)=\sum_{i\in[N]}A_i X B_i,\quad X \in \mathcal{B}(\mathcal{H}),
\end{equation}
with 
$\{ A_i \}_{i\in[N]} , \{ B_i \}_{i\in[N]} \subset \mathcal{B}(\mathcal{H})$, which always holds as long as $\mathcal H$ is finite-dimensional. For infinite-dimensional Hilbert spaces, this decomposition exists for a subclass of linear maps on $\mathcal{B}(\mathcal{H})$ called elementary operators~\cite{mathieu1989elementary}.
As with the Stinespring representation, the operator-sum representation is not unique, and different representations are related to one another~\cite[Remark 4]{Choi1975}.  
\begin{proposition}[\hspace{-1pt}{\cite[Proposition 4.25]{Heinosaari}}]
\label{freed_Kraus_prop}
Let $\Phi$ be a quantum channel with KS operators 
$\{ A_k \}_{k\in[N]}$. A collection 
$\{ A'_k \}_{k\in[N']}$ forms another KS representation of $\Phi$  if and only if there exists a unitary matrix $(U_{jk})\in \mathcal M_T(\mathbb C)$ such that 
\begin{equation}
    A'_j = \sum_{k\in[T]} U_{jk} A_k, \quad j\in[T], 
\end{equation}
with $T=\max(N,N')$. If $N \neq N^\prime$, the shorter collection is padded with zero operators so that it has length $T$.
\end{proposition}
The minimal number of operators in the KS decomposition is called the \emph{KS rank} of $\Phi$ and denoted with $\mathrm{rank}_{\mathrm{KS}}(\Phi)$.
\begin{figure}
    \centering
    \includegraphics[scale=1.2]{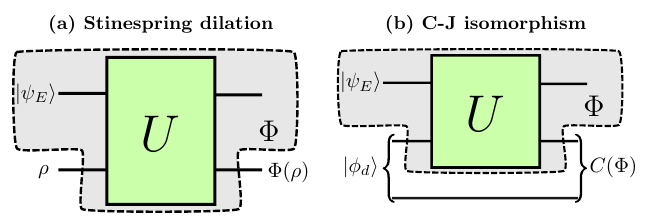}
    \caption{(a) Stinespring dilation. Any quantum channel $\Phi$ acting on $\rho$ can be realized through a unitary $U$ with an environment initialized in a pure state $\ket{\psi_{\mathrm E}}$, cf.\ Equation~\eqref{Stin_Schr}. (b) Choi--Jamio{\l}kowski isomorphism. The channel's action can be described by the trivial extension  $\Phi\otimes \mathsf{1}$ applied to the maximally entangled state $\ket{\phi_d}$ defined by~\eqref{max_ent}.} 
    \label{fig:representations_quantum_channel}
\end{figure}

The third representation is a particular one-to-one correspondence  between the space of linear maps on $\mathcal{B}(\mathcal{H})$ and the  Banach space $\mathcal{B}(\mathcal{H} \otimes \mathcal{H})$, called \emph{Choi--Jamio{\l}kowski isomorphism}~\cite{Choi1975,Jamiolko_1972}. Given a linear map $\Psi:\mathcal B(\mathcal H)\rightarrow \mathcal B(\mathcal H)$, and an orthonormal basis 
$\{ \ket{i} \}_{i\in[d]}$ of $\mathcal{H}$, the \emph{Choi operator} $C({\Psi})$ of $\Psi$ is defined as 
\begin{align}
    \label{CJ_iso}
    C({\Psi}) := (\Psi \otimes \mathsf{1})(\ketbra{\phi_d})=\frac{1}{d} \sum_{i,j\in[d]} \Psi(\ketbra{i}{j}) \otimes \ketbra{i}{j}\in\mathcal B(\mathcal H\otimes \mathcal H),
\end{align}
where 
\begin{equation}
    \label{max_ent}
    \ket{\phi_d}= \frac{1}{\sqrt{d}} \sum_{i\in[d]} \ket{i} \otimes \ket{i}
\end{equation}
is a maximally entangled state of $\mathcal{H} \otimes \mathcal{H}$.
Given a $\chi\in\mathcal B(\mathcal H\otimes \mathcal H)$, the Choi--Jamio{\l}kowski inverse $\Phi_\chi = C^{-1}(\chi):\mathcal B(\mathcal H)\rightarrow \mathcal B(\mathcal H)$ is given by
\begin{align}
    (C^{-1}(\chi))(X)=d \Tr_{2}(\chi(\mathbb I\otimes X^T)), \quad X\in\mathcal B(\mathcal H).
\end{align}
All the defining properties of quantum channels and UCP maps and, notably, complete positivity can be easily characterized via this isomorphism~\cite[Proposition 9]{Chr_review_2022}.

\begin{theorem}[Choi--Jamio{\l}kowski characterizations]
\label{Choi_charac}
Let $\Psi$ be a linear map on $\mathcal{B}(\mathcal{H})$ with Choi operator $C({\Psi})$. Then
\begin{itemize}
    \item[\textnormal{i.}] $\Psi$ is completely positive if and only if $C({\Psi}) \geq 0$;
    \item[\textnormal{ii.}] $\Psi$ is trace-preserving if and only if $\Tr_{1}(C({\Psi}))=\frac{\mathbb{I}}{d}$;
    \item[\textnormal{iii.}] $\Psi$ is unital if and only if $\Tr_{2}(C({\Psi}))=\frac{\mathbb{I}}{d}$.
\end{itemize}
\end{theorem}
Observe that the Choi--Jamio{\l}kowski isomorphism depends on the chosen basis 
$\{ \ket{i} \}_{i\in[d]}$ in $\mathcal{H}$, in contrast with 
the basis independent statements in Theorem~\ref{Choi_charac}. For a basis-independent generalization of the Choi--Jamio{\l}kowski isomorphism see~\cite{de_Pillis_1967,Chr_review_2022}. Note that Theorem~\ref{Choi_charac} implies a channel-state duality between quantum channels on $\mathcal{B}(\mathcal{H})$ and quantum states in $\St(\mathcal{H}\otimes \mathcal{H})$ satisfying $\Tr_{1}(C({\Phi}))=\mathbb{I} / d$~\cite{ben_zyc_geom}. Furthermore,  $\mathrm{rank}_{\mathrm{KS}}(\Phi)$ equals the rank of $C({\Phi})$~\cite[Theorem 2.1]{wolf2012quantum} and, consequently, $\mathrm{rank}_{\mathrm{KS}}(\Phi) \leq  d^2$.
Besides the extremely simple characterization of complete positivity, the Choi--Jamio{\l}kowski isomorphism is also employed to define superoperator analogues of various quantities initially defined for states~\cite{chruscinski2006partially}. See also~\cite{mendl2009unital,Horsman2017,saxena2020dynamical} for other applications of this isomorphism. 

\subsection{Spectral properties and asymptotic subspaces of quantum channels}
\label{spec-qc-subsec}

In this section we will look at the spectral properties of quantum channels, which have been studied analytically within the broader context of (completely) positive maps since the seventies~\cite{evanskrohn,albeverio}. Recently, random quantum channels have been also explored from the spectral viewpoint in various works~\cite{Bruzda2009,gonzalez2018spectral,kukulski2021generating}.  
Importantly, the spectrum of quantum channels is intimately related to the dynamics induced by them. 
In particular, the spectrum encodes both the decaying behaviour of the system, i.e.\ the finite-time transient regime, and the asymptotic limit of the evolution.
By definition, the spectrum $\spec(\Psi)$ of a linear map $\Psi$ on $\mathcal{B}(\mathcal{H})$ is  the set of its eigenvalues, namely
\begin{equation}
\label{spec_def}
    \spec(\Psi)=\{\lambda\in\mathbb C\,:\, \exists\ X\neq 0 \textnormal{ s.t. } \Psi(X)=\lambda X\}.
\end{equation}

One of the key properties of quantum channels 
follows by the  \emph{Russo-Dye theorem}~\cite{russodye_66}, recalled here for the case of positive maps~\cite[Corollary 2.3.8]{bhatia}.

\begin{proposition}[Russo-Dye theorem]
\label{spec_rad_pos}
Let $\Phi$ be a positive map on $\mathcal{B}(\mathcal{H})$, and let
\begin{equation}
\label{sp_rad_op_norm_defs}
   \rho(\Phi)=\max_{\lambda \in \spec (\Phi)} \lvert \lambda \rvert,\qquad  \| \Phi \| = \sup_{ \| X \| = 1} \| \Phi(X) \|
\end{equation}
be the spectral radius and the operator norm of $\Phi$, respectively, where $\| X \|$ stands for the operator norm of $\mathcal{B}(\mathcal{H})$, see~\eqref{op_norm_def}. 
Then, we have that $\rho(\Phi) \leq \| \Phi \| = \| \Phi(\mathbb{I}) \|$. 
\end{proposition}

Therefore, the operator norm and, consequently, the spectral radius of a positive map $\Phi$ are controlled by the action of the map on the identity $\mathbb{I}$, which provides a global constraint on the behaviour of $\Phi$ on the entire space $\mathcal{B}(\mathcal{H})$.

\begin{figure}
    \centering
    \includegraphics[width=\textwidth]{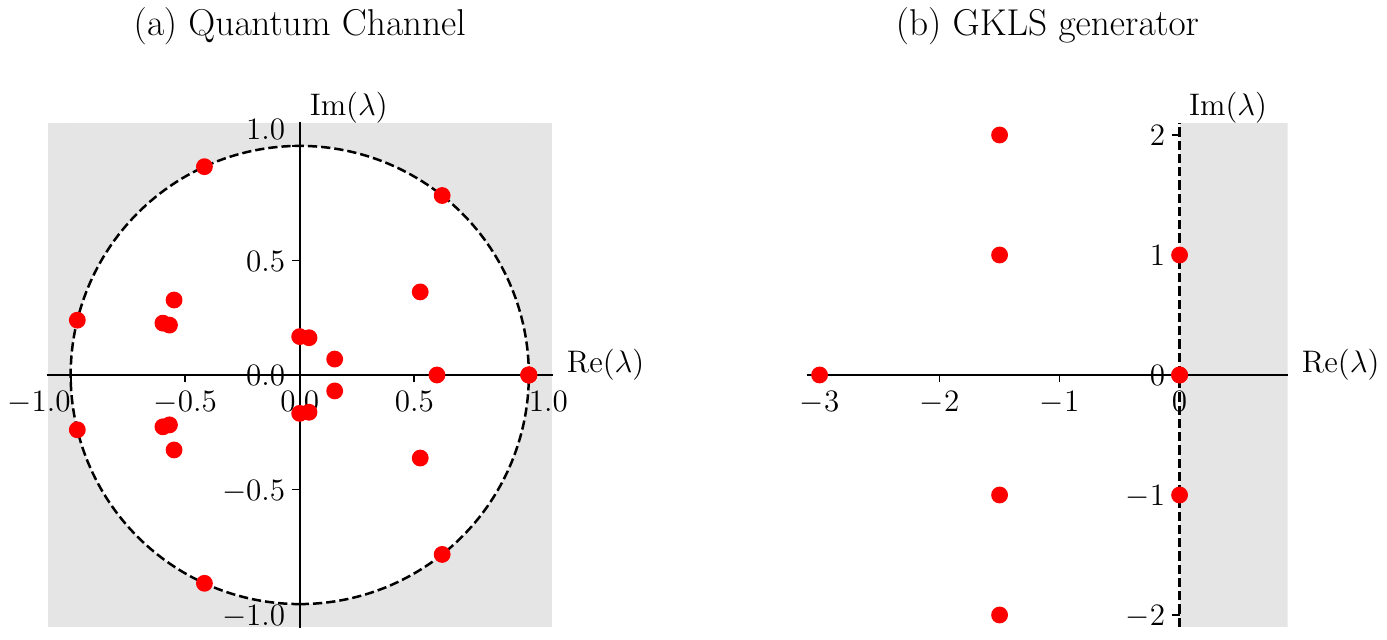}
    \caption{(a) Spectrum of a quantum channel $\Phi$. The general properties stated in Theorem~\ref{spec_qc} are shown in the figure: the spectrum
contains $\lambda_0=1$, is symmetric with respect to the real axis, and lies in the closed unit disk. The peripheral eigenvalues of $\Phi$, lying on the unit circle, marked by a dashed line, govern the long-time evolution of the QMC $\Sem$ induced by $\Phi$,
while the bulk eigenvalues of $\Phi$, lying inside the disk, describe the transient part of $\Sem$, which decays
at a rate controlled by $R(\Phi)$, cf.~\eqref{Rmax_def}. 
(b) Spectrum of a GKLS generator $\mathcal{L}$. The main spectral properties discussed in Theorem~\ref{spec_L} are visualized in the figure: the spectrum contains $\lambda_0 = 0$, is closed under complex conjugation, and lies in the closed left half-plane. The peripheral eigenvalues of $\mathcal{L}$, lying on the imaginary axis, marked by a dashed line, describe the long-time component of the QDS $\Sem$ generated by $\mathcal{L}$, while the bulk eigenvalues of $\mathcal{L}$, with negative real part, govern the transient regime of $\Sem$.}
 
    \label{fig:Spectrum}
\end{figure}
\begin{theorem}[Spectrum of quantum channels and UCP maps]
\label{spec_qc}
Let $\Phi$ be a quantum channel or a UCP map. Then, the spectrum of $\Phi$ satisfies 
\begin{itemize}
\item[\textnormal{i.}] $\spec(\Phi)=\spec(\Phi^\dagger)$;
\item[\textnormal{ii.}] $\lambda_0 = 1 \in \spec(\Phi)$;
\item[\textnormal{iii.}] $\lambda \in \spec (\Phi) \Rightarrow \overline{\lambda} \in \spec (\Phi)$;
\item[\textnormal{iv.}] $\spec (\Phi) \subseteq \{ \lambda \in \mathbb{C} \,:\,\lvert \lambda \rvert \leq 1 \}$.
\end{itemize}
These properties hold for any positive map that is also trace-preserving or unital. 
\end{theorem}
\begin{proof} Statement iii.\ follows from the fact that positivity implies Hermiticity preservation, namely $\Phi(X)^*=\Phi(X^*)$. Therefore, if $\lambda\in\C$ is an eigenvalue of $\Phi$ with eigenoperator $X$,  taking the adjoint of $\Phi(X)=\lambda X$ one gets $\Phi(X^*)=\bar{\lambda}X^*$, namely $\bar{\lambda}$ is an eigenvalue of $\Phi$ with eigenoperator $X^*$. Furthermore, since $\spec(\Phi^\dagger)=\overline{\spec(\Phi)}$, statement i.\ follows from statement iii. For a unital positive map $\Phi^\dagger$, statement ii.\ is a trivial consequence of $\Phi^\dagger(\II)=\II$, while for a trace preserving positive map $\Phi$ it follows from the fact that $\Phi^\dagger$ is unital and statement i.. Finally, statement iv.\ follows from Proposition~\ref{spec_rad_pos} and property i..
\end{proof}
The general properties of the spectrum listed in Theorem~\ref{spec_qc} are illustrated in Figure~\ref{fig:Spectrum}.
Note that the statement i.\ of Theorem~\ref{spec_qc} expresses the spectral equivalence of the Schr{\"o}dinger and Heisenberg evolutions. Furthermore, all the statements do not rely on complete positivity, but simply on positivity.

Alongside the eigenvalues, it is also necessary to study the eigenoperators of a channel in order to gain further insight into the dynamics and, in particular, into the asymptotic states, as we will discuss in Section~\ref{sec:asymptotics}.
A linear map $\Psi$ on $\mathcal{B}(\mathcal{H})$ always has a Jordan decomposition
\begin{equation}
\label{Jordan_phi}
\Psi= \sum_{\lambda\in\spec(\Psi)} (\lambda \mathcal{P}_\lambda + \mathcal{D}_\lambda),
\end{equation}
where $\lambda$'s are the distinct eigenvalues of $\Psi$, whereas $\mathcal{P}_\lambda$ and $\mathcal{D}_\lambda$ are the corresponding\emph{ spectral projections} and \emph{nilpotents}, see Appendix~\ref{Jordan_op_app} for details.

Given an eigenvalue $\lambda \in \spec(\Psi)$, we  define the \emph{generalized eigenspace} $\mathcal{M}_{\lambda}$ and the \emph{geometric eigenspace} $\mathcal N_\lambda$ corresponding to $\lambda$ as follows 
\begin{equation}
\mathcal{M}_\lambda := \Ran(\mathcal{P}_\lambda), \qquad  \mathcal{N}_\lambda := \Ker(\Psi - \lambda \mathsf{1}),
\end{equation}
for which we have $\mathcal{M}_\lambda \supseteq \mathcal N_\lambda$, and we denote by $\Ran(\cdot)$ and $\Ker(\cdot)$ the range and the kernel of an operator, respectively. The following two properties for the generalized eigenspaces of quantum channels hold.
\begin{proposition}
\label{trace_eigen_prop}
Let  $\lambda$ be an eigenvalue of the quantum channel $\Phi$ on $\mathcal{B}(\mathcal{H})$ with Jordan decomposition~\eqref{Jordan_phi}. Then
\begin{itemize}
	\item[\textnormal{i.}] If $\lambda\neq 1$, then  $\mathcal{N}_\lambda \subseteq \mathcal{M}_\lambda \subseteq \{ X \in \mathcal{B}(\mathcal{H}) \,:\, \Tr(X) = 0 \}$;
	\item[\textnormal{ii.}] $\mathcal M_{\overline{\lambda}}=\mathcal M_{\lambda}^*, \;$  $\mathcal N_{\overline{\lambda}}=\mathcal N_{\lambda}^*,$
	\end{itemize}
    where the adjoint $S^\ast$ of a set $S \subseteq \mathcal{B}(\mathcal{H})$ is defined as $S^\ast := \{ X^\ast \, : \, X \in S  \}$.
\end{proposition}
\begin{proof}
    See Appendix~\ref{proofs_app}.
\end{proof}

 As a consequence of ii., there are no nonzero Hermitian generalized eigenoperators of $\Phi$ associated with a non-real eigenvalue. Furthermore, if $m_{\lambda} =\dim(\mathcal{M}_{\lambda})$ and $n_{\lambda} = \dim(\mathcal{N}_{\lambda})$ are the \emph{algebraic} and \emph{geometric} multiplicities of $\lambda$, respectively, then property ii.\ of Proposition~\ref{trace_eigen_prop} implies that the algebraic and geometric multiplicities of all the eigenvalues of $\Phi$ and $\Phi^\dagger$ are equal~\cite[Corollary 2.22.1]{Amato_PhD_thesis}. 

It is convenient to split the spectrum $\spec(\Phi)$ of a quantum channel $\Phi$ as
\begin{equation}
\label{bulk_per_dec_spec}
\spec (\Phi)=\spec_{\mathrm{P}}(\Phi) \cup \spec_{\mathrm{B}}(\Phi) ,
\end{equation}
where 
\begin{align}
\spec_{\mathrm{P}} (\Phi) &:= \{ \lambda \in \spec (\Phi) \,:\, |\lambda|=1 \},\label{per_spec}
\\
\spec_{\mathrm{B}} (\Phi) &:= \{ \lambda \in \spec (\Phi) \,:\, |\lambda|<1 \} \label{bul_spec}
\end{align}
are respectively the \emph{peripheral} and \emph{bulk} spectra of $\Phi$.

Note that $\spec_{\mathrm{P}} (\Phi) \neq \emptyset$, since $\lambda_0 = 1$ is always an eigenvalue of $\Phi$ according to statement ii.\ of Theorem~\ref{spec_qc}. On the other hand, it may happen that the bulk spectrum is empty, $\spec_{\mathrm{B}}(\Phi)=\emptyset$, i.e.\ $\spec(\Phi) = \spec_{\mathrm{P}}(\Phi)$, which is equivalent to saying that the quantum channel is unitary, see Proposition~\ref{unit_ch} below. 

Let us now introduce the \emph{peripheral}, \emph{asymptotic}, or \emph{attractor subspace} of a quantum channel as
\begin{equation}\label{eq:attractor_suspace}
    \Attr(\Phi):=\mathrm{span}\{\mathcal M_\lambda\,:\, \lambda \in \spec_{\mathrm{P}}(\Phi)\}=
    \Ran\Bigl(\sum_{\lambda\in \spec_{\mathrm{P}}(\Phi)} \mathcal P_\lambda\Bigr).
\end{equation}
In particular, as discussed in Section~\ref{sec:asymptotics}, the evolution described by the QMC $\Sem = (\Phi^n)_{n\in\mathbb N}$ induced by the channel $\Phi$ converges in the long-time limit $n\rightarrow\infty$ towards the attractor subspace $\Attr(\Phi)$. 
Thus, the peripheral spectrum and, correspondingly, the attractor subspace govern the long-time dynamics of the system, so they describe the \emph{recurrent} part of the evolution, while the bulk part of the spectrum is related to the \emph{transient} part of the evolution, only occurring at finite times. We shall see in Proposition~\ref{unit_ch} that the latter component of the dynamics does not exist only for closed-system evolutions, where the dynamics is quasiperiodic; see~\cite{kotowski_tight_2026,gupta2026recurrence} for recent bounds on the recurrence time of closed quantum systems.
Let us now present an important property of the peripheral spectrum of quantum channels~\cite[Proposition 6.2]{wolf2012quantum}.
\begin{proposition}[Semisimplicity of peripheral eigenvalues]
    \label{periph_semisimple}
    Let $\Phi$ be a quantum channel on $\mathcal{B}(\mathcal{H})$ with Jordan decomposition~\eqref{Jordan_phi}. Then, all the peripheral eigenvalues $\lambda \in \spec_{\mathrm{P}} (\Phi)$ are semisimple, i.e., the spectral nilpotents $\mathcal{D}_\lambda$ corresponding to the peripheral eigenvalues are zero.
\end{proposition} 
First, note that an equivalent characterization of semisimple eigenvalues is that their geometric and algebraic multiplicities coincide, see Appendix~\ref{Jordan_op_app}. In addition, several constraints for the multiplicities of peripheral eigenvalues will be discussed in Theorem~\ref{cor_unit} below.
The previous result does not hold for the bulk spectrum (see~\cite[Equation~(53)]{burgarth2013ergodic} for a counterexample), but diagonalizable (and even simple) channels (see Appendix~\ref{Jordan_op_app}) are dense in the set of quantum channels~\cite{vom2024almost}. 
Now, as a consequence of Proposition~\ref{periph_semisimple} one has
\begin{equation}
	\label{attr_def_old}
	\Attr (\Phi)=  \mbox{span} \{ X \in \mathcal{B} (\mathcal{H}) \,:\, \Phi(X)=\lambda X , \, \lambda \in \spec_{\mathrm{P}} ( \Phi) \},
\end{equation}
namely the attractor subspace is spanned by the eigenoperators of $\Phi$ corresponding to the peripheral eigenvalues of the evolution. 
In particular, the attractor subspace contains the invariant elements of the dynamics, forming the \emph{fixed-point} or \emph{steady subspace} of $\Phi$
\begin{equation}
\label{fix_def}
    \Fix(\Phi) :=\{ X \in \mathcal{B}(\mathcal{H}) \,:\, \Phi(X)=X \},
\end{equation}
which is always non-trivial, i.e., $\Fix(\Phi) \neq \{ 0 \}$, as a consequence of Theorem~\ref{spec_qc}. In fact, the fixed-point subspace $\Fix(\Phi)$ of $\Phi$ always contains at least a state, see Proposition~\ref{basis_states_prop}. The states $\rho \in \Fix(\Phi)$ are called \emph{steady} or \emph{stationary states} of the quantum channel $\Phi$.

Importantly, the attractor subspace $\Attr(\Phi)$ is invariant under the action of $\Phi$
\begin{equation}
    \label{Attr_inva}
    \Phi(\Attr(\Phi))= \Attr(\Phi).
\end{equation}
Thus, within the attractor subspace, the state may still undergo a residual evolution, whose general structure will be shown in Theorems~\ref{Wolf_struc} and~\ref{thm:cycles}, but it never escapes from it.
We are now in the position to define the  attractor and the steady subspaces of the QMC $\Sem = (\Phi^n)_{n\in\mathbb{N}}$ induced by a quantum channel $\Phi$
\begin{equation}
\label{Fix_Attr_QMC}
    \Attr(\Sem) := \bigcap_{n \in \mathbb{N}} \Attr(\Phi^n) = \Attr(\Phi), \quad \Fix(\Sem) := \bigcap_{n \in \mathbb{N}} \Fix(\Phi^n) = \Fix(\Phi), 
\end{equation} 
where the second equalities in~\eqref{Fix_Attr_QMC} readily follow from  
\begin{equation}
   \Attr(\Phi)=\Attr(\Phi^n), \quad  \Fix(\Phi) \subseteq  \Fix(\Phi^n), \quad \forall n \geq 1.
\end{equation}
In words, the steady and attractor subspaces of a QMC equal the ones of the quantum channel generating it.

Finally, we can also define the attractor and steady subspaces for a UCP map $\Phi^\dagger$ as 
\begin{align}
    \label{Attr_def_Heis}
    \Attr(\Phi^\dagger)&:= \mbox{span} \{ X \in \mathcal{B}(\mathcal{H}) \,:\, \Phi^\dagger(X)=\lambda X , \, \lambda\in\spec_{\mathrm{P}}(\Phi^\dagger) \} ,
    \\
    \Fix(\Phi^\dagger )&:=\{ X \in \mathcal{B}(\mathcal{H}) \,:\, \Phi^\dagger(X)= X \},
\end{align}
where the peripheral spectrum $\spec_{\mathrm P}(\Phi^\dagger)$ is the set of eigenvalues of $\Phi^\dagger$ with unit modulus.
Also in the Heisenberg picture, they coincide with the corresponding spaces for the quantum Markov chain $\Sem^\dagger$ generated by $\Phi^\dagger$:
\begin{equation}
\label{Fix_Attr_QMC2}
    \Attr(\Sem^\dagger) := \bigcap_{n \in \mathbb{N}} \Attr((\Phi^\dagger)^n) = \Attr(\Phi^\dagger), \quad \Fix(\Sem^\dagger) := \bigcap_{n \in \mathbb{N}} \Fix((\Phi^\dagger)^n) = \Fix(\Phi^\dagger). 
\end{equation}
In particular, the fixed-point subspace  $\Fix(\Phi^\dagger)$ of $\Phi^\dagger$ involves the \emph{conserved quantities} of the QMC $\Sem^\dagger$ induced by $\Phi^\dagger$, which are defined as the Hermitian operators $X = X^\ast$ of $\Fix(\Phi^\dagger)$. See Section~\ref{sec:Noether} for their connection with symmetries. 

As a final remark, note that $\spec_{\mathrm P}(\Phi^\dagger) = \spec_{\mathrm P}(\Phi)$ by point i.\ of Theorem~\ref{spec_qc}, while it may happen that $\Attr(\Phi) \neq \Attr(\Phi^\dagger)$.

\section{GKLS generators and quantum dynamical semigroups}
\label{QDS_sec}
In the previous section, we introduced discrete-time quantum Markov chains induced by a quantum channel that describes state changes in unit time. Alternatively, we can consider a continuous-time evolution of an open quantum system, generally given by a family $\Sem = (\Phi_t)_{t\in\mathbb R^+}$ of quantum channels called a \emph{quantum dynamical map}. In this work, we focus on the homogeneous Markovian case, which corresponds to a \emph{\gls*{qds}}~\cite{alickilendi,Chr_review_2022}, and which renders the long-time behaviour much more tractable. Nevertheless, various generalized notions of Markovianity have been introduced in the continuous-time setting~\cite{Chr_review_2022}, and a few efforts have been made in order to understand the long-time limit of these more general evolutions, see~\cite{pathirana2025asymptotic,dimeglio2025asymptotic}. 

Let us now give the precise definition of a QDS~\cite[Subsection 3.2.1]{BreuerPetruccione}.

\begin{definition}[Quantum dynamical semigroup]
\label{QDS_def}
A family of quantum channels $\Sem = ( \Phi_t )_{t\in \mathbb{R}^+}$ on $\mathcal{B}(\mathcal{H})$ is  a \emph{quantum dynamical} (or \emph{Markovian}) \emph{semigroup} if and only if
\begin{itemize}
\item[\textnormal{i.}] $\Phi_{0}=\mathsf{1}$;
\item[\textnormal{ii.}] $\Phi_{t+s}=\Phi_t \circ \Phi_s$, for all $t,s\in\mathbb{R}^+$ (semigroup property);
\item[\textnormal{iii.}] 
$\lim\limits_{t \downarrow 0} \Phi_{t} = \mathsf{1}$ 
(continuity).
\end{itemize} 
\end{definition}    
\begin{remark}
Assumption \textnormal{iii.}\ in Definition~\ref{QDS_def} expresses continuity of the QDS $\Sem$ at $t=0$. Together with the semigroup property \textnormal{ii.}, it implies continuity for all $t\in\mathbb{R}^+$.
Since $\mathcal{H}$ is finite-dimensional, the space of linear maps on $\mathcal{B}(\mathcal{H})$ is finite-dimensional as well. Therefore, all norms on this space are equivalent, and the limit in \textnormal{iii.}\ may be taken with respect to any norm.
\end{remark}
\begin{remark}
    The classical counterpart of QDSs are the (continuous-time) Markov chains summarized in Appendix~\ref{classic_sec}.  
\end{remark}
Thus, $\Sem = (\Phi_{t})_{t\in\mathbb{R}^+}$ being a continuous semigroup of maps on $\mathcal{B}(\mathcal{H})$, is differentiable for all $t\in\mathbb{R}^+$, with derivative~\cite{engelnagel}
\begin{equation}
\label{master_eq}
\frac{\mathrm{d}}{\mathrm{dt}}\Phi_t = \mathcal{L} \circ \Phi_t, \quad \Phi_0 = \mathsf{1}.
\end{equation} 
The map $\mathcal{L}: \mathcal{B}(\mathcal{H})\to \mathcal{B}(\mathcal{H})$, defined via
\begin{equation}
\label{gen_def}
\mathcal{L} := \frac{\mathrm{d}}{\mathrm{dt}}\Phi_t \Big\rvert_{t=0} = \lim_{t\downarrow 0}\frac{\Phi_t - \mathsf{1}}{t},
\end{equation}
is called the \emph{generator} of the semigroup, while the differential Equation~\eqref{master_eq} governing the semigroup dynamics $\Sem$ is called \emph{Markovian master equation}. Accordingly, any map $\Phi_t$ of $\Sem$ can be expressed as $\Phi_t = \mathrm{e}^{t\mathcal{L}}$, and we will call $\Sem$ the QDS generated by $\mathcal L$. If the system is initially in the state $\rho_0$ at time $t=0$, its state $\rho(t)$ at time $t>0$ reads $\rho(t) =\mathrm e^{t\mathcal{L}}\rho_0$.
A quantum channel of the form $\Phi=\mathrm e^{\mathcal L}$, obtained through a unit-time evolution of a QDS with generator $\mathcal{L}$, will be called \emph{Markovian}~\cite{wolfdividing}.

Once the evolution, i.e.\ the QDS $\Sem$, is expressed in terms of the generator $\mathcal{L}$, one would like to characterize $\mathcal{L}$ directly from the properties of the semigroup and, subsequently, to obtain an explicit representation of the generator, analogous to the operator-sum representation~\eqref{op_sum_rep} of quantum channels. The first step is provided by the following proposition~\cite[Propositions 7.2-7.3]{wolf2012quantum}, which mirrors the axiomatic definition of a quantum channel given in Definition~\ref{qc_def}. 
\begin{proposition}
\label{char_gen_QDSs}
A map $\mathcal{L}$ on $\mathcal{B}(\mathcal{H})$ is the generator of a QDS (Definition~\ref{QDS_def}) if and only if 
\begin{itemize}
\item[\textnormal{i.}] $\Tr(\mathcal{L}(X))=0$ for all $X\in \mathcal{B}(\mathcal{H})$ (trace-annihilating property);
\item[\textnormal{ii.}] $\mathcal{L}(X^\ast) = \mathcal{L}(X)^\ast$ for all $X \in \mathcal{B}(\mathcal{H})$ (Hermiticity-preserving property);
\item[\textnormal{iii.}] $(\mathbb{I} - \ketbra{ \phi_d })C({\mathcal{L}})(\mathbb{I} - \ketbra{\phi_d}) \geq 0$;
\end{itemize}
where $\ket{\phi_d}$ is the maximally entangled vector state defined in~\eqref{max_ent} and $C(\mathcal{L})$ is the Choi operator of $\mathcal{L}$ defined in~\eqref{CJ_iso}.

\end{proposition}
Incidentally, any map fulfilling the properties ii. and iii. of Proposition~\ref{char_gen_QDSs} is called \emph{conditionally completely positive}~\cite{Evans_CCP_77}. 
The explicit form for the generator $\mathcal L$ was obtained by Gorini, Kossakowski, and Sudarshan~\cite{GKS76} and, independently, by Lindblad~\cite{Lindblad76}, and  reads
\begin{equation}
\label{diag_GKLS}
\mathcal{L}(X)=-\mathrm i[H,X] + \sum_{\alpha \in [N]} \Bigl(L_\alpha X L_\alpha^\ast - \frac{1}{2} \{ L_\alpha^\ast L_\alpha , X \} \Bigr), \quad X \in \mathcal{B}(\mathcal{H}),
\end{equation}
where $H=H^*$ is a Hermitian operator which plays the role of an effective Hamiltonian of the system, while the operators $\{L_\alpha \}_{\alpha=1}^N$ are called \emph{noise} or \emph{jump operators}. In particular, the operators $L_\alpha$'s can be chosen to be traceless and mutually orthogonal~\cite{BreuerPetruccione}, in which case $N$ is at most equal to $d^2-1$. 
We will therefore call $\mathcal{L}$ \emph{\gls*{gkls} generator} and~\eqref{diag_GKLS} the \emph{GKLS form} or \emph{representation} of $\mathcal{L}$. See~\cite{BreuerPetruccione,kimura2026kossa} for self-contained derivations of~\eqref{diag_GKLS}. 
Therefore, according to~\eqref{diag_GKLS}, any GKLS generator $\mathcal{L}$ can be written as the sum of two contributions: $\mathcal{L} = \mathcal{L}_{\mathrm{H}} + \mathcal{L}_{\mathrm{D}}$, where
\begin{equation}\label{hamiltonian_dissipative_GKLS}
\mathcal{L}_{\mathrm{H}}(X)=-\mathrm i[H , X],\quad \mathcal{L}_{\mathrm{D}}(X)=\sum_{\alpha \in [N]} \Bigl( L_\alpha X L_\alpha^\ast - \frac{1}{2} \{ L_\alpha^\ast L_\alpha , X \} \Bigr), \quad X \in \mathcal{B}(\mathcal{H})
\end{equation}
are respectively  the \emph{Hamiltonian} and \emph{dissipative} parts of $\mathcal{L}$.
The Hamiltonian part generates a unitary evolution and it is the only contribution for a closed system, while the dissipative part is physically related to decay and decoherence taking place in a purely-open-system dynamics. 
The GKLS representation~\eqref{diag_GKLS} of $\mathcal{L}$ is not unique as shown by the following result~\cite[Equations~(3.72)-(3.73)]{BreuerPetruccione}.
\begin{proposition}
\label{GKLS_freed_prop}
A GKLS generator $\mathcal L$ in the form~\eqref{diag_GKLS} is invariant under the following two classes of transformations:
\begin{itemize}
\item[\textnormal{i.}] Unitary mixing of jump operators:
\begin{align}
 H^\prime = H, \qquad 
 L_\alpha^\prime = \sum_{\beta\in[T]} 
 U_{\alpha\beta} L_\beta, \quad\alpha\in[T],
\end{align}
where $T = \max(N,N^\prime)$, with $N^\prime$ being the number of non-zero primed jump operators $L_\alpha^\prime$, $(U_{\alpha \beta})$ is a unitary matrix of order $T$, and we add zero operators to the shorter family of jump operators;
\item[\textnormal{ii.}] Affine shift of the jump operators:
\begin{align}
H^\prime = H + \frac{1}{2\mathrm i} 
\sum_{\alpha\in[N]} (\bar{z}_\alpha L_\alpha - z_\alpha L_\alpha^\ast) + x\mathbb{I},\qquad 
L_\alpha^\prime = L_\alpha+ z_\alpha \mathbb{I}, \quad \alpha\in[N],  
 \label{transf2_2}
\end{align}
with $z_\alpha \in \mathbb{C}$ and $x \in \mathbb{R}$.
\end{itemize}
\end{proposition}
The first class of transformations discussed in Proposition~\ref{GKLS_freed_prop} is related to the unitary gauge freedom of the first term of the dissipative part as in Proposition~\ref{freed_Kraus_prop}, while the second one arises from the freedom in the choice of the trace for the Hamiltonian, i.e., the zero of the energy, and the jump operators.  
Note also that the latter proposition shows that the Hamiltonian and dissipative parts of a GKLS generator are not uniquely defined, unless additional restrictions are imposed, for example requiring the jump operators to be traceless; see~\cite{GKS76,davies1980uniqueness,vom2024understanding,vom_ende_choi_inf} for related results.

In the Heisenberg picture, the Markovian continuous-time dynamics of an open quantum system is given by the (Hilbert--Schmidt) adjoint semigroup $\Sem^\dagger = (\Phi_t^\dagger)_{t\in\mathbb{R}^+}$, which is a continuous semigroup of UCP maps on $\mathcal{B}(\mathcal{H})$.
Note that, in line with the convention adopted for QMCs and if not stated otherwise, the Schr{\"o}dinger (Heisenberg) dynamics over the interval $[0,t]$ will be denoted with $\Phi_{t}$ ($\Phi_t^\dagger$) and, correspondingly, the Schr{\"o}dinger (Heisenberg)  generator of the QDS $\Sem = (\Phi_t)_{t\in\mathbb{R}^+}$ with $\mathcal{L}$ ($\mathcal{L}^\dagger$).
The generator $\mathcal{L}^\dagger$ of the Heisenberg dynamics $\Sem^\dagger$ admits the following GKLS representation
\begin{equation}
\label{GKLS_adjoint}
\mathcal{L}^\dagger (X)=\mathrm i[H,X] + \sum_{\alpha\in[N]}  
\Bigl( L_\alpha^\ast X L_\alpha - \frac{1}{2} \{ L_\alpha^\ast L_\alpha , X \} \Bigr), \quad X \in \mathcal{B}(\mathcal{H}),
\end{equation}
which follows by taking the adjoint of~\eqref{diag_GKLS}. 
The map $\mathcal{L}^\dagger$ may be also characterized in an axiomatic way, as done in Proposition~\ref{char_gen_QDSs} for $\mathcal{L}$.
\begin{proposition}
\label{CCP_Evans_prop_2}
$\mathcal{L}^\dagger$ generates an adjoint QDS if and only if it satisfies properties \textnormal{ii.}\ and \textnormal{iii.}\ of Proposition~\ref{char_gen_QDSs} (conditional complete positivity) and $\mathcal{L}^\dagger(\mathbb{I})=0$. 
\end{proposition}
\subsection{Spectral properties and asymptotic subspaces of GKLS generators}
\label{spec_lindbladian}
Let us now explore the spectral properties of a QDS, which is equivalent to studying the spectrum of its GKLS generator. As for quantum channels, the topic has been also addressed for random ensembles of GKLS generators in~\cite{denisov2019universal,tarnowski2021random,tarnowski2023random,hontarenko2026beyond}. 
First, the spectrum of any GKLS generator satisfies the basic properties listed in the next theorem. 
\begin{theorem}[Spectrum of a GKLS generator]
\label{spec_L}
Let $\mathcal{L}$ be a GKLS generator. Then, its spectrum $\spec (\mathcal{L})$ satisfies the following properties:
\begin{itemize}
\item[\textnormal{i.}] The spectra of $\mathcal L$ and $\mathcal{L}^\dagger$ coincide, $\spec(\mathcal L)=\spec(\mathcal L^\dagger)$;
\item[\textnormal{ii.}] Zero is always an element in the spectrum, $\lambda_0= 0\in\spec(\mathcal L)$;
\item[\textnormal{iii.}] The spectrum is invariant under complex conjugation, i.e., $\spec (\mathcal{L} )=\overline{\spec (\mathcal{L} )}$;
\item[\textnormal{iv.}] The spectrum is contained in the closed left half-plane, namely  $\spec(\mathcal{L})~\subseteq\{\lambda\in\mathbb{C}\,:\,\Re(\lambda)\leq 0\}$.
\end{itemize}
Moreover, the above properties are valid for any generator of continuous semigroups of positive maps that are trace-preserving or unital.
\end{theorem}
The first property follows from the Hermiticity-preservation of $\mathcal{L}$, see Proposition~\ref{char_gen_QDSs}, while the last three properties can be proved from Theorem~\ref{spec_qc} after observing that $(\mathrm e^{t\mathcal L})_{t\in\mathbb{R}^+}$ is a family of positive trace-preserving maps. 
Note that an analogue of Proposition~\ref{trace_eigen_prop} for GKLS generators may be proved along the lines of Appendix~\ref{proofs_app}, see also~\cite[Propositions~2.25-2.26]{Amato_PhD_thesis}. In addition, the spectra of GKLS generators in the Schr{\"o}dinger and Heisenberg pictures are equal even counting the eigenvalues with their algebraic or geometric multiplicities, as discussed for quantum channels after  Proposition~\ref{trace_eigen_prop}.

As for quantum channels, it is convenient to divide the spectrum $\spec (\mathcal{L})$ of a GKLS generator $\mathcal{L}$ into its \emph{peripheral} and \emph{bulk} parts 
\begin{align}
\label{spec_L_def}
\spec_{\mathrm{P}}(\mathcal{L}) & := \{ \lambda \in \spec (\mathcal{L}) \,:\, \Re ( \lambda ) = 0 \}, \\
\spec_{\mathrm{B}} (\mathcal{L}) & := \{ \lambda \in \spec (\mathcal{L}) \,:\,\Re ( \lambda ) < 0 \}, \label{specB_L}
\end{align}
where the eigenvalues of the former set are semisimple. The last result is well-known, but we provide a simple proof for completeness. 
\begin{proposition}[Semisimplicity of peripheral eigenvalues of a generator]
\label{semisim_prop}
Let $\mathcal{L}$ be a GKLS generator.\ Then, the peripheral, i.e., purely imaginary, eigenvalues $\lambda \in \spec_{\mathrm{P}} (\mathcal{L})$ are semisimple. 
\end{proposition}
\begin{proof}
According to Proposition~\ref{spec_rad_pos}, the adjoint QDS $(\Phi_t^\dagger)_{t\in\mathbb{R}^+}$ generated by $\mathcal{L}^\dagger$ satisfies $\| \Phi_t^\dagger \| = 1$, where $\| \!\cdot\! \|$ denotes the operator norm defined by~\eqref{sp_rad_op_norm_defs} for all $t\in\mathbb{R}^+$, therefore the claim readily follows from~\cite[Corollary 2.11]{engelnagel},
claim iv.\ of Theorem~\ref{spec_L}, and the equality of the algebraic and geometric multiplicities of the eigenvalues of $\mathcal{L}$ and $\mathcal{L}^\dagger$, see after Theorem~\ref{spec_L}.  
\end{proof} 

Again, this property is not generally true for all the eigenvalues, see~\cite[Example 6]{Chr_review_2022}, although the set of simple and, consequently, diagonalizable generators is dense in the set of all GKLS generators~\cite{vom2024almost}.

The attractor subspace $\Attr(\mathcal{L})$ of the GKLS generator $\mathcal{L}$ is the span of the generalized eigenspaces $\mathcal{M}_\lambda$ of the peripheral spectrum $\lambda \in\spec_{\mathrm P}(\mathcal L)$, i.e., 
\begin{equation}
\Attr(\mathcal{L}) := \mbox{span} \{ \mathcal{M}_{\lambda} \,:\, \lambda \in \spec_{\mathrm{P}}(\mathcal{L}) \},
\end{equation}
which can be characterized as
\begin{equation}
\label{Attr_L_def} 
\Attr (\mathcal{L}) = \mbox{span} \{ X \in \mathcal{B} (\mathcal{H}) \,:\, \mathcal{L}(X)=\lambda X ,\, \lambda \in \spec_{\mathrm{P}} ( \mathcal{L}) \},
\end{equation}
by Proposition~\ref{semisim_prop}. In particular, the attractor subspace $\Attr(\mathcal{L})$ of $\mathcal{L}$ includes its kernel 
\begin{equation}
	\label{Ker_L_def}
	 \Ker(\mathcal{L})  := \{ X \in \mathcal{B}(\mathcal{H}) \,:\, \mathcal{L}(X) = 0 \}.
\end{equation}
In analogy with QMCs, the attractor and steady subspaces of a QDS $\Sem = (\Phi_t)_{t\in\mathbb{R}^+}$ generated by $\mathcal{L}$ are 
\begin{equation}
\label{fix_attr_QDS}
  \Attr(\Sem) := \bigcap_{t\in\mathbb R^+} \Attr(\Phi_t) = \Attr(\mathcal{L}) , \quad   \Fix(\Sem) := \bigcap_{t\in\mathbb R^+} \Fix(\Phi_t) = \Ker(\mathcal{L}),
\end{equation}
where the second equalities arise from $\Phi_t = \mathrm{e}^{t\mathcal{L}}$ with $t \in \mathbb{R}^+$. 
In particular, $\Attr(\Sem)$ can be identified as the asymptotic subspace of the evolution (see Section~\ref{asympt_QDS_sec}), while $\Fix(\Sem)$ involves the \emph{stationary states} of the evolution $\Sem$.
Observe also that, for a given time $t>0$, we have
\begin{equation}
\Ker(\mathcal{L}) \subseteq \Fix(\Phi_t), 
\label{Fix_QDS_Ker_L}
\end{equation}
where the inclusion is strict whenever the generator $\mathcal{L}$ has peripheral eigenvalues being non-zero integer multiples of $2\pi \mathrm{i} / t$. 

Now, as in the Schr{\"o}dinger picture, given an adjoint QDS $\Sem^\dagger = (\Phi_t^\dagger)_{t\in\mathbb{R}^+}$ with generator $\mathcal{L}^\dagger$, we can introduce   attractor subspace and kernel of $\mathcal{L}^\dagger$ as
\begin{align}
\label{Ker_L_def2}
\Attr (\mathcal{L}^\dagger) &:= \mbox{span} \{ X \in \mathcal{B} (\mathcal{H}) \,:\, \mathcal{L}^\dagger(X)=\lambda X , \, \lambda \in \spec_{\mathrm{P}} ( \mathcal{L}^\dagger) \}, \\
\label{Attr_L_def2} 
 \Ker(\mathcal{L}^\dagger)  &:= \{ X \in \mathcal{B}(\mathcal{H}) \,:\, \mathcal{L}^\dagger(X) = 0 \}, 
\end{align}
with $\spec_{\mathrm{P}}(\mathcal{L}^\dagger)$ being the purely imaginary spectrum of $\mathcal{L}^\dagger$,
as well as the attractor and steady subspaces of $\Sem^\dagger$ 
\begin{align}
    \Attr(\Sem^\dagger) &:= \bigcap_{t\in\mathbb R^+} \Attr(\Phi_t^\dagger)  = \Attr (\mathcal{L}^\dagger),  \\
  	\Fix(\Sem^\dagger) &:= \bigcap_{t\in\mathbb R^+} \Fix(\Phi_t^\dagger) = \Ker(\mathcal{L}^\dagger). 
\end{align}
In particular, the kernel $\Ker(\mathcal{L}^\dagger)$ of $\mathcal{L}^\dagger$ involves the \emph{conserved quantities} of the QDS $\Sem^\dagger$ generated by $\mathcal{L}^\dagger$, defined as the Hermitian operators $X = X^\ast$ of $\Ker(\mathcal{L}^\dagger)$. See Section~\ref{sec:Noether} for their relation with symmetries.  

Finally, as with quantum channels, note that $\spec_{\mathrm{P}} ( \mathcal{L}) = \spec_{\mathrm{P}} ( \mathcal{L}^\dagger) $ as a consequence of point i. of Theorem~\ref{spec_L}, but $\Attr(\mathcal{L}) \neq \Attr(\mathcal{L}^\dagger)$ in general.

\section{Asymptotic dynamics of quantum Markov chains}
\label{sec:asymptotics}
In this section we explore the long-time dynamics of open quantum systems described by discrete-time quantum Markov chains, introduced in Section~\ref{qc_sec}. The corresponding problem for the continuous-time Markovian case will be investigated in Section~\ref{asympt_QDS_sec}. 

The present section is organized as follows: the next subsection introduces the preliminary concepts for the description of the asymptotic dynamics in the discrete-time setting.\ Then, we discuss in Section~\ref{sec:examples} several examples, illustrating the building blocks of the asymptotic structure theorem in the Schr{\"o}dinger picture. The latter result is presented in Section~\ref{sec:structure_theorems} together with its counterpart in the Heisenberg picture, Theorem~\ref{struc_Heis_th}, and with an alternative decomposition of the asymptotics in the Schr{\"o}dinger picture, which we shall refer to as the cyclic decomposition, Theorem~\ref{thm:cycles}.

\subsection{Basic notions}
\label{bas_notions_subsec}
Let $\Sem = (\Phi^n)_{n\in\mathbb{N}}$ be a discrete-time QMC in the Schr{\"o}dinger picture generated by a quantum channel $\Phi$. Then,  
the asymptotic dynamics of the system corresponds to the limit $n\rightarrow \infty$ and, from the spectral viewpoint, it is only governed by the peripheral part of the spectrum~\eqref{per_spec}. Indeed, if $X$ is an eigenoperator associated with a bulk eigenvalue $\lambda \in \spec_{\mathrm{B}} (\Phi)$, then we have for any norm  on $\mathcal{B}(\mathcal{H})$
\begin{equation}\label{eq:bound_limit}
\Phi^n (X) = \lambda ^n X  \to 0,
\qquad \text{as } {n \rightarrow \infty},
\end{equation}  
since $|\lambda|<1$ in~\eqref{bul_spec}. The same conclusion holds for generalized eigenoperators associated with bulk eigenvalues: although polynomial factors in $n$ may arise from nontrivial Jordan blocks, they are dominated by the exponential decay $|\lambda|^n$~\cite[Section 3.2.5]{horn}.
Thus, in the non-unitary case, the speed of convergence towards equilibrium is controlled by the slowest decaying mode of the quantum system, namely
\begin{equation}
\label{Rmax_def}
   R(\Phi) := \max_{\lambda \in \spec_{\mathrm{B}}(\Phi) } | \lambda |.
\end{equation}

Therefore, we conclude that only peripheral eigenvalues and the corresponding eigenoperators of $\Phi$ contribute to the long-time dynamics of the QMC $\Sem$.
More precisely, for any initial state $\rho$, the component of $\Phi^n(\rho)$ outside the attractor subspace $\Attr(\Phi)$ given by~\eqref{attr_def_old} vanishes as $n\to\infty$, and the asymptotic dynamics is entirely confined to $\Attr(\Phi)$.
We will refer to the density operators belonging to the attractor subspace as \emph{asymptotic states}.

The above reasoning can be applied to the adjoint QMC $\Sem^\dagger$ generated by a UCP map $\Phi^\dagger$, so we can identify the asymptotic subspace of $\Sem^\dagger$ as the attractor subspace $\Attr(\Phi^\dagger)$ of $\Phi^\dagger$. 

Henceforth, when discussing the long-time behaviour of a quantum channel $\Phi$ or of a UCP map $\Phi^\dagger$, we will implicitly refer to the asymptotic dynamics of the corresponding QMC $\Sem$ or $\Sem^\dagger$, respectively.

Since the evolution of the system starting from an element within $\Attr(\Phi)$ tends to return arbitrarily close to it under the action of the induced QMC, the attractor subspace of a channel $\Phi$ is also called the \emph{space of recurrences}. Explicitly, we have~\cite{wolf2012quantum}

\begin{equation}
X \in \Attr(\Phi) \quad \Leftrightarrow \quad \forall \varepsilon >0,\ \exists\, n \geq 1 \mbox{  s.t.  } \|\Phi^{n}(X)-X \| \leq \varepsilon,
\end{equation}
with respect to some norm $\| \cdot \|$ of $\mathcal{B}(\mathcal{H})$.

In particular, the \emph{limit cycles} are those elements in the attractor subspace satisfying
\begin{equation}
    \Phi^{n}(X)=X \mbox{ for some }n \geq 1.
\end{equation}
For $n=1$ one recovers the fixed-point subspace $\Fix(\Phi)$ 
given in~\eqref{fix_def}, and, in particular, the steady states of the evolution.

To proceed with the analysis, it is convenient to introduce  three maps associated with $\Phi$. The first is the spectral projection $\mathcal{P}$ corresponding to the eigenvalue $\lambda = 1$ of $\Phi$,
\begin{equation}
\label{steady_proj_channel}
    \mathcal{P} :=  \mathcal{P}_{\lambda=1}.
\end{equation}
Its range is the steady subspace $\Fix(\Phi)$, and we will refer to it as the \emph{fixed-point} or \emph{steady projection} of $\Phi$. 
The second map is the \emph{peripheral projection} of $\Phi$ defined as 
\begin{equation}
\label{per_proj_def}
\mathcal{P}_{\mathrm{P}} := \sum_{ \lambda \in \spec_{\mathrm{P}} (\Phi) } \mathcal{P}_\lambda,
\end{equation}
whose range is $\Ran(\mathcal{P}_{\mathrm{P}})=\Attr(\Phi)$. The third  is the \emph{peripheral channel} of $\Phi$, defined by
\begin{equation}
\label{per_ch_def}
\Phi_{\mathrm{P}} := \sum_{ \lambda \in \spec_{\mathrm{P}} (\Phi) } \lambda \mathcal{P}_\lambda.
\end{equation}
This map is the part of the Jordan decomposition of $\Phi$ associated with the peripheral spectrum $\spec_{\mathrm P} (\Phi)$.\ 
As shown in Proposition~\ref{periph_semisimple}, the peripheral spectrum is semisimple, so no nilpotents occur in $\Phi_{\mathrm P}$. 

The following result~\cite[Proposition 6.3]{wolf2012quantum} relates the above three maps to the quantum channel $\Phi$.
\begin{proposition}[Peripheral maps of a quantum channel]
\label{peripheral_Wolf}
Let $\Phi$ be a quantum channel on $\mathcal{B}(\mathcal{H})$. Then, $\mathcal{P}$, $\mathcal{P}_{\mathrm{P}}$ and $\Phi_{\mathrm{P}}$ may be written as
\begin{align}
&\mathcal{P}=\lim_{N\rightarrow \infty}\frac{1}{N} 
\sum_{n\in[N]}\Phi^{n} \label{P_for} , \\
&\mathcal{P}_{\mathrm{P}} = \lim_{i \rightarrow \infty} \Phi^{n_i}, \label{Pp_for} \\
&\Phi_{\mathrm{P}} = \Phi \circ \mathcal{P}_{\mathrm{P}} = \mathcal{P}_{\mathrm{P}} \circ \Phi, \label{Phip_for}
\end{align} 
for some increasing sequence $( n_i)_{i\in\mathbb{N}}$ of natural numbers. In particular, $\mathcal{P}$, $\mathcal{P}_{\mathrm{P}}$ and $\Phi_{\mathrm{P}}$ are quantum channels.
\end{proposition}
Note that $\lim_{n \to \infty} \Phi^n$ does not exist in general because of the possible presence of peripheral eigenvalues different from one, see also Proposition~\ref{prop:Noether_convergence} below. Nevertheless, according to Equation~\eqref{Pp_for}, we can always find a convergent subsequence $(\Phi^{n_{i}})_{i \in \mathbb{N}}$, with $(n_i)_{i \in \mathbb{N}}$ being related to the peripheral spectrum $\spec_{\mathrm{P}}(\Phi)$ of $\Phi$, whose limit is the peripheral projection $\mathcal{P}_{\mathrm{P}}$ of $\Phi$ defined via~\eqref{per_proj_def}. 

Moreover, given a QMC $\Sem^\dagger = ((\Phi^{\dagger})^n)_{n\in\mathbb N}$ in the Heisenberg picture generated by a UCP map $\Phi^\dagger$, the steady projection, the peripheral projection, and the peripheral UCP map can be defined  as in the Schr{\"o}dinger picture, and they are the Hilbert--Schmidt adjoints of the corresponding channels.
			We now define the special class of faithful evolutions. 
\begin{definition}[Faithful channels]
\label{faith_ch_def}
    Let $\Phi$ be a quantum channel. Then $\Phi$ is said to be faithful if and only if it admits a full-rank (invertible) stationary state,  $0 < \rho \in \mathcal{S}(\mathcal{H})$. A UCP map $\Phi^\dagger$ is called faithful if its adjoint quantum channel $\Phi$ is faithful.
\end{definition}
\begin{remark}
    The channels considered in Definition~\ref{faith_ch_def} were called ``faithful" in~\cite{albert2019asymptotics}. In~\cite{rajaramaperipheral_2}, the same class of maps is referred to as stationary. It is  worth noting that, in the mathematical literature, the term ``faithful" is sometimes used with a different meaning, see e.g.~\cite[Definition~1.4.1]{stormer2012positive}.
\end{remark}
The following characterization of faithfulness holds; see~\cite[Section 2.1]{AFK_asympt_1}.
\begin{proposition}
\label{char_faith_prop}
    Let $\Phi$ be a quantum channel on $\mathcal{B}(\mathcal{H})$ with peripheral projection $\mathcal{P}_{\mathrm{P}}$, and let
    \begin{equation}
        \mathcal{H}_{0} := \Ran(\mathcal{P}_{\mathrm{P}}(\mathbb{I})).
    \label{eq:H0def}
    \end{equation}
    Then, $\Phi$ is faithful if and only if $\mathcal{H} = \mathcal{H}_{0}$. 
\end{proposition}
The following proposition unveils the physical meaning of the subspace $\mathcal{H}_{0}$ introduced in Proposition~\ref{char_faith_prop} and, in turn, of faithfulness~\cite[Subsection 2.1]{AFK_asympt_1}.
\begin{proposition} 
\label{supp_QC_prop}
Let $\Phi$ be a quantum channel on $\mathcal{B}(\mathcal{H})$ with peripheral projection $\mathcal{P}_{\mathrm P}$, given in~\eqref{per_proj_def}, 
and attractor subspace $\Attr(\Phi) = \Ran(\mathcal{P}_{\mathrm P})$.
Let $\mathcal H_0=\Ran (\mathcal{P}_{\mathrm{P}}(\mathbb{I}))$, as in~\eqref{eq:H0def}, and consider the decomposition
\begin{equation}
    \mathcal H=\mathcal H_0^\perp\oplus\mathcal H_0.
\end{equation}
Then, every $X \in \Attr(\Phi)$ is supported on $\mathcal{H}_{0}$,
i.e.\ $X=P_0 X P_0$, where $P_0$ is the Hermitian projection onto $\mathcal{H}_0$. Equivalently,
\begin{equation}
    \Attr(\Phi)\subseteq 0\oplus\mathcal{B}(\mathcal{H}_{0}), \label{Attr_DEC}
\end{equation}
where $0$ denotes the zero operator on $\mathcal{H}_{0}^{\perp}$. Furthermore, the subspace $0\oplus\mathcal{B}(\mathcal{H}_{0})$ is invariant under $\Phi$.
\end{proposition}

Therefore, in the long-time limit, the entire probability weight supported on the subspace $\mathcal{H}_{0}^\perp$ is transferred to $\mathcal{H}_{0}$, and we thus call  $\mathcal{H}_{0}^\perp$ the \emph{transient space} of $\Phi$; see~\cite{baumgartner2012structures} for a detailed analysis of the process. As shown in Section~\ref{sec:examples}, this effect is physically related to amplitude damping or decay.

We are now in the position to define the following faithful quantum channel~\cite[Lemma 6.4]{wolf2012quantum} 
\begin{equation}\label{faith_ind_def}
  \Phi_{\mathrm{faith}}: \mathcal B(\mathcal H_0) \to \mathcal B(\mathcal H_0), \quad    \Phi_{\mathrm{faith}}(X_0):=\Phi(0 \oplus X_0) |_{\Hi_0}, \; X_0 \in \mathcal{B}(\mathcal{H}_0).
\end{equation}
Then, due to the invariance of $0 \oplus \mathcal B(\mathcal H_0)$, one has that $\Phi(0\oplus X_0)\equiv0\oplus \Phi_{\mathrm{faith}}(X_0)$. We call $\Phi_{\mathrm{faith}}$ the \emph{induced faithful channel} of $\Phi$.
In particular, combining Proposition~\ref{supp_QC_prop} with~\eqref{faith_ind_def} yields
\begin{equation}
    \Attr(\Phi)= 0\oplus \Attr(\Phi_{\mathrm{faith}}), \quad \Fix(\Phi)= 0\oplus \Fix(\Phi_{\mathrm{faith}}).
    \label{attr:phi_phit}
\end{equation}
Therefore, we can reduce the analysis of the fixed-point and attractor subspaces in the Schr{\"o}dinger picture to faithful channels. 

We now provide an explicit characterization of the fixed-point subspace of faithful UCP maps in terms of the KS representation~\eqref{op_sum_rep}; for the proof see~\cite[Theorems 6.12--6.13]{wolf2012quantum}.
\begin{proposition}\label{th:fixed_kraus}
Let $\Phi^\dagger$ be a UCP map with KS operators $\{ A_k \}_{k=1}^N$ in the representation~\eqref{op_sum_rep}. 
Let
\begin{equation}
S :=\{ A_{k} , A_{k}^{\ast} \}_{k=1}^{N},
\label{fix_point_Phi}
\end{equation}
and denote by $S^{\prime}$  its commutant.
Then, one has the inclusion
\begin{equation}
S^{\prime} \subseteq \Fix(\Phi^\dagger),
\end{equation}
with equality if $\Phi^\dagger$ is faithful according to Definition~\ref{faith_ch_def}.
\end{proposition}
We recall that the \emph{commutant} $S'$ of a set $S\subset \mathcal B(\mathcal H)$ is the set of all operators commuting with all the elements of $S$, namely,
\begin{equation}
\label{comm_def}
     S'=\{Y\in\mathcal B(\mathcal H) \,:\, [X,Y]=0\quad \forall X\in S \},
\end{equation}
where $[X,Y] := XY - YX$ with $X,Y \in \mathcal{B}(\mathcal{H})$.
We shall see in Section~\ref{sec:alg_approach} that the attractor subspace of a faithful UCP map can be similarly characterized in terms of its KS operators; cf.\ Equation~\eqref{char_Attr_Kraus_faithful}.
Furthermore, under the faithfulness assumption and according to~\cite[Subsection 4.3]{jex_st_2018}, the steady and attractor subspaces of $\Phi$ and $\Phi^\dagger$ can be related to one another in the following way:
\begin{equation}
\label{Schr_Heis_corresp}
    \Fix(\Phi^{\dagger}) =
  \sigma^{-1/2} \Fix(\Phi)\sigma^{-1/2}, \quad \Attr(\Phi^{\dagger}) =
  \sigma^{-1/2} \Attr(\Phi)\sigma^{-1/2} ,
\end{equation}
with $\sigma$ being any full-rank invariant state  of the quantum channel $\Phi$. See~\cite[Corollary 1]{jex_st_2018} for a more general correspondence based on operator monotone functions.

In particular, if $\Phi$ is a unital channel, i.e.\ $\Phi(\mathbb{I}) = \mathbb{I}$, one may choose $\sigma=\mathbb I/d$, where $d=\dim(\mathcal H)$. Hence, $\Fix(\Phi) = \Fix(\Phi^\dagger)$ and $\Attr(\Phi) = \Attr(\Phi^\dagger)$.

\subsection{Examples}\label{sec:examples}

We present now several paradigmatic examples that will provide valuable insights into the general structure of the asymptotic evolution of a QMC, which will be discussed in detail in the next  Section~\ref{sec:structure_theorems}.

\begin{figure}
    \centering
    \includegraphics[width=\textwidth]{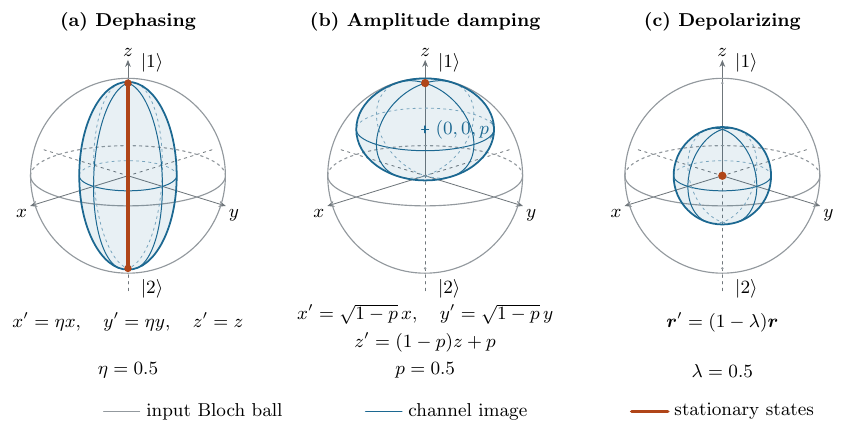}
    \caption{Images of the Bloch ball under (a) qubit dephasing, cf.~\eqref{eq:dephasing_bloch}, (b) amplitude damping, cf.~\eqref{eq:amplitude_damping_bloch}, and (c) depolarizing, cf.~\eqref{eq:depolarizing_bloch}, channels. Grey and blue surfaces bound the input and output state spaces, respectively; red marks stationary states. }
    \label{fig:Bloch}
\end{figure}

The examples below are formulated on finite-dimensional Hilbert spaces of
arbitrary dimension. Some of them admit particularly transparent
geometrical representations in the qubit case. In particular, the
dephasing, amplitude-damping, and depolarizing channels are illustrated in terms of their action on the Bloch ball (see Figure~\ref{fig:Bloch}), using the representation
\begin{equation}\label{eq:Bloch}
        \rho(\bm{r})
    =\frac{1}{2}\left(
\mathbb{I}+x\sigma_1+y\sigma_2+z\sigma_3
    \right),
    \qquad
    \bm{r} =(x,y,z)\in\mathbb{R}^3,
    \qquad
    \|\bm{r}\|\leq 1 ,
\end{equation}
where $\sigma_j$, with $j \in [3]$, are the Pauli matrices.
\subsubsection{Unitary channels}\label{sec:unitary_evolution}
We begin with the simplest case of a closed quantum system, namely, a system that does not interact with its environment. In this case the dynamics of the system 
is described by a \emph{unitary channel}
\begin{equation}
\label{unit_def}
\Phi_U(X)=UXU^\ast, \quad X\in \mathcal{B}(\mathcal{H}),\ UU^\ast = \mathbb{I},
\end{equation}
which can be always expressed via a Hamiltonian $H=H^\ast$ as
\begin{equation}
\Phi_U(X)=\mathrm{e}^{-\mathrm iH}X\mathrm{e}^{\mathrm iH},  \quad X\in \mathcal{B}(\mathcal{H}).
\end{equation}
Observe also that $\Phi_{U}$ is a Hilbert--Schmidt unitary map on $\mathcal{B}(\mathcal{H})$, i.e., $\Phi_{U}^\dagger = \Phi_{U}^{-1}$, and, consequently, it is normal, namely $[\Phi_{U} , \Phi_{U}^\dagger] = 0$, cf.\ Appendix~\ref{Jordan_op_app}.

We now show that the attractor subspace of a unitary channel is the entire space of operators over $\mathcal{H}$, i.e., $\Attr(\Phi_U)=\mathcal B(\mathcal H)$. To this purpose, write the spectral decomposition of $U$ as
\begin{equation}
    U=\sum_{k\in[N]} \omega_k P_k,
\end{equation}
where $\omega_k$ are the $N$ distinct unimodular eigenvalues of $U$, and $P_k = P_k^2=P_k^*$ the corresponding spectral  projections with $k \in [N]$. The eigenvalues of $\Phi_U$ are given by
\begin{equation}
\lambda_{k\ell} = \overline{\omega}_k \omega_\ell, \quad k,\ell \in [N],
\end{equation}
therefore $\lambda_{k\ell} \in \spec_{\mathrm{P}}(\Phi_U)$ for all $k,\ell \in [N]$, meaning that $\Attr(\Phi_U) = \mathcal{B}(\mathcal{H})$ by~\eqref{attr_def_old} and the normality of $\Phi_{U}$. 
Interestingly, the latter is one of the properties characterizing unitary evolutions, as stated in the next proposition.
\begin{proposition}
\label{unit_ch}
Let $\Phi$ be a quantum channel on $\mathcal{B}(\mathcal{H})$. Then the following statements are equivalent:
\begin{itemize}
\item[\textnormal{i.}] $\Phi$ is a unitary channel;
\item[\textnormal{ii.}] $\Attr(\Phi) = \mathcal{B}(\mathcal{H})$, namely $\spec(\Phi) = \spec_{\mathrm{P}}(\Phi)$;
\item[\textnormal{iii.}] $\Phi$ has unit determinant, $\det(\Phi)=1$;
\item[\textnormal{iv.}] $\Phi$ has unit KS rank, $\mathrm{rank}_{\mathrm{KS}}(\Phi)=1$ (see Section~\ref{repr_qc_sec});
\item[\textnormal{v.}] The inverse $\Phi^{-1}$ of $\Phi$ is a quantum channel.
\end{itemize}
\end{proposition}
\begin{proof}
First, the equivalence between i. and iii. readily follows from~\cite[Theorem 6.1]{wolf2012quantum}, if we take into account that $\Phi$ is a quantum channel by assumption.
Then, the argument before Proposition~\ref{unit_ch} shows that i. implies ii. and, moreover, an immediate consequence of iv.\ of Theorem~\ref{spec_qc} is that $|\det(\Phi)| \leq 1$ where the equality holds if and only if all the eigenvalues are peripheral, proving that i.\ and ii.\ are also equivalent,  by~\cite[Theorem 6.1]{wolf2012quantum} again. Of course, the unitarity of the channel implies iv.\ and, conversely, the property $\mathrm{rank}_{\mathrm{KS}}(\Phi)=1$ is equivalent to requiring that $\Phi(X) = A X A^\ast$ with $X \in \mathcal{B}(\mathcal{H})$ and for some $A \in \mathcal{B}(\mathcal{H})$, which in turn must be unitary by the trace-preservation of $\Phi$. Finally, the equivalence between i.\ and v.\ was proven in~\cite[Theorem III.2]{buscemi2005clean}.  
\end{proof}
In the above proposition, $\det(\Phi)$ refers to the determinant of the matrix representation $\hat{\Phi}$ of $\Phi$ with respect to some Hilbert--Schmidt orthonormal basis of $\mathcal{B}(\mathcal{H})$.
Concerning the steady subspace, observe that an element $X\in\mathcal B(\mathcal H)$ is a fixed-point of $\Phi_U$ if and only if 
\begin{equation}
\label{fix_unit_channels}
    X=\sum_{k\in[N]} 
    P_k X P_k\in\{U\}' ,
\end{equation}
where $\{U\}'$ denotes the commutant of $\{U\}$.
If we now call $\mathcal H_k := P_k \mathcal H$, then we can decompose the Hilbert space $\mathcal H$ as
\begin{equation}
\mathcal H= 
\bigoplus_{k\in[N]} \mathcal H_k,
\end{equation}
and, correspondingly, we can express $\Fix(\Phi_U)$ as
\begin{equation}
    \Fix(\Phi_U)= 
    \bigoplus_{k\in[N]} \mathcal B(\mathcal H_k)=\{U\}'.
\end{equation}
Therefore, if $\Phi_{U} \neq \mathsf{1}$, then $\Fix(\Phi_{U}) \subset \Attr(\Phi_{U}) = \mathcal{B}(\mathcal{H})$, that is,
there exists at least one asymptotic state of the evolution $\Phi_{U}$ not being stationary. 

\subsubsection{Dephasing channels}\label{sec:dephasing_channel}

Let us now discuss dephasing channels, intensively studied also over the last years~\cite{Devetak2005,winter2016operational,levick2017quantum,benatti2019hadamard,Puchaa2021,cunden2021generic,Kribs2024}, as they model the loss of coherence induced by environmental interactions~\cite{Zurek2003,BreuerPetruccione} as well as measurement operations~\cite{Kraus_71}. For any such channel, a preferred orthonormal basis, called \emph{decoherence} or \emph{pointer basis}, is left invariant by the evolution, while the coherences between some of the basis elements are suppressed. 
More precisely, given a pointer basis 
$\{ \ket{i} \}_{i\in[d]}$ of $\mathcal{H}$, a \emph{dephasing channel} $\Phi$ is defined by the condition
\begin{equation}\label{eq:diagonal_algebra}
    \Fix(\Phi) \supseteq \mathcal{D}_d := \Bigl\{ X = 
    \sum_{i\in[d]} x_i \ketbra*{i}{i} \;:\; x_i \in \mathbb{C} , \; i \in [d] \Bigr\} ,
\end{equation}
which is also equivalent to saying that~\cite[Theorem 1]{li1997special} 
\begin{equation}
\Phi(X) =
\sum_{i,j\in[d]} C_{ij}X_{ij}\ketbra{i}{j}, \quad X=
\sum_{i,j\in[d]} X_{ij}\ketbra{i}{j} \in \mathcal{B}(\mathcal{H}),
\label{eq:Schurdephasing}
\end{equation}
for some correlation matrix $C$, i.e., $C \geq 0$ with $C_{ii}=1$ for all $i \in [d]$. 
Yet another characterization of dephasing channels passes through the KS representation~\cite[Proposition 3]{Landau1993}: a map $\Phi$ is a dephasing channel if and only if it admits a KS representation whose operators $A_k$ belong to the commutative algebra $\mathcal D_d$.

Note that, if in~\eqref{eq:Schurdephasing} $|C_{ij}| <1$ for at least one couple $(i,j) \in [d] \times [d]$, then the QMC $\Sem =(\Phi^n)_{n\in\mathbb{N}}$ describes a nontrivial decoherence with respect to the pointer basis 
$\{ \ket{i} \}_{i\in[d]}$, namely $\Phi^n(\ketbra{i}{j}) \to 0$ as $n \to \infty$.
The asymptotic evolution of dephasing channels can be addressed in full generality~\cite{AFK_schur}. 

We now focus our attention on the asymptotics of a particular class of dephasing evolutions. If the KS operators form a complete family 
$\{P_k\}_{k\in[N]}$ of mutually orthogonal (Hermitian) projections in $\mathcal D_d$, then we obtain the \emph{pinching channel} 
\begin{equation}
\label{dep_channel}
    \Phi_{\mathrm{pinch}}(X):=
    \sum_{k\in[N]} P_k XP_k, \quad X \in \mathcal{B}(\mathcal{H}).
\end{equation}
Setting $\mathcal H_k=P_k\mathcal H$, we can decompose the Hilbert space as
\begin{equation}
    \mathcal H=
    \bigoplus_{k\in[N]}\mathcal H_k.
\end{equation}
Attractor and steady subspaces coincide, so there is no nontrivial asymptotic evolution, and they are of the form
\begin{equation}
\label{attr_fix_dep}    \Attr(\Phi_{\mathrm{pinch}})=\Fix(\Phi_{\mathrm{pinch}})=
\bigoplus_{k\in[N]} \mathcal B(\mathcal H_k).
\end{equation}
Note that, under a single application, the channel~\eqref{dep_channel} completely erases the coherence between the subspaces $\mathcal H_k$'s, yielding the creation of \emph{superselection sectors} or \emph{rules}~\cite{haag1964algebraic}. The system 
$\{ P_k \}_{k\in[N]}$ can be regarded as the spectral projections of an observable $A$ to be measured, with $P_k$ being the projection corresponding to the $k$-th output (eigenvalue) $a_k$. 

A second example of dephasing channel is given by the  Markovian channel $\Phi = \mathrm{e}^{\mathcal{L}}$ generated by
\begin{equation}
\label{normal_GKLS}
  \mathcal{L}(X) = 
  \sum_{\alpha\in[N]} \Bigl( L_{\alpha} X L_{\alpha}^\ast - \frac{1}{2} \{ L_\alpha^\ast L_\alpha , X \} \Bigr),  \quad X \in \mathcal{B}(\mathcal{H}),
\end{equation}
with jump operators $L_\alpha$  belonging to the diagonal algebra $\mathcal{D}_d$ in~\eqref{eq:diagonal_algebra}.  
It is easy to check that $\mathcal{D}_d \subseteq \Ker(\mathcal{L}) \subseteq \Fix(\Phi)$, so condition~\eqref{eq:diagonal_algebra} holds.

Interestingly, the attractor and steady subspaces of the evolution $\Phi = \mathrm{e}^{\mathcal{L}}$ generated by $\mathcal{L}$ in~\eqref{normal_GKLS} have again the structure~\eqref{attr_fix_dep} for some suitable Hilbert spaces $\mathcal H_k$. However, a generic initial state will converge to the asymptotic one only in the long-time limit of the corresponding QMC $\Sem$, at variance with what happens with the pinching channel~\eqref{dep_channel}.

As a particular application, think of  the standard way of describing dephasing in the Markovian qubit case, which involves the generator 
\begin{equation}
    \mathcal{L}(X) = \gamma(\sigma_3 X \sigma_3 -  X), 
\end{equation}
where $ \sigma_3$ is the third Pauli matrix, and $\gamma>0$. In the Bloch
representation~\eqref{eq:Bloch}, the corresponding evolution reads
\begin{equation}
\label{eq:dephasing_bloch}
    \mathrm{e}^{\mathcal{L}}\rho(x,y,z)
    =\rho(\eta x,\eta y,z),
    \qquad \eta=\mathrm{e}^{-2\gamma}. 
\end{equation}
Thus, in the eigenbasis $\{ \ket{i} \}_{i=1}^2$ of $\sigma_3$, populations are preserved while coherences decay: geometrically,
the Bloch ball is mapped onto an ellipsoid with semiaxes
$(\eta,\eta,1)$, leaving the segment
$\bm{z} := \{(0,0,z):|z|\leq 1\}$ fixed pointwise, see left panel of Figure~\ref{fig:Bloch}.
Thus, under the action of the QMC $\Sem$ induced by $\Phi = e^{\mathcal{L}}$ in the form~\eqref{eq:dephasing_bloch}, the image contracts onto $\bm{z}$, which represents
the diagonal density operators, in the limit $n\to\infty$. The peripheral projection is precisely the
pinching channel~\eqref{dep_channel} with
$P_j=\ketbra{j}{j}$, $j=1,2$,
which itself yields the same contraction after a single step.

\subsubsection{Amplitude damping channels}
\label{sec:amplitude_damping_channels}
The class of amplitude damping channels describes the depletion of state populations from a given subspace of the system’s Hilbert space. For example, they model the decay of an atom interacting with the electromagnetic field~\cite{BreuerPetruccione}, the interaction of a spin with a thermal bath at zero temperature~\cite{giovannetti2005information}, or the scattering of a photon in an interferometer~\cite{srikanth2008squeezed}. 

Given an orthonormal basis 
$\{ \ket{i} \}_{i\in[d]}$ of $\mathcal{H}$, an amplitude damping channel can be modeled as
\begin{equation}
\label{ampl_channel_2_def}
    \Phi_{\mathrm{ampl}}(X) = 
    \sum_{i\in\{1,2\}} A_i X A_i^\ast, \quad X \in \mathcal{B}(\mathcal{H}),\; A_1 := \ketbra{1}{1},\; A_2:= 
    \sum_{i\in[d-1]} \ketbra{i}{i+1},
\end{equation}
or through the Markovian channel
\begin{equation}
\label{ampl_channel_def}
      \Phi_{\mathrm{ampl}}^\prime(X) = \mathrm{e}^{\mathcal{L}}(X), \quad \mathcal{L}(X) = \gamma(A_2 X A_2^{\ast}-\frac{1}{2} \{ A_2^{\ast} A_2 , X \}), \quad X \in \mathcal{B}(\mathcal{H}),
\end{equation}
for some $\gamma >0$, see also~\cite{NielsenChuang,chessa2021quantum} for possible variants.

Observe that under the action of the channels~\eqref{ampl_channel_2_def} and~\eqref{ampl_channel_def} the population from the level $\ket{n+1}$ is transferred to the level $\ket{n}$ for any $n \in[d-1]$ under the action of the QMC.
Consequently, the unique stationary state is the projection onto the ground state $|1 \rangle$
and there is no dynamics in the long-time limit, so the asymptotic and steady subspaces read
\begin{equation}
	\Attr(\Phi_{\mathrm{ampl}}) = \Fix(\Phi_{\mathrm{ampl}}) = \mathbb{C} \ketbra{1}{1},
\end{equation}
and similarly for $\Phi_{\mathrm{ampl}}^\prime$. Therefore, if we 
decompose the Hilbert space as
\begin{equation}
    \mathcal H=\mathcal{H}_0^\perp \oplus \mathcal{H}_0, \qquad \mathcal{H}_0 = \mathbb{C} \ket{1}, \quad
    \mathcal{H}_0^\perp = \mathrm{span} \{\ket{i}\}_{i=2}^d, 
\end{equation}
then we conclude that, in the asymptotic limit, the system cannot be in any state supported in the $(d-1)$-dimensional transient space $\mathcal{H}_0^\perp$
of $\Sem$; see Proposition~\ref{supp_QC_prop}.
In Sections~\ref{bas_notions_subsec} and~\ref{sec:structure_theorems}, we  refer to such an evolution with a nontrivial transient space as non-faithful.
An equivalent description of these decay processes can be obtained by studying the evolution of the density operator $\rho$ within the subspace $\mathcal{H}_0^\perp$, using an optical potential~\cite{Saverio_Darek_review}, and considering a dynamics that does not preserve probability.

In the qubit case, the generator~\eqref{ampl_channel_def} takes the form
\begin{equation}
\label{Markov_ampl_qubit}
\mathcal{L}(X) = \gamma\bigl(\sigma_+ X \sigma_- - \tfrac{1}{2} \{ \sigma_{-}\sigma_+ , X \} \bigr), \quad X \in \mathcal{B}(\mathbb{C}^2),
\end{equation}
where $\sigma_{\pm} = \tfrac{1}{2} (\sigma_1 \pm \mathrm{i}\sigma_2)$ denote the ladder operators and $\gamma>0$. {
Since $\sigma_+=\ketbra{1}{2}$, with $\{ \ket{i} \}_{i=1}^2$ again being the eigenbasis of $\sigma_3$, this generator describes decay towards the ground state
$\ket{1}$. Its action in the Bloch representation is
\begin{equation}
\label{eq:amplitude_damping_bloch}
    \mathrm{e}^{\mathcal{L}}\rho(x,y,z)
    =\rho\bigl(\sqrt{1-p}\,x,\sqrt{1-p}\,y,(1-p)z+p\bigr),
    \qquad p=1-\mathrm{e}^{-\gamma}. 
\end{equation}
For $0<p<1$, the image of the Bloch ball is an ellipsoid centered at
$(0,0,p)$, with semiaxes
$(\sqrt{1-p},\sqrt{1-p},1-p)$, tangent to the unit sphere at the
north pole (see Figure~\ref{fig:Bloch}). Under the action of $\Sem$ generated by $\Phi = e^{\mathcal{L}}$ in the form~\eqref{eq:amplitude_damping_bloch}, this ellipsoid contracts to
$(0,0,1)$, corresponding to the stationary state $\ketbra{1}{1}$, in the limit $n \to \infty$.
By contrast, the discrete channel~\eqref{ampl_channel_2_def}
already implements complete damping for a qubit: it corresponds
to $p=1$ and maps every state to $\ketbra{1}{1}$ in a single step.
}
\subsubsection{Depolarizing channels}\label{sec:depolarizing_channels}

Let us consider the depolarizing channel, which provides a standard model of incoherent noise in current experimental platforms~\cite{NielsenChuang} and has consequently received considerable attention in the quantum information community~\cite{king2003capacity,crow2014classical,vovrosh2021simple,urbanek2021mitigating}. The channel is defined as
\begin{equation}\label{eq:dep_channel}
    \Phi_{\mathrm{dep}}(X)= (1-\lambda)X+ {\lambda}\mathrm{Tr}(X) \frac{\mathbb{I}}{d}, \quad X \in \mathcal{B}(\mathcal{H}),
\end{equation}
with $0<\lambda\leq 1$, $d=\dim(\mathcal{H})$. Notice that for $\lambda\neq 1$  $\Phi_{\mathrm{dep}}$ is Markovian, namely $\Phi_{\mathrm{dep}} = \mathrm{e}^{\mathcal{L}_{\mathrm{dep}}}$ with 
\begin{equation}
    \mathcal{L}_{\mathrm{dep}}(X) = \gamma \biggl(\mathrm{Tr}(X) \frac{\mathbb{I}}{d} - X\biggr), \quad \lambda = 1-e^{-\gamma} , \; X \in \mathcal{B}(\mathcal{H}).
\end{equation}
Thus, under the action of the channel~\eqref{eq:dep_channel}, an error, transforming the state into the maximally mixed one, occurs with probability $\lambda$, while the state remains invariant with probability $1-\lambda$. Under the repeated action of the QMC generated by $\Phi_{\mathrm{dep}}$, the state will be transformed into the maximally mixed one with higher and higher probability. 

Attractor and steady subspaces coincide, as in the previous two examples, and are given by the  one-dimensional space
\begin{equation}
\Attr(\Phi_{\mathrm{dep}})=\Fix(\Phi_{\mathrm{dep}})=\mathbb C \mathbb I.
\end{equation}
Consequently, the asymptotic evolution induced by the channel $\Phi_{\mathrm{dep}}$ is again trivial.

In~\eqref{eq:dep_channel} we can also replace the maximally mixed state $\mathbb I/d$ with a general state $\rho$, so that both the attractor and steady subspaces are equal to $\mathbb C\rho$. In particular,  in the qubit case, if we set $\rho = \ketbra{1}{1}$ and $\lambda = 1$, then we obtain the amplitude damping channel defined in~\eqref{ampl_channel_2_def}.  

For a qubit, the depolarizing channel can be represented in KS form~\eqref{op_sum_rep} as
\begin{equation}\label{eq:dep_channel_qubit}
    \Phi_{\mathrm{dep}}(X)=\Bigl(1-\frac{3\lambda}{4}\Bigr)X+
    \sum_{i\in[3]} \frac{\lambda}{4}\sigma_i X \sigma_i , \quad X \in \mathcal{B}(\mathbb{C}^2),
\end{equation}
where $\sigma_i$, with $i \in [3]$, are the Pauli matrices.
Therefore, we can think of the channel~\eqref{eq:dep_channel} as inducing 
a bit-flip error $\sigma_1 X \sigma_1$, or a phase-flip error $\sigma_3X\sigma_3$, or their combination $\sigma_2 X \sigma_2$, each with probability $\lambda/4$. In the Bloch representation, Equation~\eqref{eq:dep_channel_qubit} becomes
\begin{equation}
\label{eq:depolarizing_bloch}
    \Phi_{\mathrm{dep}}\bigl(\rho(\bm r)\bigr)
    =\rho\bigl((1-\lambda)\bm r\bigr).
\end{equation}
Hence, the Bloch ball undergoes an isotropic contraction to the
concentric ball of radius $1-\lambda$ (see Figure~\ref{fig:Bloch}). Under repeated application,
the Bloch vector evolves as
$\bm r\mapsto(1-\lambda)^n\bm r$ and converges to
the origin $\bm{r} = 0$, representing the maximally mixed state $\mathbb{I}/2$.

\subsection{Asymptotic structure theorems}
\label{sec:structure_theorems}
We have shown in Section~\ref{bas_notions_subsec}
that characterizing the asymptotic behaviour of a QMC $\Sem$ generated by a quantum channel $\Phi$ amounts  to determining the attractor subspace $\Attr(\Phi)$ of $\Phi$ and its action on this space. We will do this in the next three general structure theorems covering both the Schr{\"o}dinger and Heisenberg pictures of the  evolution. In particular, the third structure theorem will allow us to derive the fine structure of the fixed-point subspace of quantum channels; cf. Equation~\eqref{fix_dec_Wolf}.

\subsubsection{Schr\"odinger picture}
\label{Schr_asympt_subsec}
\begin{figure}
    \begin{center}
    \includegraphics[width=0.90\textwidth]{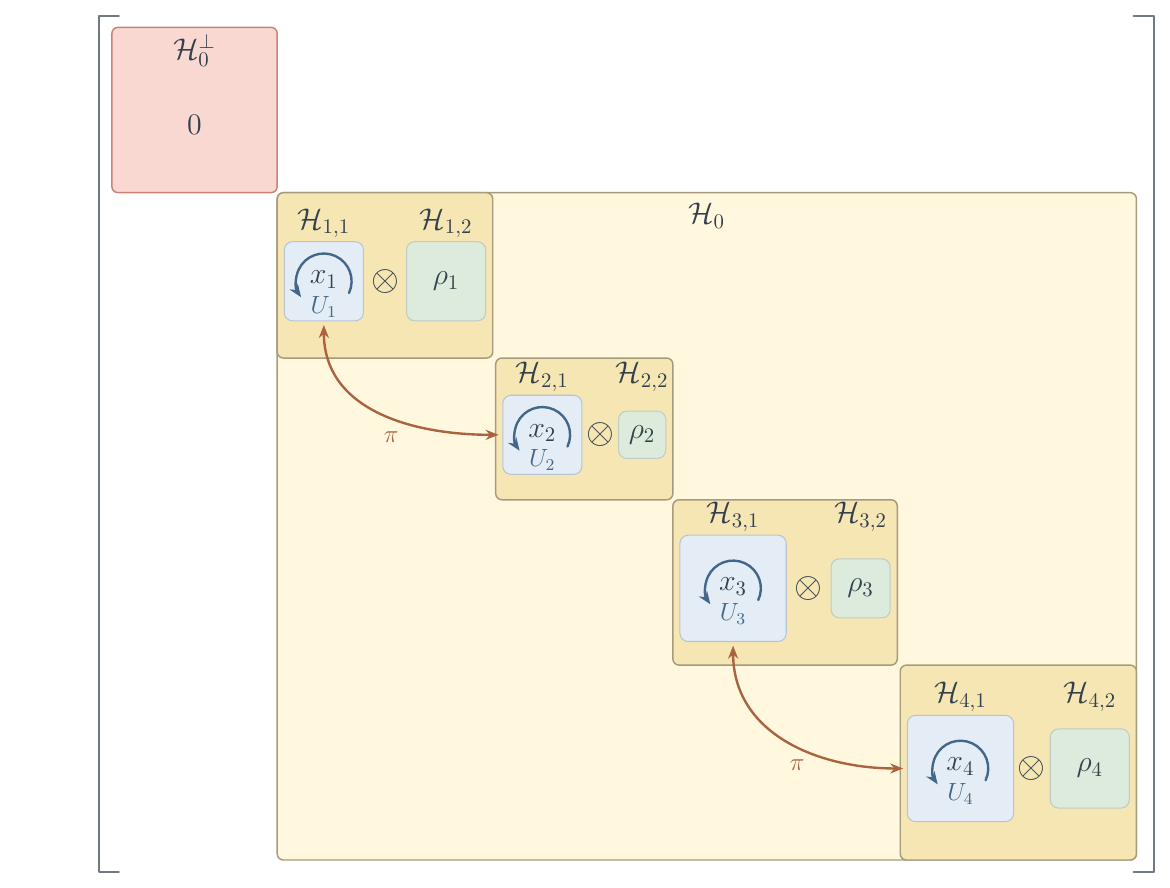}
    \end{center}
    \caption{Schematic illustration of the asymptotic structure described in Theorem~\ref{Wolf_struc} for $M=4$ and $\pi = (2,1,4,3)$. The first red box represents the zero part in~\eqref{eq:element_in_the_attractor} acting on $\mathcal{H}_{0}^\perp$, while the remaining orange ones are the nontrivial blocks in~\eqref{eq:element_in_the_attractor} acting on $\mathcal{H}_{k,1} \otimes \mathcal{H}_{k,2}$, $k \in [M]$. The unitary evolutions induced by $U_k$ on the $\mathcal{H}_{k,1}$ factor of the $k$-th sector are pictorially represented with a blue arrow, and permutations between them are represented by red arrows. Note that permutations can exchange only operators acting on the first factors (light blue blocks) with the same dimension $d_k=\dim(\Hi_{k,1})$, whereas no constraints are imposed on the dimensions $\dim(\Hi_{k,2})$ of the second factors (green blocks); cf.\ Remark~\ref{attr_inva}. Finally, the non-decaying subspace $\mathcal{H}_{0}$, cf. Equation~\eqref{eq:H0def}, is depicted by a light yellow box.
    } 
\label{fig:Wolf_structure_theorem}
\end{figure}

The first result, obtained in~\cite[Theorem 6.16]{wolf2012quantum}, concerns the long-time dynamics of a QMC in the Schr{\"o}dinger picture, and exploits the structure of the asymptotic space provided by Proposition~\ref{supp_QC_prop}; see Figure~\ref{fig:Wolf_structure_theorem} for a schematic representation.

\begin{theorem}[Structure theorem of the Schr{\"o}dinger asymptotics]
\label{Wolf_struc}
Let $\Sem=(\Phi^n)_{n\in\mathbb{N}}$ be a quantum Markov chain generated by a quantum channel $\Phi$ on $\mathcal{B}(\mathcal{H})$ with attractor subspace $\Attr(\Phi)$ and peripheral projection $\mathcal{P}_{\mathrm{P}}$.				
Then, there exist
\begin{itemize}
    \item[\textnormal{i.}] a family of Hilbert spaces $\mathcal{H}_{k,i}$, with $k\in[M]$ and $i\in \{ 1,2 \}$, that decompose the Hilbert space $\mathcal{H}$ as
    \begin{equation}
        \mathcal{H} = \mathcal{H}_0^\perp \oplus \mathcal{H}_0, \quad 	\mathcal{H}_0= 
        \bigoplus_{k\in[M]} \bigl( \mathcal{H}_{k,1}\otimes \mathcal{H}_{k,2} \bigr) ,
        \label{H_dec}
    \end{equation}
	with $\mathcal{H}_0=\Ran(\mathcal{P}_\mathrm{P}(\mathbb{I}))$, and
	\item[\textnormal{ii.}] positive-definite density operators $\rho_{k}$ on $\mathcal{H}_{k,2}$, with $k\in[M]$,
\end{itemize}
such that the attractor subspace of $\Sem$ splits as
\begin{equation}\label{attr_struc}
    \Attr(\Phi)=0\oplus
    \bigoplus_{k\in[M]} \bigl(\mathcal{B}(\mathcal{H}_{k,1}) \otimes \mathbb{C}\rho_{k}\bigr).
\end{equation}

Furthermore, there exist a permutation $\pi$ on the set $[M]$ and unitaries $U_{k} : \mathcal{H}_{\pi(k),1} \to \mathcal{H}_{k,1}$ such that, for a given $X\in\Attr(\Phi)$ in the form
\begin{equation}\label{eq:element_in_the_attractor}
     X=0\oplus
     \bigoplus_{k\in[M]} ( x_{k} \otimes \rho_{k}), 
\end{equation}
the action of $\Phi$ on $X$ reads
\begin{equation}\label{struc_phiP}
\Phi(X)=0\oplus 
\bigoplus_{k\in[M]} \bigl( U_{k}x_{\pi(k)}U_{k}^{\ast}\otimes \rho_{k}\bigr). 
\end{equation}
\end{theorem}
\begin{remark}
\label{attr_inva}
The permutation $\pi$ 
must necessarily satisfy the property $d_{\pi(k)}=d_k$ with $d_{k} :=\dim(\mathcal{H}_{k,1})$ for all $k\in[M]$, in line with the invariance of $\Attr(\Phi)$ under $\Phi$, see Equation~\eqref{Attr_inva}. 
\end{remark}

An algorithm to derive the structure~\eqref{attr_struc} of the attractor subspace of a given channel, which is polynomial in the Hilbert space dimension, has recently been presented in~\cite{fawzi2024capacities}. 
Notice that Theorem~\ref{Wolf_struc} and, in particular, the structure~\eqref{attr_struc} of the attractor subspace provides an explicit decomposition for the range $\Ran(\mathcal{P})$ of an idempotent quantum channel $\mathcal{P}$, used to address several problems in quantum information theory~\cite{kuperberg2003capacity,delsol2025emulation}. 

The general structure~\eqref{attr_struc} of the attractor subspace of a QMC is the combination of the asymptotic subspaces that emerged in the examples discussed in Section~\ref{sec:examples}. First, the zero component acting over $\mathcal H_0^\perp$ in~\eqref{attr_struc} reveals the effect of decay, as shown by the amplitude damping channel discussed in Section~\ref{sec:amplitude_damping_channels}. Next, the incoherent sum over the $M$ subspaces $\mathcal H_{k,1}\otimes \mathcal H_{k,2}$ is connected to dephasing, described in Section~\ref{sec:dephasing_channel}. The density operators $\rho_k$ describe the complete loss of information in the factors $\mathcal H_{k,2}$, which asymptotically relax to unique states $\rho_k$; generalized depolarizing channels introduced in Section~\ref{sec:depolarizing_channels} provide the simplest example of this mixing mechanism. Finally, the asymptotic dynamics~\eqref{struc_phiP} involves both unitary evolutions, discussed in Section~\ref{sec:unitary_evolution}, and, surprisingly, the emergence of permutations. The former contribution to open-system asymptotics is physically expected, while the latter is more intriguing, and it will be analyzed more in depth in Section~\ref{sec:cyclic_dec} and Section~\ref{unit_asympt_sec},
where an example of dynamics exhibiting permutations in the long-time limit is also provided.

	\subsubsection{Heisenberg picture}
\label{Heis_asympt_subsec}
Let us now state the analogue of Theorem~\ref{Wolf_struc} for a QMC in the Heisenberg picture of the evolution, which was obtained in~\cite[Theorem 3]{AFK_asympt4}; see also~\cite[Proposition 3]{albert2019asymptotics}.

\begin{theorem}[Structure theorem of the Heisenberg asymptotics]
\label{struc_Heis_th}
Let $\Sem^\dagger$ be an adjoint quantum Markov chain generated by a UCP map $\Phi^\dagger$ on $\mathcal{B}(\mathcal H)$ with attractor subspace $\Attr(\Phi^\dagger)$ and peripheral projection $\mathcal{P}_{\mathrm{P}}^\dagger$. Then, there exist
\begin{itemize}
\item[\textnormal{i}.] a family of Hilbert spaces $\mathcal H_{k,i}$, with $k\in[M]$ and $i\in \{ 1,2 \}$, that decompose the Hilbert space $\mathcal H$ as
	\begin{equation}\label{eq:dec_hilbert_heisenberg}
		\mathcal{H}=\mathcal H_0^\perp\oplus \mathcal{H}_0,\quad \mathcal H_0 = 
        \bigoplus_{k\in[M]} \bigl(\mathcal{H}_{k,1} \otimes \mathcal{H}_{k,2}\bigr),
	\end{equation}
	with $\mathcal{H}_0=\Ran( \mathcal{P}_{\mathrm{P}}(\mathbb{I}))$, and
    \item[\textnormal{ii}.] a UCP map $\Lambda:\mathcal{A} \rightarrow \mathcal B(\mathcal H_0^\perp)$, where
    \begin{equation}
         \mathcal{A} = 
         \bigoplus_{k\in[M]} \bigl(\mathcal{B} (\mathcal{H}_{k,1})\otimes \mathbb{I}_{k,2}\bigr),
    \end{equation}
\end{itemize}
such that the attractor subspace of $\Phi^\dagger$ reads
\begin{equation}
	\label{struc_attr_Heis}
	\Attr (\Phi^\dagger)=\bigl\{ \Lambda(X) \oplus X \,:\, X \in \mathcal A \bigr\}.
\end{equation}
Furthermore, there exist a permutation $\pi$ on the set $[M]$ and unitaries $U_{k} : \mathcal{H}_{k,1} \to \mathcal{H}_{\pi(k),1}$ such that, for a given $ \Lambda ( X )\oplus X \in \Attr(\Phi^\dagger)$ with
\begin{equation}
    X = 
    \bigoplus_{k\in[M]} ( x_k\otimes \mathbb{I}_{k,2})\in \mathcal{A} ,\label{A_struc_elem}
\end{equation}
the action  of $\Phi^\dagger$ is given by 
\begin{align}\label{eq:asymptotic_action_heisenberg}
   \Phi^\dagger(\Lambda(X) \oplus X)=\Lambda(X')\oplus X', \quad  X' &=  
   \bigoplus_{k\in[M]} \bigl( U_{\pi(k)}^\ast x_{\pi(k)}U_{\pi(k)} \otimes\mathbb{I}_{k,2}\bigr).
\end{align}
\end{theorem}
Notice that we defined $\Lambda$ as a completely positive map on a $C^\ast$-subalgebra $\mathcal{A}$ of $\mathcal{B}(\mathcal{H}_{0})$. Indeed, complete positivity may also be defined in this more general context~\cite[Chapter 3]{paulsen2002completely}. 
Let us now make several remarks about Theorem~\ref{struc_Heis_th}, whose algebraic consequences will be further explored in Section~\ref{alg_sec}.
First, the decomposition~\eqref{eq:dec_hilbert_heisenberg} of the Hilbert space $\mathcal{H}$ for a UCP map $\Phi^\dagger$ coincides with the one~\eqref{H_dec} of the corresponding quantum channel $\Phi$. Moreover, the unitaries $U_{\pi(k)}$ are placed on the right of $x_{\pi(k)}$ instead of on the left, according to the standard convention for the Heisenberg picture. In addition, the permutation $\pi$ appearing in the asymptotic dynamics~\eqref{eq:asymptotic_action_heisenberg} is the inverse of the one emerging in the Schr{\"o}dinger picture~\eqref{struc_phiP}, therefore $\pi$ satisfies the property discussed in Remark~\ref{attr_inva}.

   In the faithful case, since $\mathcal H_0^\perp=\{0\}$, the structure of the attractor subspace of $\Phi^\dagger$ reads
\begin{equation}
\label{attr_phi00_struc}
\Attr(\Phi^\dagger)=  
\bigoplus_{k\in[M]} (\mathcal B(\mathcal H_{k,1})\otimes \mathbb{I}_{k,2}),
\end{equation}
while the asymptotic dynamics of the UCP map $\Phi^\dagger$ is in the form
\begin{equation}
X = 
\bigoplus_{k\in[M]} (x_k\otimes \mathbb{I}_{k,2}) \mapsto \Phi^\dagger(X)= 
\bigoplus_{k\in[M]} (U_{\pi(k)}^\ast x_{\pi(k)}U_{\pi(k)}\otimes \mathbb I_{k,2}). 
\end{equation}

The non-faithful case is more delicate, and the structure of the asymptotics shown in Theorem~\ref{struc_Heis_th} can be obtained using the decomposition of idempotent UCP maps first derived in~\cite[Theorem 7.1]{gartner2012coherent}. In particular, with the notation of Equation~\eqref{eq:asymptotic_action_heisenberg}, one has $X^\prime = \Phi_{\mathrm{faith}}^\dagger(X)$, where $\Phi_{\mathrm{faith}}^\dagger$ denotes the Hilbert--Schmidt adjoint of the induced faithful channel $\Phi_{\mathrm{faith}}$ defined by~\eqref{faith_ind_def}. This also allows us to generalize the correspondence~\eqref{Schr_Heis_corresp} to the non-faithful case~\cite[Section 4]{AFK_asympt4}. 

Importantly, the structure of the asymptotic subspace given by~\eqref{eq:asymptotic_action_heisenberg} is still block-diagonal with respect to the decomposition~\eqref{eq:dec_hilbert_heisenberg} of $\mathcal{H}$, but it reveals the presence of a non-zero block supported on $\mathcal{H}_{0}^\perp$, at variance with what happens in the Schr{\"o}dinger picture.    
We illustrate the point with a simple example.
\begin{example}
\label{ex_6}
Take the idempotent UCP map $\mathcal{P}^\dagger$ on the space $\mathcal{B}(\mathbb{C}^2) =\mathcal{M}_2(\mathbb{C})$ of complex matrices of order 2
\begin{equation}
\label{1_qubit}
\mathcal{P}^\dagger(X) = 
\begin{pmatrix}
x_{11} & 0 \\
0 & x_{11}
\end{pmatrix} 
= \Tr(\rho X)\, \mathbb{I}, \quad \rho =  \begin{pmatrix}
1 & 0 \\
0 & 0
\end{pmatrix}.
\end{equation}
The map is non-faithful and, using the notation from Theorem~\ref{struc_Heis_th}, we have $\mathcal{H}_{0} \simeq \mathcal{H}_{0}^{\perp} \simeq \mathbb{C}$ and
\begin{equation}
\Lambda: \mathbb{C} \to \mathbb{C}, \quad \Lambda(x_{11}) = x_{11}.
\end{equation}
Note that $\mathcal{P}$ is both the steady and peripheral projections of the amplitude damping channel~\eqref{ampl_channel_def} in the qubit case.
\end{example}

\subsubsection{Cyclic decomposition}\label{sec:cyclic_dec}
We now  present a third structure theorem for the attractor subspace in the Schr{\"o}dinger picture. Essentially, in each disjoint cycle of the permutation $\pi$ appearing in~\eqref{struc_phiP}, the channel can be written as the tensor product between a unitary evolution and a permutation. 
In Theorem~\ref{Wolf_struc}, by contrast, both contributions act on the first tensor factor of each of the $M$ blocks in the decomposition~\eqref{struc_phiP} of the asymptotic dynamics. 
 To simplify the notation, we consider the case where the permutation $\pi$ has only one cycle, but the result can be extended to the general case in a straightforward way~\cite[Theorem 3.2]{AFK_asympt_1}.
\begin{theorem}[Cyclic decomposition]\label{thm:cycles}
    Let $\Sem$ be a QMC generated by a quantum channel $\Phi$ on $\mathcal{B}(\mathcal{H})$, and suppose that the permutation $\pi$ in Equation~\eqref{struc_phiP} is cyclic.
    Then, there exist
    \begin{itemize}
\item[\textnormal{i.}] a Hilbert space $\mathcal H_1$ and $M$ Hilbert spaces $\mathcal H_{k,2}$, with $k\in[M]$, that decompose the Hilbert space $\mathcal H$ as
        \begin{equation}\label{H:cycl_dec}
            \mathcal H=\mathcal H_0^\perp \oplus \mathcal H_0,\quad \mathcal H_0=\mathcal H_1\otimes \Bigl(  
            \bigoplus_{k\in[M]} \mathcal H_{k,2}\Bigr),
        \end{equation}
        with $\mathcal H_0 =\Ran(\mathcal P_{\mathrm{P}}(\mathbb I))$, and
\item[\textnormal{ii.}] positive-definite density operators $\rho_k\in\mathcal{S}(\mathcal H_{k,2})$, with $k\in[M]$,
    \end{itemize}
    such that the attractor subspace $\Attr(\Phi)$ of $\Phi$ reads
    \begin{equation}\label{Attr_cyc_struc}
        \Attr(\Phi)=  0\oplus \mathcal{B}(\mathcal H_1)\otimes \mathcal{K}, \qquad \mathcal{K} := \bigoplus_{k\in[M]} \mathbb{C}\rho_k.
    \end{equation}
    Furthermore, there exists a unitary $U\in\mathcal B(\mathcal H_1)$ with $\{ U^M \}^\prime = \{ U \}^\prime$ such that, given an element of $\Attr(\Phi)$ in the form
    \begin{equation}
        X=0\oplus X_1\otimes 
        \Bigl(\bigoplus_{k\in[M]}\lambda_k \rho_k\Bigr),
    \end{equation}
    the action of $\Phi$ reads
    \begin{equation}\label{eq:action_cyclic_dec}
        \Phi(X)=0\oplus U X_1 U^* \otimes 
        \Bigl(\bigoplus_{k\in[M]} \lambda_{\pi(k)} \rho_k\Bigr)
    \end{equation}
    and extends by linearity to all the elements of $\Attr(\Phi)$.
\end{theorem}
		
\begin{remark}\label{cycl_vs_Wolf_rmk}
     For a trivial $\pi=\mathrm{id}$, the structures provided by Theorems~\ref{Wolf_struc} and~\ref{thm:cycles} coincide. Several sufficient conditions for $\pi=\mathrm{id}$ will be discussed in Section~\ref{unit_asympt_sec}.
\end{remark}
		
Equation~\eqref{Attr_cyc_struc} allows us to obtain the relation between the attractor and steady subspaces of $\Phi$ 
\begin{equation}\label{eq:fix_cyclicdec}
    \Fix(\Phi)= 0 \oplus \mathrm{Fix} (\Phi_{U})\otimes \mathbb{C} {\rho}, \quad {\rho}=\frac{1}{M}
    \bigoplus_{k\in[M]}\rho_k,
\end{equation}
where $\rho$ is the unique positive-definite state in $\mathcal{K}$ invariant under the cyclic permutation $\pi$, and $\Phi_U$ is the unitary channel $\Phi_U(X_1)=U X_1 U^*$ with  $X_1 \in \mathcal B(\mathcal H_1)$. Moreover, according to~\eqref{fix_unit_channels}, the fixed-point subspace $\Fix(\Phi_{U})$ of $\Phi_U$ reads
\begin{equation}
    \Fix(\Phi_U)=\{  U \}^\prime = 
    \bigoplus_{r\in[N]} \mathcal{B}(\mathcal{H}_{1,r}),
\end{equation}
with $\mathcal{H}_{1,r}$ being the $r$-th eigenspace of $U$, so
Equation~\eqref{eq:fix_cyclicdec} becomes
\begin{equation}\label{fix_dec_Wolf}
    \Fix(\Phi) = 0\oplus 
    \bigoplus_{r\in[N]} \bigl(\mathcal{B}(\mathcal{H}_{1,r}) \otimes \rho \bigr).
\end{equation}
Alternatively, the steady subspace $\Fix(\Phi)$ of a channel $\Phi$ admits the decomposition in Theorem~\ref{Wolf_struc}, since it is the image of the steady projection~\eqref{P_for}, which is an idempotent quantum channel by Proposition~\ref{peripheral_Wolf}. See also~\cite[Theorem 6.14]{wolf2012quantum},~\cite[Theorem 2]{guan2018structure} and~\cite[Theorems 19, 21]{fagnola2019role}. 

As for the attractor subspace, Equation~\eqref{fix_dec_Wolf} provides the explicit structure of the fixed-point subspace of a quantum channel in the case of a cyclic permutation $\pi$, see~\cite[Equation (63)]{AFK_asympt_1} for the general formula.
Moreover, it is also possible to obtain an analogue of Theorem~\ref{thm:cycles} for the attractor subspace of Heisenberg evolutions, which, in turn, allows us to infer the explicit structure of the fixed-point subspace discussed in~\cite[Appendix A]{AFK_asympt4}, see also~\cite[Theorem 3]{blume2008characterizing} and~\cite[Theorem 8.3]{gartner2012coherent}.

Finally, let us show why we can interpret the cyclic permutation $\pi$ in Equation~\eqref{eq:action_cyclic_dec} as a classical evolution taking place in the long-time limit of the dynamics. Specifically, following the notation of Theorem~\ref{thm:cycles}, 
let us take into account the map 
\begin{equation}
\label{attr_Phi_pi_ell_elem}
 \Phi_{\pi}: \mathcal{K} \to \mathcal{K} \qquad X = 
 \bigoplus_{k\in[M]} \lambda_k \rho_k \mapsto \Phi_{\pi}(X) := 
 \bigoplus_{k\in[M]} \lambda_{\pi(k) } \rho_k, \quad \lambda_k \in \mathbb{C},
\end{equation}
where $\mathcal{K}$ is defined via~\eqref{Attr_cyc_struc},
which describes the component of the asymptotic evolution, cf.\ Equation~\eqref{eq:action_cyclic_dec}, acting on $\mathcal{K}$.
The action of $\Phi_{\pi}$ only consists of permuting the complex coefficients $\lambda_k$ of the direct-sum combination of the density operators $\rho_k$ and, interestingly, it is equivalent to the action of a classical Markov chain induced by $\pi$ and involving the $M$ states $\rho_k$. 
For instance, if the initial state is $\rho(0)=\rho_1$, then the chain is made up of the following transitions
\begin{equation}
\rho(0) =\rho_1\rightarrow {\rho}_{\pi^{-1}(1)}\rightarrow \dots \rightarrow {\rho}_{1},
\end{equation}
so it comes back to the initial state $\rho_1$ after $M$ steps.
Alternatively, the map $\Phi_{\pi}$ in~\eqref{attr_Phi_pi_ell_elem} describes a classical dynamics in the sense that the subspace $\rho^{-1/2} \mathcal{K} \rho^{-1/2}$, with $\rho$ being defined by~\eqref{eq:fix_cyclicdec}
is a commutative $C^\ast$-algebra; see Appendix~\ref{intro_Cast}.

\section{Asymptotic dynamics of quantum dynamical semigroups}
\label{asympt_QDS_sec}

We introduced continuous-time Markovian evolutions, or quantum dynamical semigroups (QDSs), in Section~\ref{QDS_sec} as an alternative to the discrete-time quantum Markov chains considered in Section~\ref{qc_sec} for describing the evolution of open quantum systems. In the present section, we discuss the asymptotic behaviour of QDSs, following the approach previously developed for QMCs.

\subsection{Basic notions}
\label{subsec:asympt_QDS_1}
Let $\Sem=(\Phi_t)_{t\in\mathbb{R}^+}$ be a QDS with GKLS generator $\mathcal{L}$. Then,  the long-time behaviour of the system corresponds to the limit $t\to \infty$ and is governed by the peripheral spectrum $\spec_{\mathrm{P}}(\mathcal{L})$ of $\mathcal{L}$, defined in~\eqref{spec_L_def}. 
Accordingly, the asymptotic dynamics of $\Sem$ is confined  to the attractor subspace $\Attr(\mathcal{L})$. More precisely, for any initial state $\sigma\in\mathcal{S}(\mathcal{H})$, the component of $\Phi_t(\sigma)$ outside $\Attr(\mathcal{L})$ vanishes as $t\to \infty$.

As for quantum channels, let us denote with 
\begin{equation}
\label{steady_proj_L}
  \mathcal{P} := \mathcal{P}_{\lambda_0 = 0}  
\end{equation} 
the \emph{steady projection} of $\Sem \equiv(\Phi_t)_{t\in\mathbb{R}^+}$, which is the spectral projection of $\mathcal{L}$ corresponding to the eigenvalue $\lambda _0= 0$, projecting onto the fixed-point subspace $\Fix(\Sem) = \Ker(\mathcal{L})$ of $\Sem$; see Equation~\eqref{Ker_L_def}.  Moreover, the \emph{peripheral projection} of $\mathcal{L}$ is defined as
\begin{equation}
\label{per_proj_def_L}
\mathcal{P}_{\mathrm{P}} := \sum_{\lambda \in \spec_{\mathrm{P}} (\mathcal{L})} \mathcal{P}_\lambda,
\end{equation} 
where $\mathcal{P}_\lambda$ with $\lambda \in \spec_{\mathrm{P}} (\mathcal{L})$ 
is the spectral projection of $\mathcal{L}$ corresponding to  an eigenvalue in the peripheral spectrum~\eqref{spec_L_def}. Clearly, the map $\mathcal{P}_{\mathrm{P}}$ projects onto   $\Ran(\mathcal{P}_{\mathrm{P}}) = \Attr(\mathcal{L})$. Furthermore, 
\begin{equation}
\label{per_gen_def}
\mathcal{L}_{\mathrm{P}} := \sum_{\lambda \in \spec_{\mathrm{P}} (\mathcal{L})} \lambda \mathcal{P}_\lambda = \mathcal{L}  \mathcal{P}_{\mathrm{P}}
\end{equation}
is the \emph{peripheral generator} of $\Sem$, and
\begin{equation}
\Phi_t^{\mathrm{P}} := \Phi_t  \mathcal{P}_{\mathrm{P}}, \quad t \in\mathbb{R}^+,
\end{equation} 
the corresponding \emph{peripheral quantum semigroup} associated with $\Sem$. Note that $\mathcal{L}_{\mathrm{P}}$ defined via~\eqref{per_gen_def} is not a GKLS generator in general.

We now give the analogue of Proposition~\ref{peripheral_Wolf} for QDSs, for which we also provide a proof for the reader's convenience.
\begin{proposition}[Peripheral maps of a Markovian dynamics]\label{prop:peripheral_maps_markovian}
Let $\Sem = (\Phi_t)_{t\in\mathbb{R}^+}$ be a QDS with GKLS generator $\mathcal{L}$. Then the spectral projections $\mathcal{P}$ and $\mathcal{P}_{\mathrm{P}}$ 
introduced above are quantum channels and $(\Phi_t^{\mathrm{P}})_{t\in \mathbb{R}^+}$ is a semigroup of quantum channels. 
\end{proposition}
\begin{proof}
As the semigroup $\Sem$ is mean-ergodic~\cite[Definition V.4.3]{engelnagel}, the spectral projection $\mathcal{P}$ may be written in terms of the semigroup $(\Phi_t)_{t\in \mathbb{R}^+}$ via a Ces{\`a}ro mean as~\cite[Lemma V.4.4]{engelnagel}
\begin{equation}
\label{P_for_sem}
\mathcal{P} = \lim_{T\rightarrow \infty}\frac{1}{T} \int_{0}^T \Phi_t \mbox{ dt},
\end{equation}
i.e., a continuous-time version of~\eqref{P_for} for the spectral projection corresponding to $\lambda =1$ of a quantum channel. Clearly, Equation~\eqref{P_for_sem} implies that $\mathcal{P}$ is a quantum channel, since the set of quantum channels is convex and compact. Furthermore, the peripheral projection $\mathcal{P}_{\mathrm{P}}$ coincides with the one of the Markovian channel $\Phi_1 = \mathrm{e}^{\mathcal{L}}$ by the Jordan decompositions of $(\Phi_t)_{t\in\mathbb{R}^+}$ and $\Phi_1$; 
see Appendix~\ref{Jordan_op_app} and~\cite[Appendix D]{AFK_asympt_1}. Thus $\mathcal{P}_{\mathrm{P}}$ is a quantum channel by Proposition~\ref{peripheral_Wolf}.

Finally, the maps $\Phi_t^{\mathrm{P}}$ are quantum channels for all $t\in\mathbb{R}^+$, as they are composition of quantum channels, and also the semigroup property holds since
\begin{equation}
\Phi_{t_1 + t_2}^{\mathrm{P}} =\Phi_{t_1 + t_2} \circ \mathcal{P}_{\mathrm{P}}= \Phi_{t_1} \circ \Phi_{t_2} \circ \mathcal{P}_{\mathrm{P}} = \Phi_{t_1} \circ \mathcal{P}_{\mathrm{P}} \circ  \Phi_{t_2} \circ \mathcal{P}_{\mathrm{P}} = \Phi_{t_1}^{\mathrm{P}} \circ \Phi_{t_2}^{\mathrm{P}},
\end{equation}
by the semigroup property of $(\Phi_t)_{t\in\mathbb{R}^+}$ and the invariance of $\Attr(\mathcal{L})$ under $(\Phi_t)_{t\in\mathbb{R}^+}$.
\end{proof}

Observe that the peripheral quantum semigroup $(\Phi_t^{\mathrm{P}})_{t\in\mathbb R^+}$ is not a QDS in general, as it does not satisfy condition i.\ of Definition~\ref{QDS_def} since $\Phi_0^{\mathrm{P}} = \mathcal{P}_{\mathrm{P}}$.

\subsection{Asymptotic structure theorem}
\label{subsec:asympt_QDS_2}
Let us now discuss the explicit structure of the asymptotic dynamics of the QDS $\Sem$.
First, as already observed in the proof of Proposition~\ref{prop:peripheral_maps_markovian}, the peripheral projection $\mathcal{P}_{\mathrm{P}}$ of $\Sem$ coincides with the one of the unit-time evolution $\Phi_1 = \mathrm{e}^{\mathcal{L}}$, meaning that
\begin{equation}\label{attr_QDS}
	\Attr(\Sem)=\Attr(\Phi_1).
\end{equation} 
Therefore, according to Theorem~\ref{Wolf_struc}, the attractor subspace $\Attr(\Sem)$ of the QDS $\Sem$ can be written in the form~\eqref{attr_struc}. 

Moreover, by applying Equation~\eqref{struc_phiP} and the fact that $\pi = \mathrm{id}$ in~\eqref{struc_phiP} for Markovian channels~\cite[Corollary 5.4]{AFK_asympt_1} to any channel $\Phi_t$ of the QDS $\Sem = (\Phi_t)_{t \in \mathbb{R}^+}$, we find that the evolution of any $X \in \Attr(\mathcal L)$, expressed in the form~\eqref{eq:element_in_the_attractor}, is given by
\begin{equation}\label{eq:qds_partial}
\Phi_{t}(X)=0\oplus  
\bigoplus_{k\in[M]} \bigl( U_{k}(t) x_k U_{k}^{\ast}(t) \otimes \rho_k \bigr).
\end{equation}
By the semigroup property of $(\Phi_{t})_{t\in\mathbb{R}^+}$, it turns out that the family of maps
\begin{equation}
\label{phi_kt_def}
    \Phi_{k , t}(x_k) := U_k (t) x_k U_k^\ast(t), \quad x_k \in \mathcal{B}(\mathcal{H}_{k,1}) , \; k \in [M], \; t \in \mathbb{R}^+,
\end{equation}
is a continuous semigroup of unitary channels, which can be extended to a group by defining $\Phi_{k , t}$ as in~\eqref{phi_kt_def} with $U_k (t) := U_k^\ast(-t)$ when $t < 0$. Therefore, 
we can rewrite~\eqref{eq:qds_partial} as
\begin{equation}
\Phi_{t}(X) =0\oplus 
\bigoplus_{k\in [M] } \left( \mathrm{e}^{-\mathrm i t H_k} x_k \mathrm{e}^{\mathrm i t H_k}  \otimes \rho_k \right),
\label{as-struc-QDS}
\end{equation}
for some Hermitian operators $H_k = H_k^\ast$, with $k \in [M]$.

We thus arrived at the following structure theorem for the asymptotic dynamics of QDSs~\cite[Theorem 2]{baumgartner2008analysis2}, which is the analogue of Theorem~\ref{Wolf_struc} established for QMCs.
\begin{theorem}[Asymptotic structure theorem of QDSs]
\label{asympt_QDS_thm}
Let $\Sem = (\Phi_t)_{t\in\mathbb{R}^+}$ be a QDS on $\mathcal B(\mathcal H)$, with generator $\mathcal L$ and peripheral projection $\mathcal{P}_{\mathrm{P}}$. 
Then, there exist
\begin{itemize}
\item[\textnormal{i.}] a family of Hilbert spaces $\mathcal{H}_{k,i}$, with $k\in[M]$ and $i\in\{1,2\}$, that decompose the Hilbert space $\mathcal{H}$ as
	\begin{equation}
    \label{H_dec_Mark}
		\mathcal{H} = \mathcal{H}_0^\perp \oplus \mathcal{H}_0, \quad 	\mathcal{H}_0= 
        \bigoplus_{k\in[M]} \left(\mathcal{H}_{k,1}\otimes \mathcal{H}_{k,2} \right) ,
	\end{equation}
	with $\mathcal{H}_0=\Ran(\mathcal{P}_{\mathrm{P}}(\mathbb{I}))$;
	\item[\textnormal{ii.}] positive-definite density operators $\rho_{k}$ on $\mathcal{H}_{k,2}$, with $k\in[M]$;
\end{itemize}
such that the attractor subspace $\Attr(\Sem)$ of $\mathbf{\Phi}$ reads
\begin{equation}\label{attr_struc_QDS}
\Attr(\Sem)=\Attr(\mathcal L) =0\oplus 
\bigoplus_{k\in[M]} \bigl(\mathcal{B}(\mathcal{H}_{k,1}) \otimes \mathbb{C}\rho_{k}\bigr).
\end{equation}
Furthermore, there exist Hermitian operators $H_{k}$ on $\mathcal{H}_{k,1}$ such that, for a given $X\in\Attr(\mathbf{\Phi})$ in the form
\begin{equation}
X=0\oplus 
\bigoplus_{k\in[M]}\left(x_{k} \otimes \rho_{k} \right) \in\Attr(\mathcal L),
\end{equation}
the action of $\Phi_t$ on $X$ reads
\begin{equation}\label{struc_as_QDS}
	\Phi_{t}(X)=0\oplus 
    \bigoplus_{k\in[M]}\left(\mathrm e^{-\mathrm i t H_k}x_{k}\mathrm e^{\mathrm i t H_k}\otimes \rho_{k}\right).
\end{equation}
The action of $\mathcal{L}$ on $X$ is given by
\begin{equation}
	\label{as_L_struc}
	\mathcal{L}(X)= 0 \oplus 
    \bigoplus_{k\in[M]} \left(\mathcal{L}_k(x_k) \otimes \rho_k\right),
\end{equation}
where
\begin{equation}
	\mathcal{L}_k(x_k)=-\mathrm i[H_k , x_k], \quad k\in[M].
\end{equation}
\end{theorem}

Physically, the Markovian evolution in the long-time limit is unitary, with $U(t)=\mathrm{e}^{-\mathrm{i} t H}$ being generated by the Hamiltonian on $\mathcal{H}$ in~\eqref{H_dec_Mark}
\begin{equation}
    H = 0 \oplus \bigoplus_{k\in[M]} \left({H}_{k}\otimes \mathbb{I}_{k,2} \right) , 
\end{equation}
which induces unitary evolutions on each sector $\mathcal{H}_{k,1}$ of the decomposition~\eqref{H_dec_Mark}. The possible presence of permutations, emerging in the general case~\eqref{struc_phiP}, is ruled out by the continuous-time nature of the process~\cite[Section 5.2]{baumgartner2012structures}. 

Before going on, note that we can also obtain the structure of the asymptotics of Markovian evolutions in the Heisenberg picture~\cite[Proposition 2]{albert2016geometry}, similar to the one established for QMCs in Theorem~\ref{struc_Heis_th}. 
As in the case of quantum channels in~\eqref{eq:fix_cyclicdec}, we can also derive from~\eqref{struc_as_QDS} an  expression for the steady subspace $\Fix(\Sem)$ of the QDS $\Sem$:
\begin{equation}
\Fix(\Sem) =\Ker(\mathcal{L}) =0\oplus 
\bigoplus_{k\in[M]}\left( \Ker(\mathcal{L}_{k}) \otimes \mathbb{C} \rho_k\right).
\label{fix_QDS}
\end{equation}
with $\Ker(\mathcal{L}_{k}) = \{ H_k \}^\prime$, the commutant of $H_k$. 

More explicitly, since $H_k$ is Hermitian, we can decompose the kernel $\Ker(\mathcal{L}_k)$ of $\mathcal{L}_k$ as
\begin{equation}\label{eq:dec_eigenspaces_Hk}
   \Ker(\mathcal{L}_k) =  
   \bigoplus_{r\in[n_k]} \mathcal{B}(\mathcal H_{1,r}^{(k)}),\quad k\in[M],
\end{equation}
with $\mathcal H_{1,r}^{(k)}$ being the $r$-th eigenspace of $H_k$, $r \in [n_k]$, for all $k\in[M]$. Then, the kernel of $\mathcal{L}$ can be re-written in the form
\begin{equation}\label{fix_dec_Wolf_Lind}
    \Ker(\mathcal{L}) =0\oplus 
\bigoplus_{k\in[M]}\bigoplus_{r\in[n_k]}\left(\mathcal{B}(\mathcal H_{1,r}^{(k)}) \otimes \rho_k\right),
\end{equation}
which is the analogue of the cyclic decomposition~\eqref{fix_dec_Wolf} of the fixed-point subspace of channels.
Importantly,~\eqref{fix_dec_Wolf_Lind} and~\eqref{attr_struc_QDS} unveil the relation between the steady and attractor subspaces 
of $\mathbf{\Phi}$. Equivalently, the fixed-point subspace of $\mathbf{\Phi}$ is the attractor subspace of the steady projection $\mathcal P$ defined by~\eqref{P_for_sem}, thus a decomposition of the form~\eqref{fix_dec_Wolf_Lind} immediately follows from Theorem~\ref{Wolf_struc} applied to $\mathcal{P}$. 

We conclude the section by introducing the important class of \emph{faithful GKLS generators}, whose definition mirrors Definition~\ref{faith_ch_def} of faithful quantum channels.
\begin{definition}[Faithful GKLS generators]
\label{faith_gen_def}
A GKLS generator $\mathcal{L}$ is \emph{faithful} if there exists a full-rank stationary state, 
\begin{equation}
    0 < \rho \in \mathcal{S}(\mathcal{H}) \cap \Ker(\mathcal{L}),
\end{equation}
 of the QDS $\mathbf{\Phi} =(\mathrm{e}^{t\mathcal{L}})_{t\in\mathbb{R}^+}$ generated by $\mathcal{L}$.
The adjoint GKLS generator $\mathcal{L}^\dagger$ is called \emph{faithful} if $\mathcal{L}$ is faithful.
\end{definition}
Moreover, by applying the reasoning used for quantum channels~\cite{AFK_asympt_1} in Section~\ref{bas_notions_subsec}, we  see that the faithfulness assumption for a generator $\mathcal{L}$ amounts to requiring that 
$\mathcal{H} = \mathcal{H}_{0}= \Ran(\mathcal{P}_{\mathrm{P}}(\mathbb{I}))$, where $\mathcal{P}_{\mathrm{P}}$ is the peripheral projection of $\mathcal{L}$ defined in~\eqref{per_proj_def_L}. Thus, also in the continuous-time setting, faithfulness is related to the absence of decay. 

We now characterize the steady subspace $\Fix(\Sem^\dagger)=\Ker(\mathcal L^\dagger)$ of a QDS in the Heisenberg picture, when its adjoint GKLS generator $\mathcal{L}^\dagger$ is faithful~\cite[Theorem 7.2]{wolf2012quantum}.   
\begin{proposition}
\label{fixed_L}
Let the adjoint generator $\mathcal{L}^\dagger$ be given in the adjoint GKLS form~\eqref{GKLS_adjoint}, and let
\begin{equation}
    S := \{ H \} \cup \{  L_{\alpha} , L_{\alpha}^{\ast} 
\}_{\alpha \in[N]}.
\end{equation}
Then the commutant of $S$ is contained in the kernel of $\mathcal{L}^\dagger$, namely
\begin{equation}
S^{\prime}\subseteq\Ker(\mathcal{L}^\dagger). 
\end{equation}
If, in addition, $\mathcal{L}^\dagger$ is faithful in the sense of Definition~\ref{faith_gen_def}, then the inclusion becomes an equality.
\end{proposition}
This result is the continuous-time counterpart of Proposition~\ref{th:fixed_kraus} for faithful UCP maps.

\section{Dynamics in the attractor subspace: unitarity and information protection}
\label{protecting_unitary_evolution}
In this section, we briefly discuss how unitary dynamics emerges in the asymptotic evolution of QMCs. 
Specifically, in Section~\ref{unit_asympt_sec}, we shall determine under what conditions the open-system dynamics becomes unitary when restricted to the attractor subspace.
In this case, the dynamics will be called \emph{asymptotically unitary}. 
Physically, once the dissipative transient has decayed, the long-time evolution is unitary on the attractor subspace, so that the asymptotic dynamics is effectively that of a closed system.
In particular, an evolution is asymptotically unitary if the long-time dynamics is trivial, i.e.\ its attractor subspace reduces to the steady subspace. 

Notice also that, independently of the validity or not of asymptotic unitarity, there may exist special subspaces of the attractor subspace within which coherent superpositions are preserved under the evolution and, consequently, it is possible to realize noise-protected codes~\cite{zanardi1997noiseless} or unitary gates~\cite{lidar_dec_1998}. These are the so-called \emph{decoherence-free subspaces} and \emph{noiseless subsystems}, discussed in Sections~\ref{sec:dfs} and~\ref{QEC_sec}, respectively. The notion of decoherence-free subspaces is deeply connected with the symmetries of the evolution, as detailed in Section~\ref{sec:sym_DFS}. Finally, in Section~\ref{QEC_sec}, we will also recall the concept of correctable subsystems, in which quantum information can be recovered completely by using a suitable recovery map.

\subsection{Asymptotically unitary channels}
\label{unit_asympt_sec}

We are interested in understanding under what conditions the asymptotic evolution is unitary.
We start with the formal definition of the channels fulfilling this condition.
\begin{definition}\label{unit_asy_def}
Let $\Phi$ be a quantum channel. We say that $\Phi$ is \emph{asymptotically unitary} if and only if there is a unitary $U$  on $\mathcal{H}$ such that
\begin{equation}\label{unit_asy}
    \Phi(X)=UXU^\ast, \quad \forall X\in\Attr(\Phi).
\end{equation}
\end{definition}

First, the asymptotic dynamics is unitary whenever the permutation $\pi$ is trivial, i.e.\ $\pi=\mathrm{id}$, as a consequence of~\eqref{struc_phiP}. However, the following example provides an asymptotically unitary evolution with a nontrivial permutation. 
\begin{example}
\label{ex_5}
Let us define the following qubit channel, which is a slight variant of~\cite[Equation (79)]{AFK_asympt_1}:
\begin{equation}
\label{perm_channel_ex}
\Phi(X)=\frac{1}{2}(\sigma_{1} X \sigma_{1}+\sigma_{2} X \sigma_{2}),\quad X\in\mathcal{B}(\mathbb{C}^2).
\end{equation}
The attractor subspace of $\Phi$ is given by
\begin{equation}
\label{attr_ex-4}
\Attr(\Phi)=\{ \alpha \mathbb{I}+\beta \sigma_{3} \,:\, \alpha,\beta\in\mathbb{C} \} = \mathbb{C} \ketbra{1}{1} \oplus \mathbb{C} \ketbra{2}{2},
\end{equation}
which is the result of a decoherence process and, 
furthermore, the asymptotic dynamics of $\Phi$ 
in the eigenbasis $\{ \ket{i} \}_{i=1}^2$ of $\sigma_3$ reads 
\begin{equation}
\label{asympt_ex-4}
\Attr(\Phi) \ni X= \begin{pmatrix}\alpha + \beta & 0\\
0 & \alpha - \beta\end{pmatrix} \mapsto \Phi(X)=\begin{pmatrix}\alpha - \beta & 0\\
0 & \alpha + \beta\end{pmatrix},
\end{equation}
so it is just the transposition of the two diagonal components of $X$. The asymptotic evolution~\eqref{asympt_ex-4} can be written as $\Phi(X) = \sigma_1 X \sigma_1$ for all $X \in \Attr(\Phi)$, therefore $\Phi$ is asymptotically unitary, as claimed.  
\end{example}
Note also that, in the qubit case, all quantum channels $\Phi$ are asymptotically unitary, as either $\Phi$ is a unitary channel, or $\Attr(\Phi) = \Fix(\Phi)$, or the asymptotic evolution is unitarily equivalent to that of Example~\ref{ex_5}.
This is not true in general in higher dimensions, and the unitarity of the asymptotic evolution is characterized in the following proposition~\cite[Theorem 3]{AFK_asympt_2}.
\begin{proposition}
		Let $\Phi$ be a quantum channel with attractor subspace~\eqref{attr_struc} and asymptotic evolution~\eqref{struc_phiP}. Then, $\Phi$ is asymptotically unitary if and only if $\rho_k=V_k\rho_{\pi(k)}V_k^\ast$  for some unitaries $V_k: \mathcal{H}_{\pi(k),2} \rightarrow \mathcal{H}_{k,2}$ and for all $k\in[M]$.
\end{proposition}
We collect several classes of quantum channels whose asymptotic evolution is unitary in the next proposition~\cite[Corollary 5.4, Remark 6]{AFK_asympt_1} and~\cite[Section 3]{AFK_asympt_2}.
\begin{proposition}\label{suff_unit_cor}
Let $\Phi$ be a quantum channel. Each of the following conditions is sufficient for $\Phi$ to be asymptotically unitary:
\begin{itemize}
\item[\textnormal{i.}] $\Phi$ is \emph{mixing}, i.e.,\ $\dim(\Attr(\Phi))=1$;
\item[\textnormal{ii.}] $\Phi$ is Markovian, i.e.,\ $\Phi=\mathrm{e}^{\mathcal{L}}$ for some GKLS generator $\mathcal{L}$;
\item[\textnormal{iii.}] $\Phi$ is idempotent, i.e.,\ $\Phi^{2}=\Phi$;
\item[\textnormal{iv.}] $\Phi$ is unital, i.e.,\ $\Phi(\mathbb I)=\mathbb I$.
\end{itemize} 
\end{proposition}
Notice that the first three classes of channels have an asymptotic evolution which does not involve permutations~\cite[Corollary 5.4, Remark 6]{AFK_asympt_1}, whereas the last class may exhibit permutations in the long-time limit, as illustrated by Example~\ref{ex_5}. In particular, the fact that $\pi = \mathrm{id}$ for Markovian channels was used to prove the structure Theorem~\ref{asympt_QDS_thm} of QDSs and, moreover, it allows us to interpret permutations as a contribution to the long-time evolution linked to the non-Markovianity of the channel, see also Section~\ref{sec:asympt_non_Markov}. Notice that permutations are also decoupled from the unitary part of the asymptotic dynamics according to the cyclic decomposition given in Theorem~\ref{thm:cycles}.  
The class of mixing channels appearing in Proposition~\ref{suff_unit_cor} will be further discussed in Section~\ref{qc_bounds_subsec}.

\subsection{Decoherence-free subspaces}
\label{sec:dfs}
Decoherence occurs when a quantum system loses its coherence properties due to interactions with its environment~\cite{Palma1996,Zurek2003}; see  Section~\ref{sec:dephasing_channel}. A \emph{\gls*{dfs}} is a special subspace that is protected from these effects. Zanardi and, independently, Lidar, Chuang, and Whaley systematically defined these subspaces~\cite{zanardi1997noiseless,Zanardi_dissip_1998,lidar_dec_1998}. In recent years, their role has sparked great interest in the physical community~\cite{Wang_scars_DFS_2024,Friesen2017}, particularly for their connection to the symmetries of the system~\cite{Buca2019,Tindall2020,Dubois2023}; see Section~\ref{sec:sym_DFS} and the recent review~\cite{dutta2025introduction}. The problem of determining decoherence-free subspaces of Markovian evolutions was also addressed from a complexity-theoretic viewpoint in~\cite{borras2026complexityDFS}.

We start with the formal definition of a DFS~\cite[Definitions 1-3]{shabani2005theory}.

\begin{definition}[Decoherence-free subspaces]\label{def:dfs}
Let $\Phi:\mathcal B(\mathcal H)\rightarrow \mathcal B(\mathcal H)$ be a quantum channel, and $P$ be a Hermitian projection  onto  a subspace $\mathcal X$ of $\mathcal H$. Then, 
\begin{itemize}
    \item[\textnormal{i.}] $\mathcal X$ is an imperfect initialization DFS for $\Phi$ if there exists a unitary operator $U$ on $\mathcal X$ such that for all states $\rho\in\mathcal S(\mathcal H)$:
\begin{equation}\label{eq:dfs_condition}
    P\Phi (\rho)P=UP\rho PU^*; 
\end{equation}
\item[\textnormal{ii.}] $\mathcal X$ is a perfect initialization DFS for $\Phi$ if~\eqref{eq:dfs_condition} holds for all states $\rho=P\rho P$.
\end{itemize}

\end{definition}

Note that the difference between perfect and imperfect initialization conditions is that, in the former case, only states supported on $\mathcal X$ must undergo a unitary evolution, while in the latter, which turns out to be relevant for many quantum algorithms~\cite{biham1999grover,parker2000efficient,chi2001initialization}, the initial state need not be supported on $\mathcal{X}$.

It is easy to see that, by the structure~\eqref{attr_struc} of the attractor subspace given in Theorem~\ref{Wolf_struc}, a subspace $\mathcal X$ is a perfect initialization DFS if there is a $k\in[M]$ such that $\mathcal H_{k,2} \simeq \mathbb C$ and $\pi(k)=k$ for which 
\begin{equation}
    \mathcal X=\mathcal H_{k,1}\otimes \mathcal H_{k,2}\simeq \mathcal H_{k,1}. 
\end{equation}
Interestingly, DFSs can be characterized in terms of the KS representation both in the perfect and imperfect cases~\cite[Theorem~1,~Corollary~1]{shabani2005theory}.
\begin{theorem}\label{th:dfs_charact}
    Let the quantum channel $\Phi:\mathcal B(\mathcal H)\rightarrow\mathcal B(\mathcal H)$ be given in the KS form~\eqref{op_sum_rep}. Then, a subspace $\mathcal X$ is a perfect initialization DFS if and only if all the KS operators $\{A_i\}_{i=1}^N$ can be written in the following block-matrix form with respect to the decomposition $\mathcal{H} = \mathcal{X} \oplus  \mathcal{X}^\perp$
    \begin{equation}
    \label{eq:dec_kraus_operators_perf_init}A_i=
        \begin{pmatrix}
            g_i U& D_i \\0 & A_{i,\perp}
        \end{pmatrix}, \quad i \in [N],
    \end{equation}
    for some unitary
    $U \in \mathcal{B}(\mathcal{X})$, and   $g_i \in \mathbb{C}$, $A_{i,\perp}\in\mathcal B(\mathcal X^\perp)$, and $D_i\in\mathcal B(\mathcal X^\perp,\mathcal X)$, with $i \in [N]$, satisfying the properties
    \begin{itemize}
    \item[\textnormal{i.}] 
    $\sum_{i\in[N]} |g_i|^2 = 1$, 
    \item[\textnormal{ii.}]$\sum_{i\in[N]} (D_i^\ast D_i + A_{i,\perp}^\ast A_{i,\perp}) = \mathbb{I}_{\perp}$, with $\mathbb{I}_{\perp}$ being the identity on $\mathcal{X}^\perp$, 
    \item[\textnormal{iii.}] 
    $\sum_{i\in[N]} \overline{g}_i D_i = 0$. 
    \end{itemize}
    If, in addition, $D_i=0$ for all $i\in[N]$, then  $\mathcal X$ is an imperfect initialization DFS.
\end{theorem}

Notice that the operator $D_i$ in Equation~\eqref{eq:dec_kraus_operators_perf_init} represents the decay from ${\mathcal X}^{\perp}$ to ${\mathcal X}$, and the fact that the lower-left blocks of the KS operators $A_i$ vanish ensures that the subspace $\mathcal B({\mathcal X})$ is preserved under $\Phi$. As a final comment, the previous discussion can be adapted to QDSs (Definition~\ref{QDS_def}); see~\cite{shabani2005theory,Karasik2008} for details. We also refer the reader to the reviews~\cite{Lidar2003,lidar2014review} for a more in-depth analysis of DFSs.
\subsection{Quantum error correction}
\label{QEC_sec}
Quantum error correction constitutes a central pillar of fault-tolerant quantum computation, addressing the intrinsic fragility of quantum information under decoherence and noise and, in fact, the design of efficient quantum error correcting strategies is regarded as the main obstacle towards the development of scalable quantum technologies~\cite{devitt2013quantum}. Broadly speaking, error correcting techniques can be classified into passive and active schemes. Passive methods, including perfect initialization decoherence-free subspaces, discussed in the previous subsection, and noiseless subsystems~\cite{duan1997preserving,zanardi1997noiseless,Zanardi_dissip_1998}, recalled later in this subsection, exploit noise-free subspaces, arising from the symmetries in the system-environment interaction, to encode quantum information.  
Active error correction~\cite{shor1995scheme,steane1996error,knill1997theory}, in contrast, identifies errors by redundant encoding and syndrome measurements and corrects them dynamically via a proper recovery channel, ensuring that quantum information remains intact throughout the computation.

In this subsection, we revisit some connections between quantum error correction and several subspaces used to describe the long-time evolution of quantum systems~\cite{holbrook2003noiseless,choi2006method,Choi2009,blume2008characterizing,blume2010information,rahaman2017multiplicative,kribs2005unified,kribs2006quantum}. 

First, given a Hilbert space $\mathcal H$, we say that a Hilbert space $\mathcal X$ is a \emph{subsystem} of $\mathcal H$ if we can find a Hilbert space $\mathcal Y$ such that $\mathcal X \otimes \mathcal Y$ is a subspace of $\mathcal H$. More explicitly, $\mathcal{X}$ is a subsystem of $\mathcal{H}$ if there are Hilbert spaces $\mathcal Y,\ \mathcal Z$ such that 
\begin{equation}
    \mathcal H=\mathcal X\otimes\mathcal Y \oplus \mathcal{Z}.
    \label{eq:subsysdef}
\end{equation}
 Note that, if $\dim(\mathcal{Y})=1$, $\mathcal X$ reduces to a subspace of $\mathcal H$. 

Now, given a quantum channel $\Phi$ on $\mathcal B(\mathcal H)$ modelling the action of an error, a subsystem $\mathcal X$ is called \emph{noiseless} for $\Phi$~\cite[Lemma 2]{kribs2005unified}

if there is a quantum channel $\Phi_{\mathcal Y}$ on $\mathcal B(\mathcal Y)$ such that 
\begin{equation}\label{eq:condition_noiseless}
	\Phi( \rho \otimes \sigma \oplus 0) =  \rho \otimes \Phi_{\mathcal Y}(\sigma) \oplus 0, \quad  \forall \,\rho\in \mathcal{B}(\mathcal{X}),\, \forall \,\sigma \in \mathcal{B}(\mathcal{Y}).
\end{equation}

Note that, given a noiseless subsystem $\mathcal{X}$, the channel $\Phi_{\mathcal{Y}}$ is general, namely any state $\sigma \in \mathcal{S}(\mathcal{Y})$ is mapped into an \emph{arbitrary} state $\tau \in \mathcal{S}(\mathcal{Y})$ under  $\Phi_{\mathcal{Y}}$, thus we can call $\mathcal{Y}$ a \emph{noisy subsystem} for $\Phi$. 
If $\dim(\mathcal{Y}) = 1$, then a noiseless subsystem reduces to a \emph{noiseless code}~\cite{knill1997theory}, and it is a perfect initialization DFS with unitary $U = \mathbb{I}_{\mathcal{X}}$; cf.\ Definition~\ref{def:dfs}.

As in the case of DFSs, discussed in Theorem~\ref{th:dfs_charact}, noiseless subsystems can be characterized in terms of the KS operators of the error, as shown in the following result~\cite[Theorem~1]{kribs2005unified}.
\begin{theorem}
    Let the quantum channel $\Phi:\mathcal B(\mathcal H)\rightarrow\mathcal B(\mathcal H)$ be given in the KS form~\eqref{op_sum_rep}, and let $\mathcal X$ be a subsystem of $\mathcal H$, with $\mathcal H=  \mathcal X \otimes \mathcal Y \oplus \mathcal Z $. Then, $\mathcal X$ is a noiseless subsystem for $\Phi$ if and only if all the $KS$ operators 
    $\{A_i\}_{i\in[N]}$ admit a decomposition in the form
    \begin{equation}
        A_i=\begin{pmatrix}\mathbb I_{\mathcal X}\otimes M_i  & D_i\\ 0 & A_{i,\perp}\end{pmatrix}, \quad i \in [N],
    \end{equation}
    where the operators $M_i\in\mathcal B(\mathcal Y)$, $A_{i,\perp}\in\mathcal B(\mathcal Z)$ and $D_i\in\mathcal B(\mathcal Z,\mathcal X\otimes \mathcal Y)$ with $i\in[N]$ satisfy the properties i) 
    $\sum_{i\in[N]} M_i^\ast M_i =\mathbb I_\mathcal{Y}$, ii) 
    $\sum_{i\in[N]} (D_i^* D_i + A_{i,\perp}^* A_{i,\perp})=\mathbb I_\mathcal{Z}$, and iii) 
    $\sum_{i\in[N]} (\mathbb I_\mathcal X\otimes M_i^*) D_i=0$. 
\end{theorem}
A similar result can be found in~\cite[Corollary~2]{shabani2005theory}, where a slightly different definition of noiseless subsystem and a distinction between perfect- and imperfect-initialization noiseless subsystems is introduced, along the lines of Definition~\ref{def:dfs} for DFSs.

Noiseless subsystems are closely related to the notion of \emph{correctable subsystem}~\cite{kribs2005unified}, which is at the heart of active quantum error correction. We say that the subsystem $\mathcal{X}$ of $\mathcal H$ is a correctable subsystem for the quantum channel $\Phi$ if there exists a recovery channel $\mathcal R$ such that $\mathcal{X}$ is a noiseless subsystem for $\mathcal{R} \circ \Phi$. It is worth pointing out that we can take the corresponding noisy subsystem $\mathcal Y$ to be one-dimensional. To see this, suppose that
\begin{equation}
    \mathcal R\circ \Phi(\rho\otimes \sigma \oplus 0)= \rho \otimes \Phi_{\mathcal Y}(\sigma) \oplus 0,\quad \forall \rho\in\mathcal B(\mathcal X),\, \forall \sigma \in\mathcal B(\mathcal Y),
\end{equation}
with $\mathcal H= \mathcal X\otimes \mathcal Y \oplus\mathcal Z$. Choose $\ket{\psi}\in\mathcal Y$, and define the channel $\mathcal Q:\mathcal B(\mathcal H)\rightarrow \mathcal B(\mathcal H)$ as
\begin{equation}
    \mathcal Q\begin{pmatrix}X_{00}& X_{01}\\ X_{10}&X_{11}\end{pmatrix}=\begin{pmatrix}\Tr_{\mathcal Y}(X_{00})\otimes \ketbra{\psi}& 0\\ 0&X_{11}\end{pmatrix},
\end{equation}
with $X_{00}\in\mathcal B(\mathcal X\otimes Y)$, $X_{11}\in\mathcal B(\mathcal Z)$, $X_{01}\in \mathcal B(\mathcal Z,\mathcal X\otimes \mathcal Y)$, and $X_{10}\in \mathcal B(\mathcal X\otimes \mathcal Y,\mathcal Z)$. Then, using the recovery map $\mathcal R'=\mathcal Q\circ \mathcal R$, the subspace $\mathcal X$ is a noiseless subsystem with noisy subsystem $\mathbb C \ket{\psi}$. 

We now present a characterization of correctable subsystems in terms of the KS operators of $\Phi$~\cite{nielsen_algebraic_2007}.
\begin{theorem}
    Let the quantum channel $\Phi:\mathcal B(\mathcal H)\rightarrow\mathcal B(\mathcal H)$ be given in the KS form~\eqref{op_sum_rep}, and let $\mathcal X$ be a subsystem of $\mathcal H$, with $\mathcal H=\mathcal X \otimes \mathcal Y \oplus \mathcal Z$. Let $P$ be the Hermitian projection onto $\mathcal X\otimes \mathcal Y$. Then, $\mathcal X$ is a correctable subsystem for $\Phi$ if and only if all the KS operators 
    $\{A_i\}_{i\in[N]}$ of $\Phi$ satisfy the condition 
    \begin{equation}
    P A_i^* A_j P=\mathbb I_\mathcal X\otimes M_{ij} \oplus\, 0,\quad M_{ij}\in\mathcal B(\mathcal Y), \quad i,j\in[N].
    \label{eq:necessary_condition_correctable}
    \end{equation}
\end{theorem}
In particular, condition~\eqref{eq:necessary_condition_correctable} yields the Knill-Laflamme conditions~\cite[Theorem~III.2]{knill1997theory} in the case of a \emph{correctable code}, namely when $\mathcal Y$ is one-dimensional. In line with the above discussion, it is also immediate to check that, if~\eqref{eq:necessary_condition_correctable} holds for $\mathcal{X}$ and some noisy subsystem $\mathcal{Y}$, then the same is true for $\mathcal{X}$ with \emph{any} one-dimensional noisy subsystem $\mathcal{Y}^\prime \subseteq \mathcal{Y}$.  

A correctable subsystem $\mathcal X$ is called \emph{unitarily correctable}~\cite{kribs2006quantum} if the corresponding recovery map $\mathcal{R}$ can be taken to be a unitary channel. For unital quantum channels, i.e.\ $\Phi(\mathbb{I}) = \mathbb{I}$, there is a connection between noiseless subsystems and unitarily correctable subsystems~\cite[Theorem 2]{kribs2006quantum}.
\begin{proposition}
	\label{char_noiseless_unit_correct_prop}
	Let $\Phi$ be a unital quantum channel on $\mathcal{B}(\mathcal{H})$ and $\mathcal{X}$ in~\eqref{eq:subsysdef} be a subsystem of $\mathcal{H}$. Then, the following conditions are equivalent:
	\begin{itemize}
		\item[\textnormal{i.}] $\mathcal{X}$ is a unitarily correctable subsystem for $\Phi$; 
		\item[\textnormal{ii.}] $\mathcal{X}$ is a noiseless subsystem for $\Phi^\dagger \circ \Phi$.
	\end{itemize}
\end{proposition}
According to Proposition~\ref{char_noiseless_unit_correct_prop}, the problem of finding unitarily correctable subsystems for a unital channel boils down to the search for the noiseless subsystems of $\Phi^\dagger \circ \Phi$. 
As a consequence of~\cite[Corollary 1]{choi2006method}, a subsystem $\mathcal{X} \subseteq \mathcal{H}$ in~\eqref{eq:subsysdef} is noiseless for a unital channel $\Phi$ implies that $\mathcal{B}(\mathcal{X}) \otimes \mathbb{I}$ is a subalgebra of the fixed-point subspace $\Fix(\Phi)$ of $\Phi$. 
Consequently, the unitarily correctable subsystems of $\Phi$ can be obtained from the corresponding decomposition~\eqref{fix_dec_Wolf} of $\Fix(\Phi^\dagger\circ\Phi)$.

Finally, one may also address active correction schemes for multiple occurrences of an error; for example in~\cite[Section 5]{rahaman2017multiplicative} the ``asymptotic correctability" for a unital error is linked to the structure~\eqref{attr_struc} of its attractor subspace.

\section{Geometric bounds for the asymptotic subspaces}
\label{sec:asymptotic_bounds} 
In this section, we discuss lower and upper bounds on the dimensions of the fixed-point and attractor subspaces of QMCs and QDSs, a problem that is interesting in its own right. We also recall the notions of ergodicity and mixing for both types of dynamics. In particular, we specialize the discussion to quantum channels in the next subsection, while Section~\ref{GKLS_bounds-subsec} is devoted to GKLS generators.

\subsection{Quantum channels}
\label{qc_bounds_subsec}
Let $\Phi$ be a quantum channel on $\mathcal{B}(\mathcal{H})$ with steady and attractor subspaces $\Fix(\Phi)$ and $\Attr(\Phi)$, respectively defined in~\eqref{fix_def} and~\eqref{eq:attractor_suspace}. Their dimensions are
\begin{equation}
\label{ell_def}
\ell_0 := \dim(\Fix(\Phi)) , \quad 
\ell_{\mathrm P}  := \dim  (\Attr(\Phi)) = \sum_{\lambda \in \spec_{\mathrm P} (\Phi)} \ell_\lambda, 
\end{equation}
where $\ell_\lambda$ denotes the  multiplicity of the  eigenvalue $\lambda\in\spec_{\mathrm P} (\Phi)$. Recall from Proposition~\ref{periph_semisimple} that any peripheral eigenvalue $\lambda$ of a quantum channel is semisimple, therefore $\ell_\lambda$ refers to both the algebraic and geometric multiplicities of $\lambda$. We call $\ell_{\mathrm P} $ the \emph{peripheral multiplicity} of $\Phi$. The quantity $\ell_0$, henceforth called the \emph{fixed-point multiplicity}, is the multiplicity of the eigenvalue $\lambda_0 = 1$ of $\Phi$, and therefore $\ell_{\mathrm P}  \geq \ell_0$.
Recall that a state belonging to the fixed-point subspace $\Fix(\Phi)$ is called \emph{steady} or \emph{stationary}, whereas a state in the attractor subspace $\Attr(\Phi)$ is called an \emph{asymptotic} state. 
As a consequence of Proposition~\ref{trace_eigen_prop} one has
\begin{equation} 
    \ell_0 = \dim(\Fix(\Phi)) = \dim(\Fix(\Phi^\dagger)), \quad  \ell_{\mathrm P}  = \dim(\Attr(\Phi)) = \dim(\Attr(\Phi^\dagger)).
\end{equation}
Moreover, any element of the steady and the attractor subspaces can be expressed in terms of a basis consisting of states, as the following result shows~\cite[Corollary 6.5]{wolf2012quantum}. 
\begin{proposition}[Bases of the steady and asymptotic subspaces]
\label{basis_states_prop}
Let $\Phi$ be a quantum channel on $\mathcal{B}(\mathcal{H})$ with steady subspace $\Fix(\Phi)$ and attractor subspace $\Attr(\Phi)$. Then there exist states
$\rho_i\in\mathcal{S}(\mathcal{H})$, $i\in[\ell_{\mathrm P}]$, such that
\begin{equation}
\Fix(\Phi) = \mathrm{span} \{ \rho_i \}_{i\in[\ell_0]}, \quad 
\Attr(\Phi) = \mathrm{span} \{ \rho_i \}_{i\in[\ell_{\mathrm P}]}. 
\end{equation}
The states $\rho_i$ are stationary for $i\in[\ell_0]$, whereas they are
asymptotic but non-stationary for
$i\in\{\ell_0+1,\dots,\ell_{\mathrm P}\}$. In particular, every quantum channel $\Phi$ always admits a stationary state $\rho$, i.e., $\Phi(\rho) = \rho$.
\end{proposition}
\begin{proof}
Let 
$\{ \sigma_i \}_{i\in[d^2]} \subset \mathcal{S}(\mathcal{H})$ be a basis of $\mathcal{B}(\mathcal{H})$ consisting of density operators. Since every $X \in \mathcal{B}(\mathcal{H})$ can be written as
\begin{equation}
    X = \sum_{i\in[d^2]} c_i \sigma_i , \qquad c_i \in \mathbb{C},
\end{equation}
we have
\begin{align}
\mathcal{P}(X) = \sum_{i\in[d^2]} c_i \mathcal{P}(\sigma_i),  
\qquad
\mathcal{P}_{\mathrm P} (X) = 
\sum_{i\in[d^2]} c_i \mathcal{P}_{\mathrm P} (\sigma_i). 
\end{align}
Therefore,
\begin{equation}
\Fix(\Phi) = \mathrm{span} \{ \mathcal{P}(\sigma_i) \}_{i\in[d^2]}, \quad 
\Attr(\Phi) = \mathrm{span} \{ \mathcal{P}_{\mathrm P}(\sigma_i) \}_{i\in[d^2]}. 
\end{equation}
By Proposition~\ref{peripheral_Wolf}, both $\mathcal{P}$ and
$\mathcal{P}_{\mathrm P}$ are quantum channels. Hence
$\mathcal{P}(\sigma_i)$ and $\mathcal{P}_{\mathrm P}(\sigma_i)$
are density operators for all $i\in[d^2]$.

Since $\dim(\Fix(\Phi))=\ell_0$, we can select
$\ell_0$ linearly independent elements among the stationary states $\mathcal{P}(\sigma_i)$, namely $\rho_k= \mathcal{P}(\sigma_{i_k})$ with $k\in[\ell_0]$, and the claim for $\Fix(\Phi)$ follows. 
Moreover, since
$\Fix(\Phi)\subseteq\Attr(\Phi)$, we can select from the states
$\mathcal{P}_{\mathrm P}(\sigma_i)$ further
$\ell_{\mathrm P}-\ell_0$ elements,
$\rho_k=\mathcal{P}_{\mathrm P}(\sigma_{i_k})$,
with $k\in\{\ell_0+1,\dots,\ell_{\mathrm P}\}$, so as to extend the
basis of $\Fix(\Phi)$ to a basis of $\Attr(\Phi)$.
Hence, $\Attr(\Phi) = \mathrm{span} \{ \rho_i \}_{i\in[\ell_{\mathrm P}]}$ and by construction the additional states are asymptotic but non-stationary.

Finally, since $1$ is always an eigenvalue of a quantum channel by
point~ii.\ of Theorem~\ref{spec_qc}, one has $\ell_0\geq1$, and hence
$\Phi$ admits at least one stationary state.
\end{proof}

This result implies that the fixed-point multiplicity $\ell_0$ and the peripheral multiplicity $\ell_{\mathrm P} $ are the number of linearly independent stationary and asymptotic states of a channel $\Phi$, respectively. 
If $X\in\Attr(\Phi)$ is an eigenoperator of $\Phi$ with eigenvalue $\lambda\neq1$, then 
it can always be expressed as a linear combination of asymptotic states. However, note that due to Proposition~\ref{trace_eigen_prop}, none of these states can be eigenoperators of $\Phi$ corresponding to  $\lambda\neq 1$.

Now, by Theorem~\ref{spec_qc}, $\lambda _0= 1$ is always an eigenvalue of $\Phi$, and so we obtain the following lower bound for $\ell_0$ and $\ell_P$
\begin{equation}\label{l_b_m0}
\ell_{\mathrm{P}} \geq \ell_{0} \geq 1.
\end{equation}
These bounds are sharp, as they can both be saturated.
A channel satisfying the condition $\ell_0=1$, i.e., with a one-dimensional steady subspace, is referred to as \emph{ergodic}, while a channel fulfilling the equality $\ell_{\mathrm{P}}=1$, that is, the attractor subspace is one-dimensional, is called \emph{mixing}. Historically, the notions of ergodic and mixing channels were formally introduced in~\cite{streater2009statistical,burgarth2007generalized,burgarth2013ergodic} in order to go beyond the invertibility assumption for the unique stationary state, and continue to draw a lot of attention in various contexts~\cite{singh2024ergodic,aravinda2024ergodic,hahn2024bath}. For instance, note that a random quantum channel naturally constructed through the Ginibre ensemble has been shown to be generically mixing~\cite{Bruzda2009}.

By definition, an ergodic channel $\Phi$ has a unique stationary state $\rho \in \mathcal{S}(\mathcal{H})$, meaning that
\begin{equation}
    \mathcal{P}(\sigma) = \lim_{N \rightarrow \infty} \frac{1}{N} 
    \sum_{n\in[N]} \Phi^n(\sigma) = \rho, \quad\forall \sigma \in \mathcal{S}(\mathcal{H}),
\end{equation}
	where the expression~\eqref{P_for} for the steady projection $\mathcal{P}$ of $\Phi$ was employed. In other words, the average state obtained after $N$ steps of the QMC $\Sem=(\Phi^n)_{n \in \mathbb{N}}$ generated by $\Phi$, namely the ergodic mean of the evolution, converges to $\rho$ in the limit $N \rightarrow \infty$, independently of the initial state $\sigma$. 
    
    On the other hand, when $\Phi$ is mixing, we have
\begin{equation}
\label{asympt_mix_ch}
\mathcal{P}_{\mathrm{P}}(\sigma) = \lim_{n \rightarrow \infty} \Phi^n(\sigma) = \rho, \quad\forall \sigma \in \mathcal{S}(\mathcal{H}),
\end{equation}
which follows from the expression~\eqref{Pp_for} for the peripheral projection $\mathcal{P}_{\mathrm P} $ and the fact that $\spec_{\mathrm P} (\Phi) = \{ 1 \}$~\cite[Theorem 6.7]{wolf2012quantum}. The condition~\eqref{asympt_mix_ch} can be taken as the definition of a mixing channel~\cite[Definition 2]{burgarth2013ergodic} and, in words, it guarantees that \emph{any} initial state $\sigma$ converges to the unique stationary state $\rho$ of the channel. 

Example~\ref{ex_5}, for which $\dim(\Fix(\Phi))=1$ and $\dim(\Attr(\Phi))=2$, shows that an ergodic channel is not mixing in general, see~\cite[example 9]{hahn2024bath} for other examples. Interestingly, an ergodic channel whose unique stationary state is pure, such as the amplitude damping channel discussed in Section~\ref{sec:amplitude_damping_channels}, is also mixing~\cite[Theorem 8]{burgarth2007generalized}. 
Observe also that the depolarizing channel introduced in Section~\ref{sec:depolarizing_channels} is mixing and, in fact, it is \emph{primitive}, namely it is mixing and its unique steady state is invertible, implying that the channel is faithful; cf.\ Definition~\ref{faith_ch_def}. Instead, an ergodic channel with a faithful steady state, such as the one presented in Example~\ref{ex_5}, is called \emph{irreducible}~\cite[Section 2]{wolf2012quantum}. The terminology which we just introduced is inspired by the theory of classical Markov chains, briefly reviewed in Appendix~\ref{classic_sec}.  

Finally, note that the structure of the asymptotics for an ergodic quantum channel can be easily obtained from the general results discussed in Section~\ref{sec:structure_theorems}. We have
\begin{equation}
\label{attr_struc_irred}
\Attr(\Phi) = 0\oplus 
\bigoplus_{k\in[M]} \mathbb{C} \rho_k,
\end{equation}
that is $d_k = \dim(\mathcal{H}_{k,1}) = 1$ for all $k \in [M]$. This easily follows by combining Theorem~\ref{Wolf_struc} and the defining condition $\ell_0 =1$. Moreover, the action of the channel on an element $X \in \Attr(\Phi)$ reads
\begin{equation}
\label{ergodic_asympt_dyn}
    X=0\oplus
    \bigoplus_{k\in[M]}c_k\rho_k\mapsto \Phi(X)=0\oplus  
    \bigoplus_{k\in[M]} c_{\pi(k)}\rho_k,
\end{equation}
where $\pi$ is a cyclic permutation.
Consistently, this implies that
\begin{equation}
\Fix(\Phi) = \mathbb{C} \rho, \quad \rho = 0\oplus \frac{1}{M} 
\bigoplus_{k\in[M]} \rho_k,
\end{equation}
which also follows from~\eqref{eq:fix_cyclicdec} with $\mathcal{H}_{1} \simeq \mathbb{C}$. Moreover, it readily follows from~\eqref{ergodic_asympt_dyn} that the peripheral eigenvalues of $\Phi$ are the $M$-th roots of unity, that is $\lambda_\ell = \exp{2 \pi \mathrm{i} \ell / M}$ with $\ell =0, \dots , M-1$, and are simple. 
In conclusion, the asymptotic dynamics does not involve unitaries, and it is fully induced by a cyclic permutation $\pi$, as illustrated in Example~\ref{ex_5} for a qubit. Therefore, the long-time dynamics in~\eqref{ergodic_asympt_dyn} is classical in the sense explained in the last part of Section~\ref{sec:cyclic_dec}.

We conclude the subsection by discussing sharp upper bounds for the fixed-point and peripheral multiplicities of nontrivial quantum channels.  
\begin{theorem}[Asymptotic multiplicities of quantum channels]
\label{cor_unit}
Let $\Phi \neq \mathsf{1}$ be a nontrivial quantum channel on $\mathcal{B}(\mathcal{H})$ with $d = \dim(\mathcal{H})$. Then, the fixed-point and peripheral multiplicities $\ell_0$ and $\ell_P$ of $\Phi$ satisfy
\begin{itemize}
    \item[\textnormal{i.}]  $\ell_{0} \leq d^{2}-2d+2$ and $\ell_{\mathrm P}  = d^2$, if $\Phi$ is unitary;
    \item[\textnormal{ii.}]  $\ell_0 \leq \ell_{\mathrm{P}} \leq d^{2}-2d+2$, if $\Phi$ is non-unitary. 
\end{itemize}
Moreover, all the bounds are sharp.
\end{theorem}

The claim for nontrivial unitary channels easily follows from the analysis done in Section~\ref{sec:unitary_evolution}; see~\cite[Theorem 1]{AF_bounds1} for details. The non-unitary case can be addressed by using the asymptotic structure theorem~\ref{Wolf_struc} or the KS representation~\eqref{op_sum_rep} of quantum channels, as shown in~\cite[Theorem 2]{AF_bounds1}. See~\cite{AF_bounds1} for the proof of the sharpness of the bounds.

\subsection{GKLS generators}
\label{GKLS_bounds-subsec}
Let $\Sem = (\Phi_t)_{t\in \mathbb{R}^+}$ be a QDS with GKLS generator $\mathcal{L}$
and let
\begin{equation}
m_0 := \dim(\Ker(\mathcal{L})), \quad  m_{\mathrm P}  := \dim  (\Attr(\mathcal{L})),
\end{equation}
be the dimensions of the kernel and  attractor subspaces of $\mathcal{L}$, respectively. 

Note that, as in Proposition~\ref{basis_states_prop} for quantum channels, $\Ker(\mathcal{L})$ admits a basis of stationary states, while $\Attr(\mathcal{L})$ admits a basis of asymptotic states. Consequently, $m_0$ and $m_{\mathrm P}$ are the numbers of linearly independent stationary and asymptotic states of the QDS $\Sem$, respectively.
Moreover, we have
\begin{equation}
\label{low_bound_L}
m_{\mathrm{P}} \geq m_{0} \geq 1,
\end{equation}
by Theorem~\ref{spec_L}. A QDS saturating the second inequality in~\eqref{low_bound_L}, namely $m_{0} = 1$, is called \emph{ergodic}~\cite[Section~6]{burgarth2013ergodic}. Importantly, it has been proven in~\cite[Theorem 1]{Schirmer_relax} that
\begin{equation}
 \label{Schirm_Wang_attract}
 m_{0} = 1 \Rightarrow m_{\mathrm{P}}=1,
 \end{equation} 
 while $\ell_{\mathrm{P}} =1$ does not follow in general from $\ell_0 = 1$ because of the possible presence of permutations in the asymptotic evolution; cf.\ Equation~\eqref{struc_phiP}. 
This means that, for any ergodic QDS $\Sem$, the \emph{relaxing} property  
\begin{equation}
\label{glob_attr_cond}
\mathcal{P}_{\mathrm{P}}(\sigma) = \lim_{t\rightarrow \infty} \Phi_t(\sigma) = \rho, \quad \forall \sigma \in \mathcal{S}(\mathcal{H})
\end{equation}
holds, with $\rho\in\mathcal{S}(\mathcal{H})$ being the unique stationary state of the dynamics.

The amplitude damping Markovian dynamics 
in~\eqref{ampl_channel_def} is a paradigmatic example of an ergodic QDS.
In fact, ergodic continuous-time Markovian evolutions are ubiquitous in many physical scenarios~\cite{alickilendi}, and they are also employed for various quantum algorithms such as dissipative quantum computation~\cite{Verstraete2009} and quantum reservoir engineering~\cite{Zoller_res_eng,kraus2008preparation,diehl2008quantum,Verstraete2009}, that will be discussed in Section~\ref{QRE_subsec}. Moreover, they have been analyzed numerically by means of variational methods over the last years~\cite{nagy2018driven,vicentini2019variational,nagy2019variational,hryniuk2024tensor,mortimer2025certifying}.

Similarly to quantum channels, an ergodic QDS whose unique stationary state  is invertible is called \emph{irreducible}. Note that we are adopting the notion of irreducibility introduced by Davies  in~\cite{davies1970quantum} for QDSs, which is stronger than Evans' irreducibility~\cite{evans1977_irred}; see~\cite{zhang2024criteria} for a comparison of the two concepts.  

We now review several results relating irreducible QDSs and the GKLS form of their generators.
First, let us discuss a characterization of irreducibility, which was first proved in~\cite{Frigerio_77} and immediately follows from Proposition~\ref{fixed_L}.
\begin{corollary}
\label{Frigerio_first_cor}
Let $\Sem$ be a QDS with generator $\mathcal{L}$ in the GKLS form~\eqref{diag_GKLS}.  Set \begin{equation}
   T := \{ H \} \cup \{  L_{\alpha} , L_{\alpha}^{\ast} 
\}_{\alpha \in[N]} .
\end{equation} 
Then, $\Sem$ is irreducible if and only if $\Sem$ is faithful and the commutant of $T$ is trivial, that is $T^\prime = \mathbb{C} \mathbb{I}$. 
\end{corollary}
\begin{proof}
    First, observe that, since $m_{0} = \dim(\Ker(\mathcal{L})) = \dim(\Ker(\mathcal{L}^\dagger))$, irreducibility of $\Sem$ implies that $\Ker(\mathcal{L}^\dagger) = \mathbb{C} \mathbb{I}$. If $\Sem$ is irreducible, then $\Sem$ is faithful by definition, so $\Ker(\mathcal{L}^\dagger) = \mathbb{C} \mathbb{I}$ is equivalent to $T^\prime = \mathbb{C}\mathbb{I}$ by Proposition~\ref{fixed_L}, as claimed. Conversely, if $\Sem$ is faithful and $T^\prime = \mathbb{C} \mathbb{I}$, then $\Ker(\mathcal{L}^\dagger) = \mathbb{C}\mathbb{I}$ again by Proposition~\ref{fixed_L} and, consequently, $\Sem$ admits a unique invertible stationary state, namely $\Sem$ is irreducible.  
\end{proof}
Let us now turn our attention to a more explicit characterization of irreducibility in terms of the GKLS representation of the generators~\cite[Theorem 5.2]{zhang2024criteria}; cf.\ also~\cite[Theorem 1]{yoshida2024uniqueness}. See~\cite[Theorem 5.1]{zhang2024criteria} for yet another characterization of irreducible QDSs.
\begin{theorem}
\label{irred_char_thm}
  Let $\Sem$ be a QDS on $\mathcal{B}(\mathcal{H})$ with  generator $\mathcal{L}$ in the GKLS form~\eqref{diag_GKLS}. Set 
  \begin{equation}
    S := \{ L_{\alpha} 
  \}_{\alpha \in[N]} \cup \{ K \}, \qquad  \text{with } K:= \mathrm{i}H + \frac{1}{2} 
  \sum_{\alpha \in[N]} L_{\alpha}^\ast L_{\alpha}.  
  \end{equation}
   Then, $\Sem$ is irreducible if and only if $\mathcal{A}(S) = \mathcal{B}(\mathcal{H})$, where  $\mathcal{A}(S)$ is the algebra  generated by $S$.  
\end{theorem}
 It is worth pointing out that $\mathcal{A}(S)$ is the space generated by $S$ under multiplication and complex linear combination, implying that $\mathcal{A}(S)$ is an algebra but \emph{not} a $C^\ast$-algebra in general. Interestingly, an easy consequence of Theorem~\ref{irred_char_thm}---see~\cite[Corollary 5.5]{zhang2024criteria} for details---is the following sufficient condition for irreducibility found by Frigerio in~\cite[Theorem 3.2]{Frigerio_78}.
 \begin{corollary}
 \label{Frigerio_second_cor}
     Let $\Sem$ be a QDS  with generator $\mathcal{L}$ in the GKLS form~\eqref{diag_GKLS}. Set $T := \{ L_{\alpha} \}_{\alpha \in [N]}$ and suppose that 
     \begin{itemize}
     \item[\textnormal{i.}] $\mathrm{span}(T)$ is self-adjoint, namely $X^\ast \in \mathrm{span}(T)$ whenever $X \in \mathrm{span}(T)$;
     \item[\textnormal{ii.}] The commutant of $T$ is trivial, i.e., $T^\prime = \mathbb{C} \mathbb{I}$. 
     \end{itemize}
     Then, $\Sem$ is irreducible.
 \end{corollary}
 Notice that, at variance with Corollary~\ref{Frigerio_first_cor} and Theorem~\ref{irred_char_thm}, the latter result \emph{only} relies on the jump operators $L_{\alpha}$, $\alpha \in[N]$, of $\mathcal{L}$ and, consequently, the dissipative part $\mathcal{L}_{\mathrm{D}}$ of $\mathcal{L}$. In particular, this implies that the hypotheses in Corollary~\ref{Frigerio_second_cor} are only sufficient for irreducibility.

As a final remark, notice that it is possible to derive sharp upper bounds for the multiplicity $m_{0}$ in the case of Hamiltonian continuous-time dynamics and for the multiplicity $m_{\mathrm{P}}$ (and thus $m_{0}$) for non-Hamiltonian ones, analogous to the constraints discussed in Theorem~\ref{cor_unit} for quantum channels. See~\cite[Theorems 3-4]{AF_bounds1} for the precise claims and their proofs.  

 \section{Convergence to steady states}
 \label{conv_sec}
 So far, we have focused on the structure of the fixed-point and attractor subspaces of open-quantum system evolutions, as well as on their asymptotic dynamics. 
A natural next step is to investigate how the asymptotic state depends on the initial state of the system and, in particular, the conditions under which a quantum system relaxes to (thermal) equilibrium~\cite{linden2009quantum,smirne2012role,wenderoth2023quantifying}, a problem that also played an important role in the development of open-quantum-system theory; see Section~\ref{histo_sec}. 
The problem of inferring the asymptotic state or trajectory of a quantum system from partial or complete information on its initial state has also been examined in~\cite{novotny2023jaynes}, where several versions of Jaynes's principle are formulated for general discrete-time and continuous-time Markovian evolutions.

 It is also important to investigate the inverse problem: given an open-system evolution, which initial states converge to a given steady state in the long-time limit? This question leads to the notion of attraction domains of stationary states discussed in the next subsection. In Section~\ref{QRE_subsec} we will then briefly discuss quantum reservoir engineering, namely the problem of realizing a prescribed state as the steady state of a Markovian dynamics. Finally, in Section~\ref{sec:ckks_bound}, we will address the issue of the convergence rates of open-system evolutions and, in particular, present a bound on these  rates due to Chru{\'s}ci{\'n}ski, Kimura, Kossakowski and Shishido~\cite{Chru_CKK_1}. Throughout this section, we will  focus mainly on QDSs.

\subsection{Domains of attraction}
\label{dom_attr_subsec}
In this subsection we shall explore domains of attraction of steady states, recently addressed in~\cite{zhang2023attraction,fagnola2024basins} for continuous-time Markovian evolutions. Given a QDS $\Sem = (\mathrm{e}^{t\mathcal{L}})_{t \in \mathbb{R}^+}$ with GKLS generator $\mathcal{L}$, we define the 
\emph{domain of attraction} $\mbox{DoA}(\rho_{\mathrm{s}})$ of a steady state $\rho_{\mathrm{s}} \in \mathcal{S}(\mathcal{H}) \cap \Ker(\mathcal{L})$ as
\begin{equation}
\label{dom_attr_def}
    \mbox{DoA}(\rho_{\mathrm{s}}) := \left\{ \rho \in \mathcal{S}(\mathcal{H}) \, : \, \lim_{t \rightarrow \infty} \mathrm{e}^{t\mathcal{L}}(\rho) = \rho_{\mathrm{s}} \right\}. 
\end{equation}
Similarly, for a QMC $\Sem = (\Phi^n)_{n \in \mathbb{N}}$ generated by a quantum channel $\Phi$, the domain of attraction of a steady state   $\rho_{\mathrm{s}} \in \mathcal{S}(\mathcal{H}) \cap \Fix(\Phi)$ is defined as
\begin{equation}
\label{dom_attr_def_discrete}
    \mbox{DoA}(\rho_{\mathrm{s}}) := \left\{ \rho \in \mathcal{S}(\mathcal{H}) \, : \, \lim_{n \rightarrow \infty} \Phi^{n}(\rho) = \rho_{\mathrm{s}} \right\}. 
\end{equation}

Note that the limits in~\eqref{dom_attr_def} and~\eqref{dom_attr_def_discrete} need not exist for arbitrary initial states, since $\mathcal{L}$ may have nonzero peripheral eigenvalues, and $\Phi$ may have peripheral eigenvalues other than one, see Proposition~\ref{prop:Noether_convergence} below. 
In either case, convergence to $\rho_{\mathrm{s}}$ is equivalent to the condition $\mathcal{P}_{\mathrm{P}}(\rho) = \rho_{\mathrm{s}}$, where $\mathcal{P}_{\mathrm{P}}$ denotes the peripheral projection of the corresponding generator $\mathcal{L}$ or channel $\Phi$; see~\eqref{per_proj_def} and~\eqref{per_proj_def_L}. Thus,
\begin{equation}
    \mbox{DoA}(\rho_{\mathrm{s}}) = \left\{ \rho \in \mathcal{S}(\mathcal{H}) \, : \, \mathcal{P}_{\mathrm{P}}(\rho) = \rho_{\mathrm{s}} \right \}.
    \end{equation}
Indeed, the component outside the peripheral subspace decays to zero, while the peripheral component converges to $\rho_{\mathrm{s}}$ if and only if it equals $\rho_{\mathrm{s}}$ initially. 
This characterization shows that $\operatorname{DoA}(\rho_{\mathrm{s}})$ is a closed, convex subset of the compact state space $\mathcal{S}(\mathcal{H})$, and hence is compact. It is also nonempty, since $\rho_{\mathrm{s}} \in \operatorname{DoA}(\rho_{\mathrm{s}})$.
  
 Now we focus on the domains of attraction of quantum dynamical semigroups. According to~\eqref{Schirm_Wang_attract}, when the Markovian evolution $\Sem$ admits a unique steady state $\rho_{\mathrm{s}}$, then $\mbox{DoA}(\rho_{\mathrm{s}}) = \mathcal{S}(\mathcal{H})$, namely $\rho_{\mathrm{s}}$ is globally attractive. In the case of a degenerate fixed-point subspace, the analysis of domains of attraction becomes much more involved and, indeed, already for classical dynamical systems the exact derivation of attraction domains is unfeasible in many cases and numerical approaches are often adopted to obtain partial results~\cite{zarei2018arc}. To begin with, observe that in the case $m_{0}  = \dim(\Ker(\mathcal{L})) \geq 2$ the states converging to a given steady state $\rho_{\mathrm{s}}$ are infinite by the convexity of $\mathrm{DoA}(\rho_{\mathrm{s}})$. A general result characterizing attraction domains of Markovian evolutions is the following~\cite[Theorem 1]{zhang2023attraction}.
\begin{proposition}
\label{DoA_char_Zhang}
Let $\Sem = (\Phi_t)_{t \in \mathbb{R}^+}$ be a QDS with GKLS generator $\mathcal{L}$ and $\rho_{\mathrm{s}} \in \Ker(\mathcal{L}) \cap \mathcal{S}(\mathcal{H})$ a steady state of $\Sem$. Then $\rho \in \mathrm{DoA}(\rho_{\mathrm{s}})$ if and only if
\begin{equation}
     \Tr(\eta_{\ell} \rho) = \Tr(\eta_{\ell} \rho_{\mathrm{s}}) \quad \forall \ell \in [m_{\mathrm P}] ,
\end{equation}
for any system of linearly independent Hermitian operators $\{\eta_\ell
\}_{\ell\in[m_{\mathrm P}]}$ spanning $\Attr(\mathcal{L}^\dagger)$, where $m_{\mathrm P}  = \dim  (\Attr(\mathcal{L}))$.
\end{proposition}
 So we infer that domains of attraction may be characterized in terms of a set of $m_{\mathrm P} $ expectation values of asymptotic observables $\{ \eta_\ell 
 \}_{\ell\in[m_{\mathrm P}]}$. 
 In the case of a non-unique steady state, it is also interesting to quantify the volume occupied by the attraction domains of all the steady states. Interestingly, it turns out that, if $m_{\mathrm P}  \geq 2$, then $\mu (\mbox{DoA}(\rho_{\mathrm{s}})) = 0$ for any steady state $\rho_{\mathrm{s}} \in \Ker(\mathcal{L})$ for a class of measures $\mu$ on trace-one Hermitian operators on $\mathcal{H}$, see~\cite[Theorem 2]{zhang2023attraction} for details, and the proof of~\cite[Theorem 3]{zhang2023attraction} for an explicit construction of a measure with this property. 
 
 A more explicit structure of the domains of attraction may be easily obtained by employing the decomposition of the attractor subspace~\eqref{attr_struc_QDS} for faithful Markovian evolutions~\cite[Theorem 4.1]{fagnola2024basins}.
\begin{proposition}
    Let $\Sem = (\Phi_t)_{t\in \mathbb{R}^+}$ be a faithful QDS with generator $\mathcal{L}$ and attractor subspace $\Attr(\mathcal{L})$ of the form~\eqref{attr_struc_QDS}, and let $\rho_{\mathrm{s}} \in \Ker(\mathcal{L})$ be a steady state of $\Sem$. Then, denoting by $P_k$ the Hermitian projection onto $\mathcal{H}_{k,1} \otimes \mathcal{H}_{k,2}$ with $k\in [M]$, any state $\rho \in \mathrm{DoA}(\rho_{\mathrm{s}})$ can be written in the form
    \begin{equation}
    \label{struc_DoA}
        \rho = \bigoplus_{k\in[M]}\bigl(\Tr_{k,2}(P_k \rho_{\mathrm{s}} P_k)  \otimes \omega_k\bigr) + X_0,
    \end{equation}
    for some states $\omega_k \in \mathcal{S}(\mathcal{H}_{k,2})$ with $k \in [M]$, and a Hermitian operator
    \begin{equation}
        X_0 \in \mathrm{span}\{\mathcal{M}_{\lambda}\,:\,\lambda \in\spec_\mathrm B(\mathcal{L})\},
    \end{equation}
    where $\mathcal{M}_\lambda$ is the generalized eigenspace corresponding to the bulk eigenvalue $\lambda \in \spec_{\mathrm{B}}(\mathcal{L})$ in~\eqref{specB_L}.
\end{proposition}

Notice that the operator $X_0$ involved in~\eqref{struc_DoA} must be traceless since $\Tr(\rho) = \Tr(\rho_{\mathrm{s}}) = 1$, and this is in agreement with assertion i.\ of Proposition~\ref{trace_eigen_prop}.

\subsection{Quantum reservoir engineering}
\label{QRE_subsec}
Characterizing domains of attraction is closely related to the following central problem:

    \textbf{Q.}
    \emph{ Given a state $\sigma \in \mathcal{S}(\mathcal{H})$, how to construct a Markovian dynamics 
    converging to $\sigma$ as its unique steady state?}

This question has been extensively studied from the theoretical perspective~\cite{ticozzi2008quantum,ticozzi2009analysis,ticozzi2010stabilizing,Ticozzi2012,guo2025designing,souza2024lindbladian}, including in the context of \emph{quantum reservoir engineering}~\cite{Zoller_res_eng,kraus2008preparation,diehl2008quantum,Verstraete2009}, dissipative quantum computation~\cite{Verstraete2009,sinayskiy2012efficiency}, and dissipative error correction~\cite{kastoryano2013precisely,reiter2017dissipative}. See~\cite{harrington2022engineered,fazio2025many} for overviews of this topic. In particular, quantum reservoir engineering, which has been experimentally realized on various platforms~\cite{lin2013dissipative,ma2019dissipatively,gertler2023experimental,mi2024}, aims at preparing or stabilizing a (typically entangled) state as the stationary state of a properly engineered Markovian dynamics. For example, this technique can be used to prepare  working-medium states of quantum thermal machines~\cite{brask2015autonomous}, ground states of many-body Hamiltonians in condensed matter and statistical mechanics~\cite{Verstraete2009,kraus2008preparation}, and initial states of quantum algorithms. Furthermore, quantum reservoir engineering has motivated extensive research into the entanglement properties of stationary states in multipartite systems since the late nineties~\cite{plenio1999cavity,beige2000entangling,schneider2002entanglement,hartmann2006steady}.

It is not hard to find a simple solution of the above question \textbf{Q.} 
by using the GKLS generator $\mathcal{L}(X)=\Tr(X) \sigma - X$, that generates the relaxing dynamics
\begin{equation}
\label{contra_evol}
    \Phi_t(X) =\mathrm e^{{t \mathcal{L}}}(X)= \mathrm e^{-t} X + (1-\mathrm e^{-t}) \Tr(X) \sigma, \quad t \in \mathbb{R}^+, \; X \in \mathcal{B}(\mathcal{H}),
\end{equation}
whose unique steady state is given by $\sigma$. Unfortunately, this evolution is not easy to implement in concrete experiments. Specifically, in many-body systems, such as spin chains, we are interested in realizing dissipative Markovian evolutions with quasi-local jump operators. 

We present here a solution to \textbf{Q.} satisfying these constraints for pure states of a qudit system~\cite{Ticozzi2012}. We start by properly defining the notion of quasi-locality in our setting. Consider an $n$-body system of qudits described by a Hilbert space $\mathcal{H} = (\mathbb{C}^d)^{\otimes n}$, and a family $\{\mathcal N_1,\dots,\mathcal N_T\}$ of subsets of $[n]$, namely
\begin{equation}
\mathcal N_\ell\subseteq\{1,2,\dots, n\},\quad\forall \ell\in[T],
\end{equation}
henceforth called \emph{neighbourhoods}. 
For instance, one can associate with each qudit a neighbourhood made of the adjacent qudits. We say that a jump operator $L$ on $\mathcal{H}$ is quasi-local with respect to this family of neighbourhoods if there exists an index $\ell\in [T]$ such that $L = L_{\mathcal{N}_\ell} \otimes \mathbb{I}_{\overline{\mathcal{N}}_\ell}$, where $L_{\mathcal{N}_\ell}$ acts on the subsystems labelled by $k \in \mathcal{N}_\ell$ while $\mathbb{I}_{\overline{\mathcal{N}}_\ell}$ is the identity on the remaining $\overline{\mathcal{N}}_\ell = [n] \setminus \mathcal{N}_\ell$ qudits. Similarly, a Hamiltonian $H = H^\ast$ on $\mathcal{H}$ is called quasi-local if it can be written as a sum of quasi-local operators
\begin{equation}\label{eq:local_hamiltonian}
    H = 
    \sum_{\ell\in[T]} H_\ell, \quad H_\ell = H_{\mathcal{N}_\ell} \otimes \mathbb{I}_{\overline{\mathcal{N}}_\ell}, \quad \ell \in [T].
\end{equation}

A pure quantum state $\ketbra{\Psi}{\Psi}$ is called \emph{\gls*{dqls}} if there exist quasi-local jump operators $L_\alpha$ with $L_\alpha \ket{\Psi} = 0$, $\alpha \in [N]$, such that $\ketbra{\Psi}$ is the unique steady state of the relaxing Markovian dynamics generated by 
\begin{equation}
\mathcal{L}(X) =  
\sum_{\alpha\in[N]} \Bigl( L_{\alpha} X L_{\alpha}^\ast - \frac{1}{2} \{ L_{\alpha}^\ast L_{\alpha} , X \} \Bigr), \quad X \in \mathcal{B}(\mathcal{H}).
\end{equation}

A related notion, also connected to the analysis in~\cite{kraus2008preparation,Verstraete2009}, is that of \emph{parent Hamiltonian}. Specifically, a pure state $\ketbra{\Psi}$ is said to be a frustration-free ground state if it is the ground state of a quasi-local Hamiltonian $H$ satisfying the property
\begin{equation}
    \langle \Psi |H_\ell| \Psi \rangle = \lambda_{\mathrm{min}}(H_\ell), \quad \forall \ell \in [T],
\end{equation}
where $\lambda_{\mathrm{min}}(H_\ell)$ is the minimum eigenvalue of $H_\ell$, and the Hamiltonian is called parent if it admits a unique ground state which is frustration-free.  
Interestingly, DQLS pure states and parent Hamiltonians can be characterized in the same way~\cite[Theorem 3.3, Corollary 3.4]{Ticozzi2012}.

\begin{proposition}
    Let $\ketbra{\Psi}{\Psi}$ be a pure state, and let $\{ \mathcal{N}_\ell 
    \}_{\ell\in[T]}$ be a family of neighbourhoods. Then, the following conditions are equivalent
    \begin{itemize}
        \item[\textnormal{i.}] $\ketbra{\Psi}{\Psi}$ is a DQLS state;
        \item[\textnormal{ii.}] $\ketbra{\Psi}$ admits a parent Hamiltonian;
        \item[\textnormal{iii.}] The range of $\ket{\Psi}$ satisfies the following property \begin{equation}\label{eq:condition_dqls}
        \Ran(\ketbra{\Psi}) = \bigcap
        \limits_{\ell\in[T]} \Ran(\rho_{\mathcal{N}_\ell} \otimes \mathbb{I}_{\overline{\mathcal{N}}_\ell}), 
    \end{equation}
    \end{itemize}
      where $\rho_{\mathcal{N}_\ell} := \Tr_{\overline{\mathcal{N}}_\ell}(\ketbra{\Psi})$ is the partial trace of $\Psi$ over the subsystems labelled by the complement of $\mathcal N_\ell$.
\end{proposition}
   
According to~\cite{kraus2008preparation,Verstraete2009}, any injective matrix product state~\cite{perez2008} as well as any stabilizer~\cite{gottesman1997stabilizer} and, in particular, graph~\cite{hein2006} state may be realized as the steady state of a relaxing dissipative Markovian dynamics with quasi-local jump operators for \emph{some} locality constraint, see also~\cite{ticozzi2010stabilizing}. Furthermore, engineering of mixed states, discussed in full generality in~\cite{ticozzi2010stabilizing}, has been addressed in the literature with a particular focus on thermal states~\cite{guo2024new,Chen2024} for their use in quantum algorithms~\cite{holmes2022quantum,rouze2025efficient}.

So far, we have constructed Markovian evolutions that reach the desired state only \emph{asymptotically}. However, preparing a target state in finite time from an arbitrary initial state is important for applications such as state preparation and information reset, see~\cite{cai2025far} for attempts in this direction.

\subsection{Rate of convergence}
\label{sec:ckks_bound}
 The peripheral part of the spectrum of quantum channels~\eqref{per_spec} and GKLS generators~\eqref{spec_L_def}, introduced in Sections~\ref{spec-qc-subsec} and~\ref{spec_lindbladian} respectively, 
can be fully characterized by using the structure Theorems~\ref{Wolf_struc} and~\ref{asympt_QDS_thm}, see also~\cite[Theorem 9]{wolf2010inverse}. However, a complete characterization of the bulk spectra of discrete- and continuous-time quantum dynamical semigroups remains out of reach. In particular, this would be the key step in order to understand how fast an open quantum system reaches an equilibrium state, which has been intensively addressed for channels and semigroups both by means of entropic quantities or norm distances~\cite{kastoryano2013quantum,mullerhermes2016entropy,gao2022complete,gao_complete_2025,fang2025mixing} and with spectrum-theoretic tools~\cite{terhal2000problem,temme2013lower,szehr2015spectral,mori2023symmetrized}. 
Under this latter perspective, convergence rates are encoded into the moduli of the eigenvalues $\lambda \in \spec(\Phi)$ for a QMC generated by a channel $\Phi$ and into the \emph{relaxation rates}  
\begin{equation}
\label{rel_rates}
  \gamma_{\lambda}:=\abs{\Re(\lambda)}=-\Re(\lambda),   
\end{equation}
with $\lambda \in \spec(\mathcal{L})$ for a QDS $\Sem$ with generator $\mathcal{L}$. In particular, an important figure of merit in quantifying the convergence towards equilibrium is the \emph{spectral gap}
\begin{equation}
   g(\Phi) := 1-\max_{\lambda \in \spec_{\mathrm{B}}(\Phi)}|\lambda| = 1-R(\Phi), \qquad g(\mathcal{L}) := \min_{\lambda \in \spec_{\mathrm{B}}(\mathcal{L})} \gamma_\lambda,
\end{equation}
defined for the QMC induced by a non-unitary channel $\Phi$ and the QDS generated by a non-Hamiltonian GKLS generator $\mathcal{L}$, respectively. Recall that $\spec_{\mathrm{B}}(\Phi)$ and $\spec_{\mathrm{B}}(\mathcal{L})$ are the bulk spectra of $\Phi$ and $\mathcal{L}$ respectively given by~\eqref{bul_spec} and~\eqref{specB_L}, and the parameter $R(\Phi)$ introduced in~\eqref{Rmax_def} represents the slowest decaying mode of the evolution generated by $\Phi$.  
Remarkably, the rates $\gamma_{\lambda}$ given by~\eqref{rel_rates} can be measured in experiments and, in fact, this has been done for two-level systems~\cite{Abragam_1961,Slichter_1990}.

We now present a recent constraint on the relaxation rates of a QDS, henceforth called \emph{\gls*{ckks} bound}, raised in~\cite{Chru_CKK_1,Chru_CKK_2} and subsequently proved in~\cite{muratore2024universal}.
\begin{theorem}[CKKS bound]
\label{CKKS_conj_thm}
Let $\Sem = (\Phi_t)_{t \in \mathbb{R}^{+}}$ be a QDS with generator $\mathcal{L}$. Then the relaxation rates of $\Sem$  satisfy
\begin{equation}
\label{CKKS_conj_eq}
\gamma_{\lambda} \leq \frac{1}{d} \sum_{\lambda'\in\spec(\mathcal L)} m_{\lambda^\prime} \gamma_{\lambda'}, \quad \forall \lambda \in\spec(\mathcal L),
\end{equation}
where $m_{\lambda^\prime}$ denotes the algebraic multiplicity of $\lambda'$, and $d$ is the Hilbert space dimension.
\end{theorem}  
Equation~\eqref{CKKS_conj_eq} states that no individual relaxation rate can exceed $1/d$ times the sum of all relaxation rates, counted with algebraic multiplicity. The bound is tight in every dimension~\cite{Chru_CKK_1}.
See~\cite{Chru_CKK_1} for details.

We notice that, unlike the spectral properties listed in Theorem~\ref{spec_L}, the CKKS bound does not hold in general for generators of semigroups of positive trace-preserving maps. It therefore imposes a spectral constraint beyond those implied by positivity alone.
In this sense, the bound~\eqref{CKKS_conj_eq} is a genuinely quantum spectral constraint for Markovian evolutions. 
See~\cite[Conjecture~2]{Chru_CKK_1} for a related conjecture concerning quantum channels, and~\cite{chruscinski2024universal,Ende2025,muratore2025universal} for further developments.

\section{Symmetries and asymptotic dynamics}
\label{sec:symmetries}
The nontrivial structure of the asymptotic subspace highlighted in Theorem~\ref{Wolf_struc} is highly nongeneric, reflecting the symmetries of the dynamics, which are ultimately determined by the system and its coupling to the environment. In fact, unstructured random quantum channels are generically mixing~\cite{Bruzda2009}, and therefore converge to a unique stationary state without any residual asymptotic dynamics, as shown in Section~\ref{qc_bounds_subsec}. By contrast, when present, symmetries provide one of the main mechanisms that protect nontrivial asymptotic structures against the generic tendency of the dynamics to be mixing. 
\subsection{Symmetries, decoherence-free subspaces, and noiseless subsystems}
\label{sec:sym_DFS}
Historically, the role of symmetries in open system dynamics became prominent in the study of decoherence-free subspaces~\cite{Palma1996,duan1997preserving,zanardi1997noiseless,Zanardi_dissip_1998,lidar_dec_1998},  discussed in Section~\ref{sec:dfs}, where it was realized that collective coupling of several qubits to a common environment gives rise to permutation symmetries, leading to decoherence-free subspaces and noiseless subsystems. Let us illustrate the general mechanism with the following example, adapted from~\cite{Kempe2001,zanardi1997noiseless}.
\begin{example}[Collective decoherence]
\label{ex:collective}
Consider $n$ qubits, $\Hi=(\C^{2})^{\otimes n}$, whose dynamics is described by a GKLS generator~\eqref{diag_GKLS} with $H=0$ and collective jump operators
\begin{equation}    L_\alpha=\sqrt{\gamma}\,S_\alpha,\qquad
S_\alpha=\frac{1}{2}\sum_{i\in[n]}\sigma_\alpha^{(i)},\qquad
\alpha\in[3],\quad\gamma>0,
\end{equation}
where $\sigma_\alpha^{(i)} = \mathbb{I}_2 \otimes \mathbb{I}_2 \otimes \dots \otimes \sigma_\alpha \otimes \dots \otimes  \mathbb{I}_2$ denotes the operator acting nontrivially only on the $i$-th qubit with the Pauli matrix $\sigma_\alpha$. Every  permutation $\pi\in\mathfrak{S}_n$ of the qubits is represented on $\Hi$ with a unitary $U_\pi$ commuting with all the $S_\alpha$. More precisely, as a consequence of Schur--Weyl duality~\cite{WEYL2016,FultonHarris1991}, the algebras generated by collective $SU(2)$ rotations and by qubit permutations are mutual commutants, yielding the decomposition
\begin{equation}
\label{eq:Schur_Weyl}
(\mathbb{C}^2)^{\otimes n}
=\bigoplus_j\bigl(\mathbb{C}^{n_j}\otimes\mathbb{C}^{2j+1}\bigr),
\qquad
n_j=\frac{(2j+1)\,n!}{(n/2+j+1)!\,(n/2-j)!},
\end{equation}
where $j=n/2,n/2-1,\ldots$ down to $0$ ($\,1/2\,$) when $n$ is even (odd). The factors $\mathbb{C}^{n_j}$ carry inequivalent irreducible representations $u_j$ of $\mathfrak{S}_n$, while $\mathbb{C}^{2j+1}$ are their multiplicity spaces, carrying the spin-$j$ representations of $SU(2)$. The two actions take the complementary forms
\begin{equation}
    U_\pi=\bigoplus_j \bigl( u_j(\pi)\otimes\mathbb{I}_{2j+1} \bigr),
\quad
S_\alpha =\bigoplus_j
\bigl(\mathbb{I}_{n_j}\otimes s_\alpha^{(j)}\bigr), 
\end{equation}
where
\begin{equation}
    u_j(\pi)\in \mathcal{B}(\mathbb{C}^{n_j}), \qquad s_\alpha^{(j)} = s_\alpha^{(j)\ast}\in \mathcal{B}(\mathbb{C}^{2j+1}), \quad \alpha \in [3].
\end{equation}
Thus, the noise acts only on the spin factors $\mathbb{C}^{2j+1}$, leaving the permutation representation factors $\mathbb{C}^{n_j}$ noise-protected. These will be, respectively, the factors $\mathcal{H}_{j,2}$ and $\mathcal{H}_{j,1}$ of the asymptotic decomposition~\eqref{H_dec_Mark}. 
The asymptotic structure~\eqref{attr_struc_QDS} reads
\begin{equation}
\label{eq:collective_attr}
\Attr(\mathcal{L})=\Ker(\mathcal{L})=\{S_\alpha \}_{\alpha\in[3]}^\prime=\bigoplus_j
\bigl(\mathcal{B}(\mathbb{C}^{n_j})\otimes \mathbb{C}\mathbb{I}_{2j+1}\bigr).
\end{equation}
In particular, denoting by $P_j$ the Hermitian projection onto the $j$-th direct-sum block in~\eqref{eq:Schur_Weyl}, every state satisfies
\begin{equation}    \lim_{t\to \infty}\mathrm{e}^{t\mathcal{L}}(\rho)
=\bigoplus_j
\Bigl(\Tr_{\mathbb{C}^{2j+1}}(P_j\rho P_j)\otimes\frac{\mathbb{I}_{2j+1}}{2j+1}\Bigr).
\end{equation}
The dynamics therefore erases coherences between different $j$-sectors and depolarizes each spin factor, while preserving the full reduced state on the noiseless subsystem associated with it. Equation~\eqref{eq:collective_attr} is precisely the structure of Theorem~\ref{asympt_QDS_thm}, with $\mathcal{H}_0^\perp=\{0\}$, $\mathcal{H}_{j,1}=\mathbb{C}^{n_j}$, $\mathcal{H}_{j,2}=\mathbb{C}^{2j+1}$ and $\rho_j=\mathbb{I}_{2j+1} / (2j+1)$.

For $n=3$, the $j=1/2$ sector $\mathbb{C}^2\otimes\mathbb{C}^2$ supports a noiseless qubit~\cite{KnillLaflammeViola2000,Kempe2001}. For $n=4$, the $j=0$ sector $\mathbb{C}^2\otimes\mathbb{C}$ is the two-dimensional singlet subspace, which is a decoherence-free subspace~\cite{zanardi1997noiseless}. This illustrates the emergence of noise-protected subsystems and subspaces, discussed  in Sections~\ref{sec:dfs} and~\ref{QEC_sec}.

Residual unitary dynamics can be obtained by adding a Hamiltonian of the form 
\begin{equation}
\label{Ham_Ising}
H=\sum_{k<\ell}J_{k\ell}\bm{\sigma}^{(k)}\cdot\bm{\sigma}^{(\ell)}
=\bigoplus_j \bigl(h_j\otimes\mathbb{I}_{2j+1}\bigr), \quad h_j = h_j^\ast \in \mathcal{B}(\mathbb{C}^{n_j}),
\end{equation}
with $J_{k\ell} \in \mathbb{R}$ and $\bm{\sigma}^{(k)} = (\sigma_1^{(k)} , \sigma_2^{(k)} , \sigma_3^{(k)})$. The block-diagonal structure follows from the fact that $H$ commutes with the collective spin operators $S_\alpha$, $\alpha \in [3]$.
The attractor subspace~\eqref{eq:collective_attr} is unchanged, while the $j$-th first factor evolves under the unitary dynamics
$x_j\mapsto\mathrm{e}^{-\mathrm{i}t h_j}x_j\mathrm{e}^{\mathrm{i}t h_j}$ generated by $h_j$ as in~\eqref{struc_as_QDS}~\cite{Bacon2000,Kempe2001}.
Such a Hamiltonian can break part of the permutation symmetry while preserving these noise-protected factors.
\end{example}

The mechanism illustrated in the previous example can be extended to the general algebraic description of symmetry-protected subsystems. Let $g\mapsto U_g$ be a unitary representation of a finite or compact group $G$ commuting with every KS operator $A_j$ of $\Phi$, or with the Hamiltonian $H$ and every jump operator $L_\alpha$ of the  GKLS generator $\mathcal{L}$ of a QDS. Each unitary $U_g$ is a \emph{strong symmetry} according to Definition~\ref{def:symmetries} below. In the isotypic decomposition~\cite{FultonHarris1991}
\begin{equation}
\label{eq:isotypic}
\mathcal{H}=\bigoplus_{\lambda\in[K]} \bigl(\mathcal{V}_\lambda\otimes\mathcal{M}_\lambda\bigr),
\qquad
U_g=\bigoplus_{\lambda\in[K]} \bigl(u_\lambda(g)\otimes\mathbb{I}_{\mathcal{M}_\lambda}\bigr),
\end{equation}
where $u_\lambda$ are the inequivalent irreducible representations of $G$, each acting on $\mathcal{V}_\lambda$, and $\mathcal{M}_\lambda$ are their multiplicity spaces, with $\lambda \in [K]$. Schur's lemma~\cite{hall2013quantum} forces the KS operators for a QMC (resp.\ the jump operators and the Hamiltonian for a QDS) to act trivially on $\mathcal{V}_\lambda$, exactly as in Example~\ref{ex:collective} with $H = 0$. Each stationary state $\varrho_\lambda$ of the induced dynamics on $\mathcal{M}_\lambda$ therefore yields a stationary sector
$\mathcal{B}(\mathcal{V}_\lambda)\otimes\mathbb{C}\varrho_\lambda$, implying the lower bounds~\cite{Zhang2020nonabelian}
\begin{equation}
\label{low_rep_bounds}
\ell_{\mathrm{P}} \geq \ell_0 \geq\sum_\lambda\dim(\mathcal{V}_\lambda)^2, \qquad \Bigl( \mbox{resp.} \; m_{\mathrm{P}} \geq m_0\geq\sum_\lambda\dim(\mathcal{V}_\lambda)^2 \Bigr) ,
\end{equation}
where we adopted the notation of Section~\ref{sec:asymptotic_bounds}. This protected part exhausts the attractor subspace when each dynamics on the multiplicity space $\mathcal{M}_\lambda$ relaxes to a unique faithful state and all coherences between inequivalent sectors decay; under these conditions,~\eqref{eq:isotypic} coincides with the asymptotic decomposition~\eqref{H_dec}. Symmetry alone does not impose these additional relaxation properties, so the equality in the bounds~\eqref{low_rep_bounds} need not be saturated.

The characterization of noiseless subsystems through the commutant of the interaction algebra, i.e.\ the algebra generated by the KS operators $A_j$ of $\Phi$ and their adjoints, is developed in~\cite{KnillLaflammeViola2000,Zanardi2000stabilizing,holbrook2003noiseless,kribs2005unified,choi2006method}; see~\cite{shabani2005theory,Karasik2008,ticozzi2008quantum} for the corresponding conditions for Markovian dynamics. The more general case in which the protected factors support unitary motion, illustrated above by the Hamiltonian $H$ in the form~\eqref{Ham_Ising}, is described by the structure of the decoherence-free algebra below in Theorem~\ref{struct_th_dfa} for the Markovian case~\cite{deschamps2016structure}.

Symmetries can also be engineered by dynamical decoupling and the quantum Zeno effect~\cite{Zanardi1999symmetrizing,ViolaKnillLloyd1999,FacchiPascazio2002,Zenoreview2008,zeno_paolosaverio,hahn2024bath}, and symmetry-based protection has been demonstrated on several platforms~\cite{Kwiat2000,Kielpinski2001,Viola2001,Mohseni2003,Monz2009,ZhuDFS_2019,xu2025long,zhangDFS2026}, including parity-protected cat qubits~\cite{Mirrahimi2014,Touzard2018}. See~\cite{Lidar2003,lidar2014review} for reviews, and~\cite{Dubois2023,Wang_scars_DFS_2024} for symmetry-induced and many-body decoherence-free dynamics.

\subsection{Symmetries, conservation laws, and relaxation}
\label{sec:Noether}
For a closed system, a Hermitian observable $Q$ is conserved if and only if $[Q,H]=0$ or, equivalently, the unitary group $\mathrm{e}^{-\mathrm{i}\theta Q}$ with $\theta \in \mathbb{R}$ leaves the Hamiltonian $H$ invariant. For open systems, covariance, conservation of expectation values, and commutation with the dynamical operators are distinct properties~\cite{albert2014symmetries,MarvianSpekkens2014,Cirstoiu2020}. In particular, this leads to the notions of \emph{weak} and \emph{strong symmetries}, which have attracted considerable attention since the seventies~\cite{Scutaru1979,Holevo1993,Holevo1996,keyl1999optimal,cattaneo2025strong,cattaneo2026symmetric,memarzadeh2022group}. In this subsection we first introduce all these concepts, then state the conditions under which we recover a Noether-type correspondence, and finally address the convergence to equilibrium in terms of conserved quantities.

\begin{definition}[Conserved quantities]
\label{def:conserved}
A Hermitian observable $Q\in\mathcal{B}(\mathcal{H})$ is a \emph{conserved quantity} of the QMC $(\Phi^n)_{n\in\mathbb{N}}$ generated by $\Phi$ if $\Tr[Q\Phi^n(\rho)]=\Tr(Q\rho)$ for every state $\rho \in \mathcal{S}(\mathcal{H})$ and every $n\in\mathbb{N}$. Similarly, for a QDS $(\mathrm{e}^{t\mathcal{L}})_{t \in \mathbb{R}^+}$ generated by $\mathcal{L}$, the corresponding condition reads $\Tr[Q\mathrm{e}^{t\mathcal{L}}(\rho)]=\Tr(Q\rho)$ for every state $\rho \in \mathcal{S}(\mathcal{H})$ and $t \in \mathbb{R}^+$. Equivalently,
\begin{equation}
\label{eq:conserved}
Q\in\Fix(\Phi^\dagger)
\qquad
\bigl(\text{resp. }Q\in\Ker(\mathcal{L}^\dagger)\bigr).
\end{equation}
\end{definition}

\begin{definition}[Weak and strong symmetries~\cite{BucaProsen2012,albert2014symmetries,baumgartner2008analysis2,sala2024spontaneous}]
\label{def:symmetries}
For a unitary $U\in\mathcal{B}(\mathcal{H})$, let $\Phi_U(X):=UXU^\ast$ with $X \in \mathcal{B}(\mathcal{H})$ be the corresponding unitary channel. Then, $U$ is
\begin{enumerate}
\item[\textnormal{i.}] a \emph{weak symmetry},  if
\begin{equation}
\label{cov_cond}
\Phi\circ\Phi_U=\Phi_U\circ\Phi,
\qquad
\bigl(\text{resp. }\mathcal{L}\circ\Phi_U=\Phi_U\circ\mathcal{L}\bigr).
\end{equation}
In this case, we will say that $\Phi$ (resp. $\mathcal{L}$) is \emph{weakly symmetric} or \emph{covariant} with respect to $U$. 
\item[\textnormal{ii.}] a \emph{strong symmetry}, if it commutes with all KS operators $A_j$, $j \in [N]$, in the representation~\eqref{op_sum_rep} of $\Phi$, or with the Hamiltonian $H$ and all jump operators $L_\alpha$, $\alpha \in [N]$, in the GKLS form~\eqref{diag_GKLS} of $\mathcal{L}$:
\begin{equation}
\label{strong_sym}
    [U,A_j]=0,\quad j\in[N],
\qquad
\bigl(\text{resp. }[U,H]=[U,L_\alpha]=0,\quad \alpha\in[N] \bigr).
\end{equation}
\end{enumerate}
\end{definition}
Notice that Definition~\ref{def:symmetries} covers both discrete and continuous symmetries. The latter are described by one-parameter groups $U_\theta = e^{-\mathrm{i} \theta Q}$, with $\theta \in \mathbb{R}$, 
generated by some Hermitian operator $Q = Q^\ast$.
Importantly, a strong symmetry is also a weak one, but the converse is not true, as illustrated by the examples below.
In fact, by Proposition~\ref{freed_Kraus_prop}, covariance of $\Phi$ may be recast in terms of its KS representation~\eqref{op_sum_rep} as
\begin{equation}\label{eq:unitary_invariance_combination}
UA_jU^\ast=\sum_{k\in[N]}v_{jk}A_k, \quad j \in [N],
\end{equation}
for some unitary matrix $(v_{jk})$, and thus it allows linear combinations of the KS operators. To clarify the consequences of weak symmetries, consider the unitary $U=\sum_{k\in[N]}\omega_k P_k$ with distinct unimodular eigenvalues $\omega_k$ and corresponding spectral projections $P_k$, $k \in [N]$. A channel $\Phi$ that is weakly symmetric  with respect to $U$ preserves the eigenspaces of $\Phi_U$, namely
\begin{equation}
    X \in \Ker(\Phi_U - \lambda \mathsf{1}) \Rightarrow \Phi(X) \in \Ker(\Phi_U - \lambda \mathsf{1}), \qquad  \lambda \in \mathbb{C},
\end{equation}
where each eigenspace
\begin{equation}
    X \in \Ker(\Phi_U - \lambda \mathsf{1}) \Leftrightarrow X = 
    \sum_{k, \ell} \delta_{\omega_k\overline{\omega_\ell}, \lambda} P_k X P_\ell
\end{equation}
collects operator blocks $P_k\mathcal{B}(\mathcal{H})P_\ell$ having the same eigenvalue $\lambda = \omega_k\overline{\omega_\ell}$. In particular, since $|\omega_k| = 1$ for any $k \in [N]$, each diagonal block $P_k \mathcal{B}(\mathcal{H}) P_k$ belongs to the eigenspace corresponding to $\lambda = 1$; however, it is \emph{not} left invariant in general. Moreover,  covariance alone generally guarantees neither steady-state degeneracy nor conservation of its generator.  

Let us illustrate these general observations about weak symmetries with some examples. 
\begin{itemize}
    \item First, consider the depolarizing channel~\eqref{eq:dep_channel} with $\lambda=1$, which is weakly symmetric with respect to the group $U_\theta = e^{-\mathrm{i} \theta \sigma_3}$, $\theta \in \mathbb{R}$, generated by the third Pauli matrix $\sigma_3$. Its attractor subspace is one dimensional, $\Attr(\Phi_{\mathrm{dep}}) = \mathbb{C} \mathbb{I}$ and  $\Phi_{\mathrm{dep}}^{\dagger}(\sigma_3)=0$. Then, with the spectral projections of $\sigma_3$, one has
\begin{equation}
    P_{\pm} := \frac{\mathbb{I} \pm \sigma_3}{2},\qquad  \Phi_{\mathrm{dep}}^\dagger(P_{\pm}) = \frac{\mathbb{I}}{2}.
\end{equation}
Hence, neither the generator nor its spectral projections are conserved.  
\item Similarly, the qubit amplitude damping generator~\eqref{Markov_ampl_qubit}, with $\gamma=1$ and jump operator $\sigma_+ = \ketbra{1}{2}$, where $\{ \ket{j} \}_{j=1}^2$ is the eigenbasis of $\sigma_3$, has the unique stationary state $P_+ = \ketbra{1}{1}$, i.e., $\Attr(\mathcal{L}) = \mathbb{C} P_+$, and is again covariant with respect to the group $U_\theta=\mathrm{e}^{-\mathrm{i}\theta\sigma_3}$, $\theta \in \mathbb{R}$, although
\begin{equation}
\mathcal{L}^\dagger(\sigma_3)=\mathbb{I}-\sigma_3\neq0 \Rightarrow \mathcal{L}^\dagger(P_{\pm}) = \pm\, P_-.
\end{equation}
\item 
Finally, let us take into account the unital channel $\Phi$ defined by~\eqref{perm_channel_ex}, which is covariant with respect to the parity symmetry $U = \sigma_3$, but the fixed-point subspace reduces to $\Fix(\Phi^\dagger) = \Fix(\Phi) = \mathbb{C} \mathbb{I}$ by~\eqref{asympt_ex-4} and
\begin{equation}
    \Phi^\dagger(\sigma_3) = -\sigma_3 \Rightarrow 
    \Phi^\dagger(P_{\pm}) = P_{\mp}.  
\end{equation}
Instead, a quantum channel $\Phi$ (resp.\ GKLS generator $\mathcal{L}$) having a strong symmetry $U$ preserves each block $P_k\mathcal{B}(\mathcal{H})P_\ell$ separately, and conserves each spectral projection $P_k$ of $U$, that is
\begin{equation}
\begin{split}
\Phi^\dag(P_k) &= P_k \quad \hspace{34.5pt}  (\mbox{resp. } \mathcal{L}^\dag(P_k) = 0), \; \mbox{ for all } k \in [N], \\
\Phi(P_k X P_\ell) &= P_k \Phi(X) P_\ell  \quad (\mbox{resp. } \mathcal{L}(P_k X P_\ell) = P_k \mathcal{L}(X) P_\ell), \; \mbox{ for all } k,\ell \in [N]. 
\end{split}
\end{equation}
\end{itemize}
Therefore, $\Phi \vert_{P_k \mathcal{B}(\mathcal{H}) P_k}$ (resp.\ $\mathcal{L} \vert_{P_k \mathcal{B}(\mathcal{H}) P_k}$) is a quantum channel (resp.\ GKLS generator) for all $k \in [N]$, and so each diagonal block $P_k \mathcal{B}(\mathcal{H}) P_k$ supports a stationary state, implying that
\begin{equation}
\label{low_bound_strong_sym}
    \dim(\Fix(\Phi))\geq N
\qquad
\bigl(\text{resp. }\dim(\Ker(\mathcal{L}))\geq N\bigr),
\end{equation}
because these states have orthogonal supports; see~\cite[Theorem~A.1]{BucaProsen2012} for QDSs. The lower bounds~\eqref{low_bound_strong_sym} are nontrivial when $N \geqslant 2$, namely $U \neq e^{\mathrm{i} \varphi} \mathbb{I}$ for all $\varphi \in \mathbb{R}$. The qubit pinching channel~\eqref{dep_channel}, with $P_k = \ketbra{k}{k}$, $k \in [2]$, is an example exhibiting strong symmetries, in fact any unitary $U = \sum_{k=1}^2 \omega_k P_k$, with $|\omega_k| = 1$, $k \in [2]$, fulfills condition~\eqref{strong_sym}. 

We shall see in Theorem~\ref{thm:Noether} that if $\mathrm{e}^{-\mathrm{i}\theta Q}$ is a strong symmetry for every $\theta\in\mathbb{R}$, then $Q$ is conserved. The converse implication can fail: consider the following GKLS generator on $\mathcal{B}(\mathbb{C}^3)$  
\begin{equation}
\label{qutrit_ampl_damp}
    \mathcal{L}(X) = \sum_{k = 1}^2 \Bigl( L_k X L_k^\ast - \frac{1}{2} \{ L_k^\ast L_k , X \} \Bigr), \quad L_k =\sqrt{\gamma_k}\ketbra{k}{3}, \;\gamma_k>0, \; k\in[2],
\end{equation}
and let $\Gamma=\gamma_1+\gamma_2$ be the total decay rate of $\ket{3}$ towards $\ket{1}$ and $\ket{2}$. The jump operators and their adjoints generate $\mathcal{B}(\mathbb{C}^3)$ as a $C^\ast$-algebra, so the only strong symmetries are scalar unitaries in~\eqref{strong_sym}. Nevertheless, the Hermitian operators
\begin{equation}
\label{eq:branching_conserved}
J_k=\ketbra{k}{k}+\frac{\gamma_k}{\Gamma}\ketbra{3}{3},
\qquad
\mathcal{L}^\dagger(J_k)=0, \qquad k \in [2]
\end{equation}
are conserved, and they generate the groups $U_\theta^{(k)} = e^{-\mathrm{i} \theta J_k}$, where $\theta \in \mathbb{R}$ and $k \in [2]$, with respect to which $\mathcal{L}$ is covariant, namely~\eqref{cov_cond} holds.  
However, their squares are not conserved:
\begin{equation}
\mathcal{L}^\dagger(J_k^2)
=\frac{\gamma_1\gamma_2}{\Gamma}\ketbra{3}{3}\neq 0, \quad k \in [2].
\end{equation}

In fact, a conserved quantity is not even guaranteed to generate a weak symmetry group of the evolution: take the qutrit idempotent channel 
\begin{equation}
  X = \begin{pmatrix}
        x_{11} & x_{12} & x_{13} \\
        x_{21} & x_{22} & x_{23} \\
        x_{31} & x_{32} & x_{33}
    \end{pmatrix} \mapsto  \mathcal{P}(X) = \begin{pmatrix}
        x_{11} + x_{33} & x_{12} & 0 \\
        x_{21} & x_{22} & 0 \\
        0 & 0 & 0
    \end{pmatrix}, 
\end{equation}
whose conserved quantities read 
\begin{equation}
    Q = Q^\ast \in \Fix(\mathcal{P}^\dagger) \Leftrightarrow Q = \begin{pmatrix}
        Q_0 & 0 \\
        0 & q_{11}
    \end{pmatrix}, \; Q_0^\ast = Q_0 = (q_{ij})_{i,j=1}^2  \in \mathcal{B}(\mathbb{C}^2), \; q_{11} \in \mathbb{R}.
\end{equation}
It is easy to check that $\mathcal{P}$ is not generally weakly symmetric with respect to the group $U_\theta = e^{-i\theta Q}$, $\theta \in \mathbb{R}$, when $q_{12} \neq 0$ and, as in example~\eqref{qutrit_ampl_damp}, $\mathcal{P}^\dagger(Q) = Q$ does not imply that $\mathcal{P}^\dagger(Q^2) = Q^2$ in general.

This distinction between conservation of an observable and that of its square leads to the appropriate quantum Noether-type result, as in classical Markov chains~\cite{BaezFong2013}.
See also~\cite{styliaris2020symmetries} for a different approach to the extension of Noether's theorem in the open-system case.

\begin{theorem}[Noether's theorem for open systems~\cite{arias2002fixed,Gough2015Noether,Gheondea2018}]
\label{thm:Noether}
Let $\Phi$ be a quantum channel with KS representation~\eqref{op_sum_rep} and let $S = \{A_i,A_i^\ast\}_{i\in[N]}$. Similarly, let $\mathcal{L}$ be a GKLS generator with representation~\eqref{diag_GKLS} and let 
$T=\{H\}\cup\{L_\alpha,L_\alpha^\ast\}_{\alpha\in[N]}$.
Then, if $X\in\mathcal{B}(\mathcal{H})$, we have 
\begin{equation}
    X,X^\ast X,XX^\ast\in\Fix(\Phi^\dagger)
\quad\Longleftrightarrow\quad
X\in S^\prime,
\end{equation}
and
\begin{equation}
    X,X^\ast X,XX^\ast\in\Ker(\mathcal{L}^\dagger)
\quad\Longleftrightarrow\quad X\in T^\prime.
\end{equation}
In particular, a unitary $U$ is conserved for $\Phi$ (resp. $\mathcal{L}$), namely $\Phi^\dagger(U)=U$ (resp.\ $\mathcal{L}^\dagger(U)=0$) if and only if it is a strong symmetry, namely $U \in S^\prime$ (resp. $U \in T^\prime$). For Hermitian $Q$, the unitaries $\mathrm{e}^{-\mathrm{i}\theta Q}$ are strong symmetries for all $\theta \in \mathbb{R}$ if and only if both $Q$ and $Q^2$ are conserved.
\end{theorem}
\begin{proof}
To treat the discrete- and continuous-time cases simultaneously, set $\mathcal{D}=\Phi^\dagger-\mathsf{1}$ and $K_j=A_j$, the KS operators of a quantum channel $\Phi$, or $\mathcal{D}=\mathcal{L}^\dagger$ and $K_j=L_j$ the jump operators of a GKLS generator $\mathcal{L}$, with $j \in [N]$. Equations~\eqref{op_sum_rep}--\eqref{norm_con} and~\eqref{GKLS_adjoint} give
\begin{equation}
\label{eq:dissip}
\sum_{j\in[N]}[K_j,X]^\ast[K_j,X]
=\mathcal{D}(X^\ast X)-X^\ast\mathcal{D}(X)-\mathcal{D}(X^\ast)X,
\end{equation}
whose right-hand side is known, in the GKLS case, as the
Lindblad dissipation function~\cite{Lindblad76}. Suppose first that $X$, $X^\ast X$, and $XX^\ast$ are conserved,
namely they belong to $\Ker(\mathcal{D})$.
Since $\mathcal{D}$ is Hermiticity-preserving,
we also have $\mathcal{D}(X^\ast)=0$. This implies  
\begin{equation}
   [K_j,X]=[K_j^\ast,X]=0, \quad j \in [N], 
\end{equation}
since the left-hand side of~\eqref{eq:dissip} is the sum of positive operators.
In the QDS case, the dissipative part of
$\mathcal{L}^\dagger(X)$ therefore vanishes, and the
conservation condition further yields
$\mathcal{L}^\dagger(X)=\mathrm{i}[H,X]=0$.
Consequently, $X$ belongs to $S^\prime$ or $T^\prime$,
respectively. 

Conversely, if $X$ belongs to the corresponding commutant,
then so do $X^\ast X$ and $XX^\ast$. Their conservation
follows directly from the operator-sum
representation~\eqref{op_sum_rep} or the GKLS
representation~\eqref{diag_GKLS}, proving the first two
equivalences.

Moreover, if $U$ is unitary, then $U^\ast U=UU^\ast=\mathbb{I}$,
which is always conserved. The preceding equivalences
thus show that conservation of $U$ alone is necessary
and sufficient for $U$ to be a strong symmetry. 

Finally, let $Q=Q^\ast$. Since
$Q^\ast Q=QQ^\ast=Q^2$, the first part of the theorem
implies that conservation of both $Q$ and $Q^2$ is
equivalent to the commutation of $Q$ with all the
operators defining the dynamics. In turn, this implies
that $U_\theta=\mathrm{e}^{-\mathrm{i}\theta Q}$ is a
strong symmetry for every $\theta\in\mathbb{R}$.
Conversely, if every $U_\theta$ is a strong symmetry,
differentiating the corresponding commutation relations
at $\theta=0$, and using
\begin{equation}
\left.\frac{\mathrm{d}}{\mathrm{d}\theta}
\mathrm{e}^{-\mathrm{i}\theta Q}\right|_{\theta=0}
=-\mathrm{i}Q,
\end{equation}
shows that $Q$ commutes with all the operators defining
the dynamics. The first part of the theorem then
guarantees the conservation of both $Q$ and $Q^2$.
\end{proof}

The last part of Theorem~\ref{thm:Noether} asserts that a Hermitian operator $Q$ generates a strong symmetry group if and only if its expectation value $\mathrm{E}_\rho(Q) := \Tr(\rho Q)$ and its variance 
\begin{equation}
   \mathrm{Var}_{\rho}(Q) := \Tr(\rho Q^2) - \Tr(\rho Q)^2
\end{equation}
over any state $\rho \in \mathcal{S}(\mathcal{H})$ are constant at all times, namely
\begin{equation}
    \mathrm{E}_{\Phi^n(\rho)}(Q) =  \mathrm{E}_\rho(Q), \; \mathrm{Var}_{\Phi^n(\rho)}(Q) = \mathrm{Var}_{\rho}(Q),   \; \mbox{for all } n \geq 1, \, \rho \in \mathcal{S}(\mathcal{H}),
\end{equation}
and similarly in the QDS case. 

Recall that, under the condition of faithfulness,  Propositions~\ref{th:fixed_kraus} and~\ref{fixed_L} imply that
\begin{equation}
\Fix(\Phi^\dagger)=S^\prime,
\qquad
\Ker(\mathcal{L}^\dagger)=T^\prime,
\end{equation}
where we used the notation of Theorem~\ref{thm:Noether}.
Therefore, the space of conserved quantities is a $C^\ast$-algebra spanned by strong symmetries, and conservation of a Hermitian observable is now equivalent to the fact that it generates a strong (and, consequently, weak) symmetry group of the evolution. Nevertheless, even in the faithful case, we do not obtain the equivalence between weak and strong symmetries, as shown above with the channels~\eqref{eq:dep_channel} with $\lambda = 1$ and~\eqref{perm_channel_ex}. 

Whether or not the evolution is faithful, conserved quantities determine the stationary components selected by the initial state~\cite{albert2014symmetries,albert2016geometry,albert2019asymptotics,jex_st_2012}. Choose a basis $\{\rho_k\}$ of stationary states, as in Proposition~\ref{basis_states_prop}, and the dual conserved observables $\{J_k\}$ satisfying $\Tr(J_k\rho_\ell)=\delta_{k\ell}$. Then, the steady projection, defined by~\eqref{P_for} and~\eqref{P_for_sem} for a quantum channel $\Phi$ and a QDS $\Sem$, respectively, can be written as 
\begin{equation}
    \mathcal{P}(X)=\sum_{k} \Tr(J_kX)\rho_k, \quad X \in \mathcal{B}(\mathcal{H}).
\end{equation}
For example, in the case of the QDS generated by~\eqref{qutrit_ampl_damp}, $\mathcal{L}(\ketbra{k}{k}) = 0$ and $\Tr(J_k \ketbra{k}{k}) = 1$, thus $\Tr(J_k \rho)$ gives the asymptotic population of $\ket{k}$, including the contribution from decay of $\ket{3}$, with $k \in [2]$.

Note that the steady projection $\mathcal{P}$ is obtained as a Ces\`aro limit of the evolution; cf.\ Equations~\eqref{P_for} and~\eqref{P_for_sem}. Instead, as mentioned in Section~\ref{sec:asymptotics}, a long-time limit of the dynamics only exists provided that no additional peripheral modes survive, namely $\spec_{\mathrm{P}}(\Phi) = \{ 1 \}$ (resp. $\spec_{\mathrm{P}}(\mathcal{L}) = \{ 0 \}$). See Proposition~\ref{prop:Noether_convergence} below for an algebraic characterization of this condition under faithfulness.

\section{Asymptotic information transmission}
\label{asympt_transm_sec}
Quantum channels describe how an open quantum system is affected by the environment, and, in particular, the information it encodes. Therefore, it is of paramount importance to introduce notions describing the maximum amount of classical or quantum information which is faithfully transmitted by a noisy channel. The problem has been addressed since the sixties in many works~\cite{Gordon1962,lebedev1963maximum,gordon1964noise,holevo1972mathematical,holevo1973some}. 

This section is devoted to the analysis of information transmission in the long-time limit, rather than in  the more commonly studied many-copy limit. For the latter we refer the reader, for example, to~\cite{caruso_2014_review,Holevo2012,watrous2018theory}. Specifically, in the next subsection, we will discuss entanglement breaking channels, which destroy entanglement between the transmitted system and any reference system, together with several related notions. Then, in Section~\ref{asympt_cap_subsec}, we will introduce the asymptotic quantum and classical capacities of quantum channels, and relate them to the asymptotic structure described in Theorem~\ref{Wolf_struc}.
\subsection{Entanglement breaking channels}
\label{EB_subsec}
Entanglement is a fundamental resource in many quantum information and communication protocols. Understanding how it is affected by quantum dynamics  is therefore of key importance. In this context, Horodecki \emph{et al} introduced the class of \emph{entanglement breaking quantum channels}~\cite{horodecki2003entanglement,ruskai2003qubit}; see also~\cite{holevo1998quantum}. A quantum channel $\Phi$ is called entanglement breaking if $(\Phi\otimes \mathsf{1})(\rho)$ is a separable state for every state $\rho \in \mathcal{S}(\mathcal{H} \otimes \mathcal{H})$, where $\mathsf{1}$ denotes the identity channel. Thus, a single application of $\Phi$ to one subsystem destroys all entanglement between two subsystems of equal dimension.
Several properties of entanglement breaking channels were established
in~\cite{horodecki2003entanglement}. However, their role in quantum dynamics has continued to attract considerable interest over the last twenty years~\cite{lami2015entanglement,lami2016entanglement,rahaman2018eventually,christandl2019composed,hanson2020eventually}.  

A first necessary condition for a channel to be entanglement breaking is that its steady subspace has the form~\cite[Theorem 3]{fukuda2015quantum}
\begin{equation}
\label{comm_fix_cond}
\Fix(\Phi) = 0 \oplus 
\bigoplus_{k\in[M]} \mathbb{C} \rho_k = 0 \oplus \Fix(\Phi_{\mathrm{faith}}),
\end{equation}
where $\Fix(\Phi_{\mathrm{faith}})$ is the steady subspace of the induced faithful channel $\Phi_{\mathrm{faith}}$ defined by~\eqref{faith_ind_def}.
Consequently, the fixed-point subspace $\Fix(\Phi_{\mathrm{faith}}^\dagger)$ of $\Phi_{\mathrm{faith}}^\dagger$ is a commutative $C^\ast$-subalgebra of $\mathcal{B}(\mathcal{H}_{0})$ according to~\eqref{Schr_Heis_corresp}, see also~\cite[Theorem 9.3]{gartner2012coherent}. 

Note that~\eqref{comm_fix_cond} can also be obtained by applying Theorem~\ref{Wolf_struc} to the fixed-point projection $\mathcal P$ of $\Phi$. 

Observe that, if $\Phi$ is entanglement breaking, then so is $\Phi^n$, for any $n \geq 2$. The converse does not hold in general: a channel  that is not entanglement breaking, may become so after finitely many iterations. 
A quantum channel $\Phi$ is called \emph{eventually entanglement breaking} if $\Phi^{n_0}$ is entanglement breaking for some integer $n_0\geq 1$~\cite{rahaman2018eventually,hanson2020eventually,bhat2026complete}

A quantum channel $\Phi$ is called \emph{asymptotically entanglement breaking} if its peripheral projection $\mathcal{P}_{\mathrm{P}}$, defined in~\eqref{per_proj_def}, is entanglement breaking~\cite{lami2016entanglement}. This class of channels admits a characterization in terms of the attractor subspace: $\Phi$ is asymptotically entanglement breaking if and only if the elements of  $\Attr(\Phi)$ commute pairwise~\cite[Theorem 24]{lami2016entanglement}. 
Equivalently, the asymptotic subspace $\Attr(\Phi^\dagger)$, which is related to $\Attr(\Phi)$ by~\eqref{Schr_Heis_corresp} under the faithfulness assumption, is a commutative $C^\ast$-algebra with respect to the Choi--Effros product~\eqref{CE_def}. 
By Theorem~\ref{Wolf_struc}, this condition is equivalent to 
\begin{equation}
\label{comm-attr_cond}
\Attr(\Phi) = 0 \oplus 
\bigoplus_{k\in[M]} \mathbb{C} \rho_k,
\end{equation}
i.e., $\mathrm{dim}(\mathcal H_{k,1}) = 1$ for all $k\in[M]$. 

Interestingly, condition~\eqref{comm-attr_cond} expresses the absence of nontrivial decoherence-free subspaces, and also characterizes the phenomenon of Zeno Hamiltonian suppression studied in~\cite{hahn2024bath}. Its connection with the eventually quantum scrambling property will be discussed in the next subsection.  

\subsection{Asymptotic capacity of a quantum channel}
\label{asympt_cap_subsec}

A central question in the study of open quantum systems is what information survives a noisy evolution. When a system is repeatedly subjected to the same noise, as in a quantum Markov chain $\Sem$ generated by a channel $\Phi$, its long-time behaviour is governed by the attractor subspace $\Attr(\Phi)$, characterized in Theorem~\ref{Wolf_struc}. It is therefore natural to ask how much classical or quantum information can be
stored in, and perfectly retrieved from, the system in the asymptotic regime. 

We start by analyzing classical capacity, which quantifies the amount of classical information that can be reliably transmitted through a quantum channel. To encode this information, each symbol $x$, drawn from a classical alphabet $X$, is associated with a quantum state via the encoding map
\begin{equation}
   x\in X \longmapsto \rho_x \in \mathcal S(\mathcal H).
\end{equation}
This information can be recovered without error if and only if the states $\rho_x$ have mutually orthogonal supports~\cite{Holevo2012}. The interaction with the environment, described by the channel $\Phi$, may destroy this orthogonality. We therefore define the \emph{zero-error classical capacity} $C_0(\Phi)$ as the maximum number of bits that can still be perfectly distinguished after a single use of the channel, namely
\begin{align}\label{eq:classical_capacity}
  C_0(\Phi) &:=  \log_2 \bigl(\max (C)\bigr),  \nonumber \\
  C := \{ n\geq 1 \,:\, \exists\, \rho_1, \dots , \rho_n \in &\,\mathcal{S}(\mathcal{H}) \mbox{ with } \Phi(\rho_i)\perp\Phi(\rho_j) \;\forall i \neq j \}. 
\end{align}
Here, $\Phi(\rho_i)\perp\Phi(\rho_j)$ means that the output states have orthogonal supports, $\Tr(\Phi(\rho_i)\Phi(\rho_j))=0$, which in turn implies $\rho_i\perp\rho_j$. Moreover, the input states $\rho_x$ may be chosen to be pure~\cite{Duan2013}.

Similarly, we can quantify how much quantum information a channel can perfectly preserve. Given a quantum channel $\Phi$, its \emph{zero-error quantum capacity} is defined as 
\begin{align}\label{eq:quantum_capacity}
    Q_0(\Phi)&:=\log_2 \bigl(\max (Q)\bigr), \nonumber\\
    Q:=\{\dim (\mathcal{X})\,:\,\mathcal{X}\subseteq\mathcal{H} &\mbox{ is a correctable code for } \Phi\}.
\end{align}
Thus, $Q_0(\Phi)$ quantifies the maximum number of qubits that can be encoded in a subspace and perfectly recovered after a single use of the channel; cf.~Section~\ref{QEC_sec}. 

Capacities of quantum channels are traditionally defined in terms of \emph{parallel} uses $\Phi^{\otimes n}$, which model the transmission of many independent systems through the same noisy channel~\cite{Holevo2012,watrous2018theory}. By contrast, the capacities~\eqref{eq:classical_capacity} and~\eqref{eq:quantum_capacity} involve a single use of the channel and are often called one-shot capacities. Here, we shall investigate the \emph{sequential} scenario, in which a single system undergoes the same noisy evolution $\Phi$ a large number of times, namely $\Phi^n$ in the limit $n\to\infty$~\cite{singh2024zero}.

As shown in the following theorem~\cite[Theorem 1.5]{singh2024zero}, these asymptotic classical and quantum capacities are determined by the structure of the attractor subspace~\eqref{attr_struc} described in Theorem~\ref{Wolf_struc}.

\begin{theorem}\label{th:datta}
Let $\Phi$ be a quantum channel whose attractor subspace $\Attr(\Phi)$ has the structure~\eqref{attr_struc}. Then, the zero-error classical and quantum capacities of the $n$-fold composition $\Phi^n$ converge to
\begin{align}\label{eq:dattabounds}
    \lim_{n\rightarrow \infty}C_0(\Phi^n)=  \log_2
    \biggl(\sum_{k\in[M]} d_k \biggr),\quad
    \lim_{n\rightarrow \infty} Q_0(\Phi^n)= \log_2 \biggl(\max_{k\in[M]} d_k \biggr),
\end{align}
where $d_k = \dim (\mathcal H_{k,1})$ with $k \in [M]$.
\end{theorem}

A quantum channel $\Phi$ is called \emph{eventually classically scrambling} or \emph{eventually quantum scrambling} if there is an integer $n\geq 1$ such that $C_0(\Phi^n)=0$ or $Q_0(\Phi^n)=0$, respectively. Theorem~\ref{th:datta}, together with~\cite[Theorem 1.4]{singh2024zero}, implies that a channel $\Phi$ is eventually classically scrambling if and only if $M=1$ and $d_1=1$ in~\eqref{attr_struc}, or equivalently, if and only if $\Phi$ is mixing; see Section~\ref{qc_bounds_subsec}. Similarly, $\Phi$ is eventually quantum scrambling if and only if $d_k=1$ for all $k\in[M]$ in~\eqref{attr_struc}, namely $\Phi$ is asymptotically entanglement breaking; see Section~\ref{EB_subsec}. Therefore, any eventually classically scrambling channel $\Phi$ is also eventually quantum scrambling, but the converse is not true, take e.g.\ any ergodic but non-mixing channel such as the one in Example~\ref{ex_5}.

Theorem~\ref{th:datta} concerns zero-error capacities, but one can also allow a nonzero error in recovering the information. At fixed error tolerance, Fawzi, Rahaman, and Taheri derive an exact formula for the infinite-time classical capacity, together with upper and lower bounds for the corresponding quantum capacity~\cite[Theorem 1]{fawzi2024capacities}. 

The convergence rate towards the asymptotic capacities has also been studied quantitatively. In the zero-error setting, both capacities stabilize after at most $d^2$ channel iterations~\cite[Theorem 1.4]{singh2024zero}. For approximate capacities, Singh and Datta establish  convergence after $O(d^2\log d)$ iterations under suitable assumptions~\cite{Singh2025}. Moreover, explicit convergence bounds have been also obtained in~\cite{fawzi_fast_2026} for channels satisfying a quantum detailed balance condition~\cite{gao2022complete}.

\section{Asymptotic dynamics: an algebraic approach}
\label{sec:alg_approach}
 In this section we  analyze the long-time behaviour of an open quantum system from an algebraic perspective as an alternative to the spectral approach addressed in Section~\ref{sec:asymptotics} for QMCs and Section~\ref{asympt_QDS_sec} for QDSs. See~\cite{AFK_asympt5} for a recent review on the topic. 
 
 In particular, we will work in the Heisenberg picture, which is the natural framework for studying multiplicative properties of the evolution, as it acts on the space of observables, which is usually assumed to be an algebra~\cite{araki1999mathematical}.
The motivation for seeking an algebraic structure in the asymptotic regime stems from the expectation--discussed in Section~\ref{unit_asympt_sec}--that the dynamics of the system becomes effectively unitary at long times and therefore, from an algebraic viewpoint, a homomorphism (see Appendix~\ref{intro_Cast}).  
 
 In Section~\ref{sec:dfa} we introduce the concept of a decoherence-free algebra both for QMCs and QDSs, providing an algebraic counterpart to the attractor subspace introduced in Sections~\ref{spec-qc-subsec} and~\ref{spec_lindbladian}. We then explore, in Section~\ref{alg_sec}, how the attractor subspace itself can be endowed with an algebraic structure.

\subsection{Decoherence-free algebras}\label{sec:dfa}
We now discuss the notion of \emph{\gls*{dfa}} as a tool for studying the asymptotic behaviour of a quantum system. This algebraic approach extends naturally  to infinite-dimensional systems~\cite{Frigerio_Verri_82,fagnola2019role,fidaleo2022spectral}, encompassing both  atomic~\cite{fagnola2017,fagnola2019role} and non-atomic~\cite{Sasso2023,AFK_asympt5} von Neumann algebras. In particular, various efforts have been devoted to the asymptotic properties of Gaussian quantum Markov semigroups~\cite{girotti2025gaussian,agredo2022decoherence,agredo2021gaussian}, generated by formal GKLS generators with a quadratic Hamiltonian and linear jump operators in terms of bosonic creation and annihilation operators.  
Nevertheless, several attempts have also been made to describe long-time evolutions in the infinite-dimensional case using spectral approaches~\cite{rajarama2022peripheral,rajaramaperipheral_2,bhat2025peripheral}. 

Historically, the notion of DFA, which already appeared in~\cite{evans1977_irred}, was later developed within the mathematical community by Blanchard and Olkiewicz~\cite{blanchard2003decoherence}, see also~\cite{bulinskii1995some}. 
This notion is closely related to that of a DFS discussed in Section~\ref{sec:dfs}, but the focus is now on an algebra of operators rather than a vector space.

In words, in both dynamical settings, the DFA is the largest subalgebra of operators on which the dynamics preserves operator products at all times, i.e., the evolution map acts as a $*$-homomorphism. 
Specifically, let $\Sem^\dagger$ be an adjoint  QMC  generated by a UCP map $\Phi^\dagger$. Its DFA is defined by~\cite{carbone2020period}
\begin{equation}
\label{eq:decoherence_free_algebra_UCP}
  \begin{split}
    \mathcal{N}(\Phi^\dagger) :=\{X\in\mathcal B(\mathcal H)\,:\, \ &(\Phi^{\dagger})^n (X^*X)=(\Phi^{\dagger})^n(X)^*(\Phi^{\dagger})^n(X) ,\\
	&(\Phi^{\dagger})^n(XX^*)=(\Phi^{\dagger})^n(X)(\Phi^{\dagger})^n(X)^*,\forall n \geq 1 \}.  
    \end{split}
\end{equation}

Consistently with our notation for the steady and attractor subspaces, we will often refer to the DFA of $\Sem^\dagger$ simply as the DFA of $\Phi^\dagger$.

Remarkably, we can express the DFA $\mathcal{N}(\Phi^\dagger)$ of a UCP map $\Phi^\dagger$ in terms of its KS operators $\{A_i\}_{i \in [N]}$ appearing in the representation~\eqref{op_sum_rep}. According to~\cite[Proposition 3]{carbone2020period}, one has
\begin{equation}
\label{char_N_Kraus}
    \mathcal{N}(\Phi^\dagger) = \{ A_{k_1}\dots A_{k_n} A_{\ell_1}^\ast \dots A_{\ell_n}^\ast ,  \; k_j , \ell_j \in[N] ,\; j \in [n] ,\; n \in \mathbb{N}  \}^\prime,
\end{equation}
with the prime denoting the commutant.
Similarly, let  $\Sem^\dagger =(\Phi_t^\dagger)_{t\in\mathbb R^+}$ be  an adjoint  QDS, see Section~\ref{QDS_def}. Its DFA is defined as~\cite{evans1977_irred}
\begin{align}\label{eq:decoherence_free_algebra}
	\mathcal N (\Sem^\dagger) :=\{X\in\mathcal B(\mathcal H)\,:\, \ &\Phi_t^{\dagger}(X^*X)=\Phi_t^{\dagger}(X)^*\Phi_t^{\dagger}(X),
    \nonumber\\
	&\Phi_t^{\dagger}(XX^*)=\Phi_t^{\dagger}(X)\Phi_t^{\dagger}(X)^*,\forall t\in\mathbb R^+ \}.
\end{align}
Interestingly, the DFA  admits an algebraic characterization in terms of the Hamiltonian $H$ and the jump operators $\{L_\alpha\}_{\alpha\in[N]}$ appearing in the GKLS representation~\eqref{GKLS_adjoint} of its generator $\mathcal L^\dagger$. Specifically, one has~\cite[Theorem 2.1]{fagnola2008algebraic}
\begin{equation}
\label{char_N_GKLS}
    \mathcal N(\Sem^\dagger) = \{\delta_H^{(n)}(L_\alpha),\;\delta_H^{(n)}(L_\alpha^*),\; \alpha\in [N], \; n\in\mathbb N\}',
\end{equation}
where the prime denotes the commutant and the iterated commutators are defined recursively, for $X \in \mathcal{B}(\mathcal{H})$, by
\begin{align}
	\delta_H^{(0)}(X) := X,\qquad \delta_H^{(n)}(X) :=[H,\delta_H^{(n-1)}(X)],\quad n \geq 1.
\end{align}

The characterizations~\eqref{char_N_Kraus} and~\eqref{char_N_GKLS} show that the DFAs $\mathcal{N}(\Phi^\dagger)$ and $\mathcal{N}(\Sem^\dagger)$ are $C^\ast$-subalgebras (von Neumann subalgebras) of $\mathcal{B}(\mathcal{H})$. In finite dimension they therefore admit a direct-sum structure as described in Theorem~\ref{struc_theorem_finite_dimensional}.
We  have thus obtained  a structure for the decoherence-free algebra analogous to that of the attractor subspace of a faithful UCP map, given in~\eqref{attr_phi00_struc}, or of a faithful adjoint QDS.

Importantly, in the continuous-time case, the direct-sum decomposition of the DFA $\mathcal{N}(\Sem^\dagger)$ induces a corresponding decomposition of the jump operators and the Hamiltonian appearing in the GKLS form~\eqref{GKLS_adjoint} of $\mathcal{L}^\dagger$. This, in turn, yields an explicit expression for the dynamics on $\mathcal{N}(\Sem^\dagger)$, i.e., the asymptotic dynamics in the algebraic approach. 
Specifically, the following result holds~\cite[Theorem 3.2]{deschamps2016structure}; see~\cite[Proposition 8]{carbone2020period} for an analogous result for UCP maps.   
\begin{theorem}[Structure theorem for decoherence-free algebras]\label{struct_th_dfa}
	Let $\Sem^\dagger = (\Phi_t^\dagger)_{t\in\mathbb{R}^+}$ be a QDS in the Heisenberg picture generated by an adjoint GKLS generator $\mathcal{L}^\dagger$ in the form~\eqref{GKLS_adjoint}. Then, there exists a family of Hilbert spaces $\mathcal H_{k,i}$, with $k\in[M]$ and $i\in\{1,2\}$, that decompose the Hilbert space $\mathcal H$ as
	\begin{equation}\label{eq:decomposition_hilbert_dfa}
		\mathcal H= 
        \bigoplus_{k\in[M]} \left(\mathcal H_{k,1}\otimes\mathcal H_{k,2}\right),
	\end{equation}
	such that the decoherence-free algebra $\mathcal N(\Sem^\dagger)$ reads
	\begin{equation}\label{eq:structure_of_dfa}
		\mathcal N(\Sem^\dagger) = 
        \bigoplus_{k\in[M]} \left(\mathcal B(\mathcal H_{k,1})\otimes \mathbb I_{k,2}\right).
	\end{equation}
	Furthermore, the Hamiltonian $H$ and the jump operators $L_\alpha$, $\alpha \in [N]$, in~\eqref{GKLS_adjoint} admit the following block-diagonal form with respect to the decomposition~\eqref{eq:decomposition_hilbert_dfa} of $\mathcal{H}$:
	\begin{align}
		H&= 
        \bigoplus_{k\in[M]} (H_{k,1}\otimes \mathbb I_{k,2}+\mathbb I_{k,1}\otimes H_{k,2}),\quad
		L_\alpha = 
        \bigoplus_{k\in[M]} \left(\mathbb I_{k,1}\otimes L_{\alpha,k}\right),\quad \alpha \in[N],
	\end{align}
	with $H_{k,1}\in \mathcal B(\mathcal H_{k,1})$ and $H_{k,2}\in\mathcal B(\mathcal H_{k,2})$ Hermitian operators, and $L_{\alpha,k}\in \mathcal{B}(\mathcal H_{k,2})$, with $k\in[M]$ and $\alpha\in[N]$. Then, given $X\in\mathcal N(\Sem^\dagger)$ in the form
	\begin{equation}
		X=
        \bigoplus_{k\in[M]} \left(x_k\otimes \mathbb I_{k,2}\right)\in\mathcal N(\Sem^\dagger)
	\end{equation}
	the action of $\Phi^\dagger_t$ on $X$ reads
	\begin{equation}\label{eq:dfa_action}
		\Phi_t^{\dagger}(X)=\mathrm e^{\mathrm i t H}X\mathrm e^{-\mathrm i t H}= 
        \bigoplus_{k\in[M]} \left(\mathrm e^{\mathrm i  t H_{k,1}}x_k\mathrm e^{-\mathrm i t H_{k,1}}\otimes \mathbb I_{k,2}\right).
	\end{equation}
\end{theorem}
We conclude that the asymptotic dynamics taking place on $\mathcal{N}(\Sem^\dagger)$ is entirely governed by the Hamiltonian $H$ appearing in the GKLS representation of  $\mathcal{L}^\dagger$ of $\Sem^\dagger$; see also~\cite[Remark 1]{fagnola2017}. In particular, it generates a unitary evolution on each factor $\mathcal{H}_{k,1}$ with $k \in [M]$.

Finally, as an example of application of DFAs, let us discuss below an algebraic necessary and sufficient condition for the convergence towards steady states~\cite{Frigerio_Verri_82,fagnola2008algebraic,deschamps2016structure}, already addressed in Sections~\ref{sec:asymptotics},~\ref{asympt_QDS_sec}, and~\ref{sec:symmetries}. 

\begin{proposition}
\label{prop:Noether_convergence}
Let $\Phi$ (resp.\ $\mathcal{L}$) be a faithful quantum channel (resp. GKLS generator) generating a QMC (resp. QDS) $\Sem$. Then, the limit 
$\lim_{n \to \infty} \Phi^n(\rho)$ (resp.\ $\lim_{t \to \infty} \mathrm{e}^{t\mathcal{L}}(\rho)$) exists for every state $\rho \in \mathcal{S}(\mathcal{H})$ if and only if
\begin{equation}
    \mathcal{N}(\Phi^\dagger) = S^\prime
\qquad
\bigl(\text{resp. }\mathcal{N}(\Sem^\dagger)=T^\prime\bigr),
\end{equation}
where we used the notation of Theorem~\ref{thm:Noether}.
Moreover, a faithful QDS $\Sem$ is relaxing, i.e., $\lim_{t \to \infty} e^{t \mathcal{L}}(\rho) = \rho_{\mathrm{s}}$  for some stationary state $\rho_{\mathrm{s}}$ and for every initial state $\rho \in \mathcal{S}(\mathcal{H})$, if and only if $T^\prime=\mathbb{C}\mathbb{I}$.
\end{proposition}
\begin{proof}
If $\Phi$ is a quantum channel, then $\lim_{n \to \infty} \Phi^n(\rho)$ exists if and only if $\Fix(\Phi) = \Attr(\Phi)$ or, equivalently, $\Fix(\Phi^\dagger) = \Attr(\Phi^\dagger)$. Therefore, the claimed characterization follows from the equalities 
\begin{equation}
    \Attr(\Phi^\dagger) = \mathcal{N}(\Phi^\dagger), \quad \Fix(\Phi^\dagger) = S^\prime,
\end{equation}
which hold in the faithful case, cf. Proposition~\ref{th:fixed_kraus} and Section~\ref{alg_sec}. The analogous claim for QDSs can be derived exactly in the same way.
 Finally, the last part of the proposition is just Corollary~\ref{Frigerio_first_cor}.
\end{proof}

Therefore, under the assumptions of convergence towards equilibrium and faithfulness, we get a characterization of the DFA $\mathcal{N}(\Phi^\dagger)$ (resp. $\mathcal{N}(\Sem^\dagger)$) of a QMC (resp. QDS) $\Sem^\dagger$, which is simpler than the general one~\eqref{char_N_Kraus} [resp.~\eqref{char_N_GKLS}].

\subsection{Algebraic structure of the attractor subspace}
\label{alg_sec} 

In this subsection, we  explore the algebraic structure of the attractor subspace of UCP maps in the Heisenberg picture; see~\cite[Section 4]{AFK_asympt5} for analogous results in the continuous-time Markovian setting.

First, assume that the UCP map $\Phi^\dagger$ is faithful, meaning that there exists an invertible stationary state for the corresponding adjoint evolution $\Phi$, see Definition~\ref{faith_ch_def}. In this case, as a consequence of Equation~\eqref{attr_phi00_struc}, the attractor subspace is a $C^*$-algebra with respect to the composition product of $\mathcal B(\mathcal H)$; cf.\ Theorem~\ref{struc_theorem_finite_dimensional} for the general structure of a finite-dimensional $C^*$-algebra. 

When the dynamics is not faithful, the attractor subspace need not be a $C^*$-algebra with respect to the composition product, as shown in the following example by Arveson~\cite[Remark 2]{arveson1972}:
\begin{equation}
	\label{Arves_ex}
	X=\begin{pmatrix}
		x_{11}& x_{12}&x_{13}\\
		x_{21}& x_{22}&x_{23}\\
		x_{31}& x_{32}&x_{33}
	\end{pmatrix}\mapsto
	\Phi^\dagger(X)=\begin{pmatrix}
		x_{11}&0 &0\\
		0& x_{22}&0\\
		0&0 &\frac{x_{11}+x_{22}}{2}
	\end{pmatrix}.
\end{equation}

The UCP map $\Phi^\dagger$ in~\eqref{Arves_ex} is idempotent, and its range $\Ran(\Phi^\dagger) = \Attr(\Phi^\dagger)$ is not closed under composition. A UCP map $\Phi^\dagger$ for which the attractor subspace $\Attr(\Phi^\dagger)$ is an algebra under the composition product is said to be \emph{peripherally automorphic}~\cite{rajarama2022peripheral}. A UCP map $\Phi^\dagger$ is peripherally automorphic if and only if~\cite[Theorem 2.6]{rajarama2022peripheral}
\begin{equation}\label{eq:bhat_result}
	\Attr(\Phi^\dagger)\subseteq \mathcal N(\Phi^\dagger),
\end{equation}
where $\mathcal N(\Phi^\dagger)$ denotes its DFA, defined in~\eqref{eq:decoherence_free_algebra_UCP}. As noted above, any faithful UCP map is peripherally automorphic~\cite[Corollary 2.8]{rajaramaperipheral_2}, and, in fact, the inclusion~\eqref{eq:bhat_result} holds with equality in such case; see e.g.~\cite[Theorem 1]{AFK_asympt5}. Combining this equality with~\eqref{char_N_Kraus}, we obtain the following characterization of the attractor subspace of a faithful UCP map in terms of its KS operators:
\begin{equation}
    \label{char_Attr_Kraus_faithful}
    \Attr(\Phi^\dagger) = \{ A_{k_1}\dots A_{k_n} A_{\ell_1}^\ast \dots A_{\ell_n}^\ast ,  \; k_j , \ell_j \in[N] ,\; j \in [n] ,\; n \in \mathbb{N}  \}^\prime.
\end{equation}
Equation~\eqref{char_Attr_Kraus_faithful} provides the analogue of Proposition~\ref{th:fixed_kraus} for the attractor subspace of faithful evolutions. Moreover, all qubit UCP maps are known to be peripherally automorphic~\cite[Corollary 3.13]{AFK_asympt3}. See~\cite{rajarama2022peripheral,rajaramaperipheral_2} for more details on the peripheral automorphism property. Notice also that the inclusion $\mathcal{N}(\Phi^\dagger) \subseteq \Attr(\Phi^\dagger)$, namely the converse of~\eqref{eq:bhat_result}, is not true in general; see~\cite[Example 3]{AFK_asympt3} for a counterexample.

Even when a UCP map is not peripherally automorphic, its attractor subspace can still be endowed with a modified product that makes it a $C^*$-algebra, as shown in the following theorem~\cite[Theorem~2.12]{rajarama2022peripheral}. 
\begin{theorem}[Choi--Effros product]
\label{CE_thm}
Let $\Phi^{\dagger}$ be a UCP map with peripheral projection $\mathcal{P}^\dagger_{\mathrm P} $. Then, the attractor subspace $\Attr(\Phi^\dagger)$ of $\Phi^\dagger$ is a $C^\ast$-algebra with respect to the Choi--Effros product, defined as
\begin{equation}
\label{CE_def}
X \star Y := \mathcal{P}_{\mathrm P}^\dagger (XY), \quad  X,Y \in \Attr(\Phi^\dagger),
\end{equation}
 the Hermitian conjugate $\ast$, and the operator norm $\| \cdot \|$. 
Moreover, the map $\Phi^\dagger$ restricted to $\Attr(\Phi^\dagger)$ is a $*$-automorphism, namely it is invertible and
\begin{align}
\label{aut_Phi_dagger}
	\Phi^\dagger(X \star Y) = \Phi^\dagger(X) \star \Phi^\dagger(Y),\quad X,Y\in \Attr(\Phi^\dagger).
\end{align}
\end{theorem}
The product~\eqref{CE_def} is called the \emph{Choi--Effros product} as its construction follows the argument used in the proof of~\cite[Theorem 3.1]{choi1977injectivity}, concerning injective operator systems~\cite{paulsen2002completely}. Therefore, Theorem~\ref{CE_thm} shows that dissipation and decoherence not only restrict the initial algebra $\mathcal{B}(\mathcal{H})$ of observables to the attractor subspace $\Attr(\Phi^\dagger)$,  but can also deform the ordinary operator product inherited from~$\mathcal{B}(\mathcal{H})$. See also~\cite{alipour2017dynamical,chruscinski2012observables} for further analysis in this direction. For a more explicit connection between the Choi--Effros product and the structure Theorem~\ref{struc_Heis_th}, we refer the reader to~\cite[Section~4]{AFK_asympt4}.

Theorem~\ref{CE_thm} motivates the introduction of the Choi--Effros decoherence-free algebra (\emph{Choi--Effros DFA}) $\mathcal{N}_{\star}(\Phi^\dagger)$ of a UCP map $\Phi^\dagger$. It is obtained  by replacing the ordinary operator product with the Choi--Effros product, naturally extended all over $\mathcal{B}(\mathcal{H})$ by means of~\eqref{eq:decoherence_free_algebra_UCP}~\cite{AFK_asympt3,AFK_asympt4}. 

The Choi--Effros DFA is a $C^*$-algebra with respect to the ordinary operator product and a $B^*$-algebra with respect to the Choi--Effros product; see Definition~\ref{Cast_def}.  Moreover,
it contains both the attractor subspace and the DFA of the UCP map. Interestingly, the equality $\Attr(\Phi^\dagger) = \mathcal{N}_\star(\Phi^\dagger)$ holds if and only if $\Phi^\dagger$ is faithful~\cite{AFK_asympt3}. We refer the reader to~\cite{AFK_asympt3,AFK_asympt5} for further details on the interplay between spectral and algebraic approaches to the asymptotic dynamics.

As a final remark, note that complete positivity is not needed in order to obtain all the results regarding the algebraic structure of the asymptotics~\cite{wolf2010inverse,AFK_asympt3,AFK_asympt4}. To illustrate this point more precisely, recall that a \emph{Schwarz} map $\Phi^\dagger$ is a unital positive map satisfying the operator inequality
\begin{equation}
    \Phi^\dagger(X^\ast X) \geq \Phi^\dagger (X)^\ast \Phi^\dagger (X), \quad X \in \mathcal{B}(\mathcal{H}),
\end{equation}
known as the \emph{operator Schwarz inequality}. Observe that any UCP map is Schwarz, whereas the converse does not hold in general; see~\cite[Appendix A]{choi1980some}. For an overview of the notions lying between positivity and complete positivity, see~\cite[Section 3]{amato2025bridging}. 

For the discrete-time semigroup $\Sem^\dagger$ generated by a Schwarz map $\Phi^\dagger$, the attractor subspace $\Attr(\Phi^\dagger)$ and the peripheral projection $\mathcal{P}_{\mathrm{P}}^\dagger$ are defined as in~\eqref{Attr_def_Heis} and \eqref{per_proj_def}, respectively.
These objects describe the asymptotic behaviour of the semigroup, and Theorem~\ref{CE_thm},  characterizing the algebraic structure of the asymptotic dynamics, extends to Schwarz maps~\cite[Proposition 5.1, Theorem 5.2]{AFK_asympt3}.

\section{Other topics}
\label{other_top_sec}
In this final section we  briefly discuss additional topics related to the asymptotic behaviour of open quantum systems that have not been covered earlier in this review. 
\subsection{Quantum associative memories}
\label{sec:qam}
The asymptotic dynamics of classical complex systems has found a remarkable application in the field of associative memory retrieval, where patterns are encoded as the stable configurations of the evolution of a network of interacting units, suggestively called ``neurons''~\cite{Hopfield1982}. Early quantum extensions of this paradigm, giving rise to \emph{quantum associative memories}, encoded patterns in superpositions and retrieved them through quantum algorithms~\cite{ventura2000quantum,Trugenberger2001PQM}.\ Closer to our setting are dissipative approaches, including quantum stochastic walks~\cite{Schuld2014}, open quantum extensions of Hopfield networks~\cite{Rotondo2018}, and attractor quantum neural networks formulated as quantum channels with prescribed stationary states~\cite{Lewenstein2021,Lewenstein2022Corrigendum,Marconi2022}.\ 

A general framework is presented in~\cite{LabayMora2025QST} that builds on stationary states and their basins of attraction and also covers non-orthogonal patterns, measurement-based readout, and metastable retrieval. It separates the existence of stationary states from their usefulness for retrieval, linking asymptotic subspaces and domains of attraction, addressed in Sections~\ref{sec:structure_theorems} and~\ref{dom_attr_subsec}.\ A key question is the \emph{storage capacity}, the number of storable patterns relative to system size, first studied classically with statistical mechanics~\cite{gardner1988optimal,gardner1988space} and recently for quantum Hopfield networks~\cite{Bodeker23} and perceptrons~\cite{BGM22,Gratsea24,BGMN2024,BGGMP2026}. The structure of asymptotic dynamics, detailed in the present review, allows a hardware-independent treatment of storage capacity based only on the general properties of quantum channels~\cite{Lewenstein2021,Lewenstein2022Corrigendum,LabayMora2025QST}.

\subsection{Asymptotics and non-Markovianity} 
\label{sec:asympt_non_Markov}
As discussed in Section~\ref{unit_asympt_sec}, the asymptotic dynamics of an open quantum system is not unitary in general, in fact it may exhibit various non-Markovian features. In particular, the presence of permutations in the long-time evolution, see Theorem~\ref{Wolf_struc}, may be related to the lack of divisibility of the channel inducing the QMC~\cite{AFK_asympt_1}, and it may be ruled out by a suitable coarse graining of the evolution~\cite{AFK_asympt_2}. Moreover, beyond the quantum dynamical semigroup scenario, the asymptotic state may depend on the initial one even when it is unique, and this may be regarded as another clear signature of non-Markovianity~\cite{chruscinski2010long,xiong2015non}.    
Furthermore, it was shown in~\cite{ask2022non} that open quantum systems interacting with a non-Markovian environment can be exploited to generate steady states that cannot be engineered in the Markovian regime, thereby providing another potential witness of asymptotic non-Markovianity.

Interestingly, non-Markovian effects can emerge \emph{only} after a finite time threshold, giving rise to the phenomenon of \emph{hidden non-Markovianity}~\cite{tufarelli2014non,burgarth2021quantum,burgarth2021hidden} where non-Markovianity is understood as deviations from the semigroup law. On the other hand, \emph{eternal non-Markovianity}~\cite{Hall2014,Megier2017,Jagadish2025} may also take place in time-local GKLS-like evolutions when at least one decoherence rate is negative at \emph{all} times.  \subsection{Numerical methods for steady states}\label{sec:numerical_methods}
Evaluating the (possibly non-equilibrium) steady states of open-quantum-system evolutions and their properties, such as expectation values of several observables of interest~\cite{nagy2019variational,benatti2021exact,mortimer2025certifying}, is quite important both from the theoretical and practical perspectives. Although exact steady states have been found e.g.\ in several spin-chain models~\cite{benatti2021exact,benatti2022stationary,prosen2011exact,popkov2020exact}, the problem cannot be addressed analytically for most large and complex systems, thus a wide range of numerical approaches has been developed over the years such as variational techniques~\cite{nagy2018driven,nagy2019variational}, neural networks~\cite{vicentini2019variational}, matrix product density operators~\cite{cui2015variational}, iterative methods~\cite{nation2015iterative}, tensor network algorithms~\cite{kshetrimayum2017simple,hryniuk2024tensor}, and semidefinite programming~\cite{robichon2024bootstrapping,mortimer2025certifying}. Note also that the dependence of the steady state on a parameter of the model has been  studied with perturbative and exact techniques~\cite{li2014perturbative,li2016resummation,nagib2025exact}. 
\section*{Acknowledgments}
We would like to express our sincere gratitude to Remigiusz Augusiak, Fabio Benatti,  Daniel Burgarth, Francesco Buscemi, Dariusz Chru{\'s}ci{\'n}ski, Omar Fawzi, Andrzej Kossakowski, Marilena Ligab\`o, Stefano Mancini, Giuseppe Marmo, Hiromichi Nakazato, Saverio Pascazio, George Sudarshan, Shuichi Tasaki, and Kazuya Yuasa for the many stimulating discussions and fruitful collaborations over the years that have deeply influenced our understanding of open-quantum-system dynamics.

We acknowledge support from Istituto Nazionale di Fisica Nucleare (INFN) through the project ``QUANTUM'', from the Italian National Group of Mathematical Physics (GNFM-INdAM), and
from the Italian funding within the ``Budget MUR - Dipartimenti di Eccellenza 2023–2027" - Quantum Sensing and Modelling for One-Health (QuaSiModO).
DA and PF acknowledge  support from Projects PN-RIC Q-SUD (Q-SUD B99H26000380007) and PN-RIC QUANTAS (SYNERGIA B99H26000410007).
GG acknowledges financial support from the project ``Patto territoriale sistema universitario pugliese'' (PATTI, CUP F61B23000370006). 
AK acknowledges the support of the (Polish) National Science Center through the SONATINA Grant No. 2026/60/C/ST2/00246 and SONATA BIS Grant No. 2019/34/E/ST2/00369.
\newline
\newline
\textbf{AI use disclosure.} 
During the preparation of this review, the authors used ChatGPT (OpenAI) and Claude (Anthropic) to assist with language editing, manuscript organization, consistency checks of mathematical arguments and notation, and the preparation of figures. All AI-assisted material was critically evaluated and revised by the authors, who take full responsibility for the scientific content, accuracy, and integrity of the manuscript.

\section*{Data availability statement}
All data that support the findings of this study are included within the article (and any supplementary files).

\begin{appendices}

\section{Classical Markov chains}
This appendix  is devoted to a concise introduction to classical Markov chains; see~\cite{bremaud,cufaro,shyraev} for details. The discrete-time case is discussed in the next subsection, while continuous-time Markov evolutions are presented in Section~\ref{class_sem_subsec}.
\label{classic_sec}
\subsection{Discrete-time Markov chains}
Let us briefly recall homogeneous Markov chains in the discrete-time setting.
Consider a finite set of events $\Omega = \{ 1 , \dots , d \}$
and let us consider a discrete-time stochastic process  $(X_n)_{n\in\mathbb{N}}$ of random variables on $\Omega$. Note that the process is completely determined by the joint probability distributions
\begin{equation}
    \mathbb P(X_n = i_n, X_{n-1} = i_{n-1}, \dots, X_0 = i_0)
\end{equation}
that the process is in $i_n,i_{n-1},\dots,i_0\in\Omega$ at times $n,{n-1},\dots, 0$, respectively.
A stochastic process is said to be \emph{Markovian} if the probability $\mathbb{P}$ satisfies the following condition
\begin{equation}
\label{Markov-prop}
\mathbb{P}(X_n = i_n \vert X_{n-1} =i_{n-1} , \dots , X_0 = i_0) = \mathbb{P}(X_n = i_n \vert X_{n-1} = i_{n-1})
\end{equation}
for all $n\in \mathbb{N}$ and all states $i_0 , \dots , i_n$, called the \emph{Markov property}. Here, $\mathbb{P}(A \vert B)$ denotes the conditional probability of the event $A$ given $B$. Such stochastic processes are \emph{memoryless}, in the sense that the probability of the transition $i_{n-1} \mapsto i_{n}$ from time $n-1$ to time $n$ does not depend on the full history of the process. A process $(X_n)_{n\in\mathbb{N}}$ obeying~\eqref{Markov-prop} is referred to as (discrete-time) \emph{Markov chain}, and it is called \emph{homogeneous} if the probability $\mathbb{P}(X_n = i_n \vert X_{n-1} = i_{n-1})$ of the transition $i_{n-1} \mapsto i_{n}$ does not depend on $n$. Importantly, the only relevant quantities for a homogeneous Markov chain are the initial probability distribution $\bm{p}_{0}$, whose $i$-element represents the probability of the event $i\in\Omega$ in the initial state, and the $d\times d$ \emph{transition matrix}
\begin{equation}
 Q = (Q_{\ell m}), \quad Q_{\ell m}= \mathbb{P}(X_n = \ell \vert X_{n-1} = m), \; \ell , m  \in [d].
\end{equation}
Each entry of the matrix $Q$ is non-negative, as it represents a (conditional) probability and, moreover, the sum of the entries in each column is equal to one, ensuring the normalization property, namely
\begin{align}
\label{stoch-pro_1}
Q_{\ell m} &\geq 0, \quad \ell,m \in [d], \\
\sum_{\ell\in[d]} Q_{\ell m} &=1, \quad m \in [d],
\label{stoch-pro_2}
\end{align}
hence any transition matrix $Q$ is \emph{column-stochastic}.
As a consequence of Chapman--Kolmogorov conditions, the probability vector $\bm{p}_{n}$ at time $n$ of the homogeneous Markov chain will be given by 
\begin{equation}
\label{dyn_HMC}
\bm{p}_{n} = Q^n \bm p_{0}, \quad n \geq 1.
\end{equation}
The dynamics at all times is completely described by a discrete-time semigroup $(Q^n)_{n\in\mathbb{N}}$. Similarly, we introduced in Section~\ref{qc_sec} the notion of QMC, which is given by a discrete-time semigroup $(\Phi^n)_{n\in\mathbb{N}}$ induced by a quantum channel $\Phi$, cf.\ Definition~\ref{qc_def}, which may be regarded as the quantum analogue of a column-stochastic matrix. In fact, given an orthonormal basis $\{ \ket{i} \}
_{i\in[d]}$ of $\mathcal{H}$, a quantum channel induces a column-stochastic matrix, see e.g.~\cite[Section 4]{amato2025bridging} for more details.

\begin{figure}[b!ht]
\begin{center}
\includegraphics[scale=1.1]{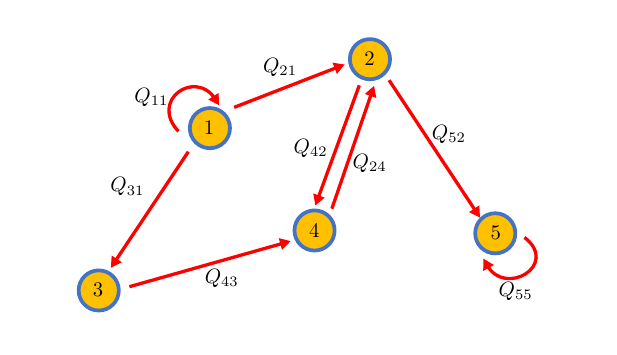}
\caption{Transition graph of a homogeneous Markov chain with 5 states: the vertices denote the states, whereas the arrow from a state $i$ to a state $j$ indicates the non-zero probability $Q_{ji}$ of the transition $i \mapsto j$.}
\label{figura1}
\end{center}
\end{figure}

Finally, as shown in Figure~\ref{figura1}, we can represent a homogeneous Markov chain with a graph, called the \emph{transition graph}, where the vertices are the states of the chain and we add an oriented edge from vertex $m$ to vertex $\ell$ if and only if $Q_{\ell m}>0$. 

\subsection{Continuous-time Markov chains}
\label{class_sem_subsec}

Continuous-time homogeneous Markov evolutions are described by a continuous family $(Q(t))_{t\in\mathbb{R}^+}$ of transition or, equivalently, column-stochastic matrices, with $Q(0)=\mathbb{I}$ being the identity matrix, and satisfying the \emph{semigroup} property  \begin{equation}
Q(t_1 + t_2) = Q(t_1)Q(t_2), \quad \forall t_1,t_2 \in\mathbb{R}^+,
\end{equation}
which formalizes the Markovian assumption. Notice that the quantum analogue of the family $(Q(t))_{t\in\mathbb{R}^+}$ we just introduced is the notion of QDS discussed in Definition~\ref{QDS_def}.
Equivalently, the evolution of the probability vector $\bm p(t)=Q(t)\bm p_0$ at time $t$, with $\bm p_0$ being the probability vector at time $t=0$, may be described by the following first-order linear differential equation
\begin{equation}\label{eq:kolmogorov_eq}
\frac{\mathrm{d}}{\mathrm{dt}}\bm p(t) = L\bm{p}(t).
\end{equation} 
The matrix $L=(L_{\ell m})$, called \emph{Kolmogorov generator}, has the explicit structure 
\begin{equation}
\label{Kolmo_struc}
L_{\ell m} = W_{\ell m} - \delta_{\ell m}W_m, \quad W_m = 
\sum_{\ell\in[d]} W_{\ell m}, \quad \ell , m \in [d],
\end{equation}
where $W_{\ell m} \geq 0$ can be regarded as the transition rate from the event $m$ to the event $\ell$, $\ell\neq m$. Equation~\eqref{eq:kolmogorov_eq} for the $\ell$-th component $p_\ell(t)$ of the probability vector $\bm p(t)$ reads
\begin{equation}
\frac{\mathrm{d} p_\ell(t)}{\mathrm{dt}} = 
\sum_{m\in[d]} [W_{\ell m}p_m(t) - W_{m\ell}p_\ell(t)], \quad \ell \in [d],
\end{equation}
which is the well-known \emph{Pauli rate equation}~\cite{vanKampen}. Finally, note that~\eqref{eq:kolmogorov_eq} is the classical version of the Markovian master Equation~\eqref{master_eq}, where the matrix $L$ is replaced by the GKLS generator $\mathcal{L}$, discussed in Section~\ref{QDS_sec}. 
In particular, 
the explicit form~\eqref{Kolmo_struc} resembles the GKLS representation~\eqref{diag_GKLS}.

\section{Jordan decomposition of an operator}
\label{Jordan_op_app}
Given an operator $A$ on an $n$-dimensional complex vector space $V$, then we can decompose $A$ as~\cite[Section~I.5.4]{kato2013perturbation}
\begin{equation}
\label{Jordan}
    A = 
    \sum_{k\in[N]} (\lambda_k P_k + D_k),
\end{equation}
where $\lambda_k$ are the distinct eigenvalues of $A$, while $P_k$ and $D_k$ are the corresponding spectral projections and nilpotents, respectively. The latter operators satisfy the following properties
\begin{align}
&P_{k} P_\ell=\delta_{k \ell} P_{k},\quad 
\sum_{k\in[N]} P_{k}=\mathbb{I}_{V}, \label{proj} \\
&P_{k} D_\ell=D_{\ell} P_{k}=\delta_{k\ell} D_{k},\quad D_{k}^{r_{k}}=0, \label{nil}
\end{align}
with $1\leq r_k \leq \rank (P_{k})$ and $\mathbb{I}_{V}$ identity on $V$. In particular, note that, if $V$ is also endowed with a scalar product $\braket{\cdot}$, it may happen that $P_k\neq P_k^*$ for some $k \in [N]$, where $\ast$ denotes the adjoint with respect to $\braket{\cdot}$.
Importantly, the ranges $\mathcal{M}_k$ of the spectral projections $P_k$, called the \emph{generalized eigenspaces} corresponding to the eigenvalues $\lambda_k$, decompose the vector space $V$, namely

\begin{equation}
\label{eigen_dec_Jordan}
V = \mathrm{span}\{\mathcal M_k \,:\,k\in[N]\}.
\end{equation}
The quantity $m_{k} = \Tr ( P_k )=\rank ( P_k )$ is called the \emph{algebraic multiplicity} of the eigenvalue $\lambda_{k}$, and, as a consequence of~\eqref{eigen_dec_Jordan}, we also have 
$\sum_{k\in[N]} m_k = n$. Instead, the \emph{geometric multiplicity} $n_k$ of $\lambda_k$ is the dimension of the subspace
\begin{equation}
\mathcal{N}_k := \Ker( A - \lambda_k \mathbb{I}_{V} ) = \{ \bm{x} \in V \,:\, A\bm{x} = \lambda_k \bm{x} \},
\end{equation}
called the (geometric) \emph{eigenspace} of $A$ corresponding to the eigenvalue $\lambda_k$. It is easy to prove that
\begin{equation}
\mathcal{N}_k \subseteq \mathcal{M}_k,
\end{equation}
from which the inequality $n_k \leq m_k$ follows.
In line with the above terminology, the elements of $\mathcal{N}_k$ and $\mathcal{M}_k$ are respectively  called \emph{eigenvectors} and \emph{generalized eigenvectors} of $A$ corresponding to $\lambda_k$. \\ \indent
If we represent $A$ as an $n \times n$ matrix with respect to some basis $\{ \bm{v}_i \}_{i\in[n]}$ of $V$, we can find its Jordan decomposition as~\cite[Theorem 3.1.11.]{horn}  
\begin{equation}
\label{Jordan_hatA}
{A} = X \Bigl( 
\bigoplus_{k\in[N]} \bigoplus_{\ell\in[n_k]} J_\ell (\lambda_k)\Bigr) X^{-1}, 
\end{equation}
for some invertible matrix $X$, i.e., ${A}$ is similar to a \emph{Jordan block matrix}, where
\begin{equation}
\label{Jordan_block}
J_\ell(\lambda_k) = \begin{pmatrix}
\lambda_k & 1 & & & \\
& \lambda_k & 1 & & \\
& & \ddots & \ddots & \\
& & & \ddots & 1 \\
& & & & \lambda_k  
\end{pmatrix}
\end{equation}
is called the $\ell$-th \emph{Jordan block} corresponding to the eigenvalue $\lambda_k$ with $\ell \in [n_k]$ and $k \in [N]$. Note that all the unspecified elements on the right-hand side of~\eqref{Jordan_block} are implicitly assumed to be zeros. The geometric multiplicity $n_k$ of $\lambda_k$ is the number of Jordan blocks corresponding to the eigenvalue $\lambda_k$, whereas the algebraic multiplicity $m_k$ of $\lambda_k$ is the sum of the sizes $d_\ell$ of the Jordan blocks. Now, it is possible to construct $d_\ell$ linearly independent generalized eigenvectors from each of the Jordan blocks  of $A$. Specifically, given $\ell \in [n_k]$ and $k \in [N]$, take the last canonical vector $\bm{y}_{\ell,0}^{(k)}$ in the  $\ell$-th block $J_{\ell}(\lambda_k)$ of the eigenvalue $\lambda_k$, and define $\bm{z}_{\ell,0}^{(k)} := X\bm{y}_{\ell,0}^{(k)}$. It is easy to check that $\bm{z}_{\ell,0}^{(k)}$ is a generalized eigenvector satisfying
\begin{equation}(A - \lambda_k \mathbb{I}_{V})^{d_\ell^{(k)}} (\bm{z}_{\ell ,0}^{(k)}) = 0, \label{gen_eigen_form_1}
\end{equation}
and from which we can define the other $d_\ell^{(k)}-1$  generalized eigenvectors of the chain as
\begin{equation}
    \label{gen_eigen_form_2}
    \bm{z}_{\ell , i}^{(k)} := (A- \lambda_k \mathbb{I}_{V}) \bm{z}_{\ell ,i-1}^{(k)} = (A- \lambda_k \mathbb{I}_{V})^{i} \bm{z}_{\ell ,0}^{(k)} \neq0, \quad i \in [d_\ell^{(k)}-1].
\end{equation}
An eigenvalue $\lambda_k$ of $A$ is said to be \emph{semisimple} if the corresponding spectral nilpotent $D_k$ is zero, i.e.\ $n_k = m_k$, meaning that all the Jordan blocks corresponding to $\lambda_k$ are one-dimensional. Therefore, in such a case, we can generically speak of the multiplicity of the eigenvalue $\lambda_k$ without ambiguity. An operator $A$ is called \emph{diagonalizable} if all the eigenvalues are semisimple
and, in this case, the Jordan decomposition of $A$ reads 
\begin{equation}
\label{Jordan_diago}
A = 
\sum_{k\in[N]} \lambda_k P_k.
\end{equation}
Now, if $V$ is a Hilbert space with inner product $\braket{\cdot}{\cdot}$, then diagonalizability implies the existence, for each eigenvalue $\lambda_k$, of two sets of  vectors $\{ \ket*{x_\ell^{(k)} } \}_{\ell=1}^{n_k}$ and $\{  \ket*{y_\ell^{(k)}} \}_{\ell=1}^{n_k}$, called \emph{right} and \emph{left eigenvectors} of $A$ corresponding to $\lambda_k$, respectively, so that
\begin{equation}
A \ket*{x_\ell^{(k)}} = \lambda_k \ket*{x_\ell^{(k)}}, \quad \braket*{y_\ell^{(k)} }{x_m^{(k)} } = \delta_{\ell m},
\end{equation}
where $\delta_{ij}$ denotes the Kronecker delta.
Therefore, the Jordan decomposition~\eqref{Jordan_diago} becomes
\begin{equation}
\label{diagonal_2}
 A = 
 \sum_{k\in[N]} \sum_{\ell\in[n_k]} \lambda_k \ketbra*{x_\ell^{(k)}}{y_\ell^{(k)}}.
\end{equation}
Moreover, an eigenvalue $\lambda_k$ is called \emph{simple}, or non-degenerate, if and only if its algebraic and geometric multiplicities satisfy  $m_k = n_k = 1$, 
 and an operator is simple whenever all the eigenvalues are. In such a case, the spectral projections $P_k$ in~\eqref{Jordan} are all rank-one operators or, equivalently, there is only a one-dimensional Jordan block corresponding to each eigenvalue $\lambda_k$. Finally, recall that an operator $A$ on a Hilbert space $V$ is said to be \emph{normal} if it can be written in the form~\eqref{Jordan_diago} with $P_k = P_k^\ast$ or, equivalently, it admits an orthonormal basis of eigenvectors. Moreover, $A$ is normal if and only if $[A , A^\ast]=0$.  

     \subsection{Proof of Proposition~\ref{trace_eigen_prop}}
\label{proofs_app}
In this appendix we provide a proof of Proposition~\ref{trace_eigen_prop}, giving information about the generalized eigenoperators of quantum channels. See~\cite[Propositions~2.19-2.20]{Amato_PhD_thesis} for an alternative proof.

Let $\Phi$ be a quantum channel. The key observation is the following consequence of~\eqref{gen_eigen_form_1}--\eqref{gen_eigen_form_2}: given $X_\lambda \in \mathcal{M}_{\lambda}$, geometric eigenspace of $\Phi$ corresponding to $\lambda \in \spec(\Phi)$, then there always exists an integer $r \geq 1$ such that 
\begin{equation}
\label{gen_eigen_prop}
    (\Phi - \lambda \mathsf{1})^r(X_\lambda) = 0.
\end{equation}
In particular, $X_\lambda$ is an eigenvector of $\Phi$ with eigenvalue $\lambda$ when $r=1$.  
That being said, property i.\ easily follows from~\eqref{gen_eigen_prop} and the trace-preservation of $\Phi$:  
\begin{equation}
    (\Phi - \lambda \mathsf{1})^r(X_\lambda) = \sum_{i=0}^r (-\lambda)^{r-i}\Phi^i(X_\lambda) = 0 \Rightarrow \Tr((\Phi - \lambda \mathsf{1})^r(X_\lambda)) = 0,
\end{equation}
where we used the assumption $\lambda \neq 1$ in the last equality. Concerning property ii., take again $X_\lambda \in \mathcal{M}_\lambda$ and observe that by~\eqref{gen_eigen_prop} and the Hermiticity-preservation of $\Phi$  
\begin{equation}
    (\Phi - \lambda \mathsf{1})^r(X_\lambda) =  \left[\sum_{i=0}^r (-\overline{\lambda})^{r-i}\Phi^i(X_\lambda^\ast)\right]^\ast = (\Phi - \overline{\lambda} \mathsf{1})^r(X_\lambda^\ast)^\ast = 0,
\end{equation}
namely $X_\lambda^\ast \in \mathcal{M}_{\overline{\lambda}}$. Similarly, given $X_{\overline{\lambda}} \in \mathcal{M}_{\overline{\lambda}}$, we have
\begin{equation}
    (\Phi - \overline{\lambda} \mathsf{1})^r(X_{\overline{\lambda}}) = 0 \Rightarrow (\Phi - \lambda \mathsf{1})^r(X_{\overline{\lambda}}^\ast) = 0,
\end{equation}
i.e., $X_{\overline{\lambda}}^\ast \in \mathcal{M}_\lambda$, so the proof is complete.
\section{Basic notions of \texorpdfstring{$C^{\ast}$}{C*}-algebras }
\label{intro_Cast}
In this section, we briefly review the theory of C*-algebras, which constitutes a cornerstone in the mathematical formulation of quantum theory~\cite{araki1999mathematical}. A comprehensive treatment lies beyond the scope of this work~\cite{Takesaki1970,conway2019course,bratteli2012operator}; instead, we focus on the essential concepts needed for our purposes. 
In particular, the mathematical concepts discussed below are the fundamental tools for the algebraic description of the asymptotic dynamics discussed in Section~\ref{sec:alg_approach}.

	First, a  complex associative algebra $\mathcal{A}$, or simply an \emph{algebra}, is a vector space over the complex field $\mathbb{C}$ equipped with an associative bilinear product $\cdot$. An algebra is said to be \emph{unital} if it contains the identity $e$. Let us now endow $\mathcal{A}$ with an involution and a topological structure.
	
	\begin{definition}[$\ast$-algebra]
	A $\ast$-algebra is an algebra $\mathcal{A}$ equipped with an involution $a \in \mathcal{A}  \mapsto a^\ast \in \mathcal{A}$ satisfying the following properties for all $a,b\in \mathcal{A}$ and $\alpha\in\mathbb{C}$ 
		\begin{align}
          (a+b)^\ast &= a^\ast + b^\ast \\
          (\alpha a)^\ast &= \bar{\alpha} a^\ast, \\
			(a^\ast)^\ast &=a,\\
			(a \cdot b)^\ast &= b^\ast \!\cdot a^\ast,
		\end{align}
		where $\bar{\alpha}$ denotes the complex conjugate of $\alpha$.
	\end{definition}
	\begin{definition}[$B^\ast$-algebra and $C^\ast$-algebra]
		A Banach algebra $\mathcal{A}$, or $B$-algebra, is an algebra equipped with a norm $\| \cdot \|$ which makes $\mathcal{A}$ a Banach space, and satisfies the submultiplicative property
		\begin{equation}
			\label{eq:balgebrainequality}
			\| a \cdot b \| \leq \| a \| \| b \|, \quad a,b \in \mathcal{A}.
		\end{equation}
		A Banach $\ast$-algebra $\mathcal{A}$, or $B^\ast$-algebra, is an involutive B-algebra, where the involution $\ast$ is an isometry over $\mathcal{A}$, namely $\| a^\ast \| = \| a \|$ for all $a \in \mathcal{A}$. A ${C}^\ast$-algebra $\mathcal{A}$ is a ${B}^\ast$-algebra which satisfies the equality
		\begin{equation}
			\label{C*-id}
			\| a^\ast \!\cdot a \| = \| a \|^2, \quad a \in \mathcal{A},
		\end{equation}  
		which is called $C^\ast$-identity.
        \label{Cast_def}
	\end{definition}
A standard example of $C^\ast$-algebra is the space of bounded operators over a Hilbert space. Specifically, the space $\mathcal{B}(\mathcal{H})$ of bounded operators on a Hilbert space $\mathcal{H}$ is a ${C}^\ast$-algebra with respect to the composition product, the adjoint involution $\ast$, and the operator norm 
	\begin{equation}
	\label{op_norm_def}
			\| X \| = \sup_{ \|\psi \| = 1} \| X \ket{\psi} \|.
	\end{equation}

We can now discuss the following structure theorem for finite-dimensional subalgebras of $\mathcal B(\mathcal H)$~\cite[Theorem III.1.1]{davidson1996c}.

\begin{theorem}[Structure theorem for finite-dimensional $C^*$-algebras]\label{struc_theorem_finite_dimensional}
Let $\mathcal A$ be a $C^\ast$-subalgebra of $\mathcal{B}(\mathcal{H})$, with $\mathcal H$ finite-dimensional Hilbert space. Then, there exist Hilbert spaces $\mathcal H_0$ and $\mathcal H_{k,i}$, with $k\in[M]$ and $i \in \{ 1,2 \}$, that decompose the Hilbert space as
\begin{equation}
    \mathcal H=\mathcal H_0\oplus 
    \bigoplus_{k\in[M]} \left(\mathcal H_{k,1}\otimes \mathcal H_{k,2}\right),
\end{equation}
such that the algebra $\mathcal A$ reads 
\begin{equation}
\label{direct_sum_struc_alg}
     \mathcal A=0\oplus 
     \bigoplus_{k\in[M]} \left(\mathcal{B} (\mathcal{H}_{k,1})\otimes \mathbb{I}_{k,2}\right).
\end{equation}
In particular, when the identity $\mathbb I$ of $\mathcal B(\mathcal H)$ is an element of $\mathcal A$, we have that $\mathcal H_0=\{ 0 \}$ and the algebra is said to be non-degenerate.
\end{theorem}
Notice that any finite-dimensional $C^\ast$-algebra is unital with identity $0 \oplus \mathbb{I}_{\perp}$, where $\mathbb{I}_{\perp}$ is the identity over $\mathcal{H}_0^\perp$.

Let us also recall that a $C^\ast$-algebra $\mathcal{A}$ is called \emph{commutative} if $a \cdot b = b \cdot a$ for all $a,b \in \mathcal{A}$. 
Finally, a linear map $\phi$ on the $\ast$-algebra $\mathcal A$ is said to be a \emph{$\ast$-homomorphism} whenever
\begin{equation}
\phi(a \cdot b ) =\phi(a) \cdot \phi(b), \quad \phi(a^\ast) =\phi(a)^\ast, \quad a,b \in \mathcal{A},\label{hom_1}
\end{equation} 
and, in particular, $\phi$ is a \emph{$\ast$-automorphism} if it is also invertible.

\newpage
	\section{Guide for the reader}
    \label{sec:symbols}
\section*{\textbf{Abbreviations}}
\vspace{-0.9cm}
\printglossary[type=\acronymtype, style = longleft , title={}]

 \renewcommand{\arraystretch}{1.2}
\vspace{-0.3cm}
\begin{xltabular}{\linewidth}{@{}l>{\raggedright\arraybackslash}X@{}}

\caption{List of symbols and notations used throughout the text.}
\label{tab:symbols}\\

\toprule
\textbf{Symbol} & \textbf{Meaning / Definition} \\
\midrule
\endfirsthead

\toprule
\textbf{Symbol} & \textbf{Meaning / Definition} \\
\midrule
\endhead

\bottomrule
\endfoot

\bottomrule
\endlastfoot

$\mathbb{N}$ & Set of non-negative integers \\[4.8pt]

$\mathbb{R}^+$ & Set of non-negative real numbers \\[4.8pt]

$:=$
  & Definition symbol (“is defined as”) \\[4.8pt]

$\overline{z}$
  & Complex conjugate of $z \in \mathbb{C}$ \\[4.8pt]

$[M]$
  & Integer index set $\{1,2,\ldots,M\}$ of length $M$ \\[4.8pt]

$\Hi$
  & Finite-dimensional Hilbert space \\[4.8pt]

$\dim(V)$
  & Dimension of a vector space $V$ \\[4.8pt]

$\mbox{span}(S)$
  & Complex linear span of a subset $S$ of a vector space $V$ \\[4.8pt]

$\mathcal{B}(\Hi)$
  & Algebra of bounded linear operators on $\mathcal{H}$ \\[4.8pt]

  $S'$
  & Commutant of a set $S \subseteq \mathcal{B}(\mathcal{H})$, cf.~\eqref{comm_def} \\[4.8pt]

$\St(\Hi)$
  & Set of density operators (states) on $\mathcal{H}$ \\[4.8pt]

$\| \cdot \|$
  & Norm on a Banach space $\mathcal{B}$ 
   \\
   [4.8pt]

$X > 0\; (X \geq 0)$
  & $X$ positive (semi)definite operator on $\mathcal{H}$ \\[4.8pt]

$\Ker(\cdot)$, \; $\Ran(\cdot)$
  & Kernel and range of an operator \\[4.8pt]

  $\Tr(\cdot)$
  & Trace of an operator \\[4.8pt]

$\mathrm{Tr}_i(\cdot)$
  & Partial trace over the Hilbert space $\mathcal{H}_i$ \\[4.8pt]

$\mathbb{I}, \; \mathsf{1}$
  & Identity operators on $\Hi$ and $\Bo(\Hi)$, respectively \\[4.8pt]

$X^\ast$
  & Adjoint of $X \in \mathcal{B}(\mathcal{H})$ \\[4.8pt]

$\Psi^\dagger$
  & Hilbert--Schmidt adjoint of a linear map $\Psi$ on
    $\mathcal{B}(\mathcal{H})$ \\[4.8pt]

$\Psi \circ \Xi$
  & Composition of two linear maps $\Psi$ and $\Xi$ on
    $\mathcal{B}(\mathcal{H})$ \\[4.8pt]

$\Phi$
  & Quantum channel on $\mathcal{B}(\mathcal{H})$,
    see Definition~\ref{qc_def} \\[4.8pt]

$\Phi^\dagger$
  & Unital completely positive (UCP) map on
    $\mathcal{B}(\mathcal{H})$ \\[4.8pt]

$\Sem$
  & Quantum Dynamical Semigroup (QDS)
    $(\Phi_t)_{t\in\mathbb{R}^+}$ or Quantum Markov Chain (QMC) $(\Phi^n)_{n\in\mathbb{N}}$ on $\mathcal{B}(\mathcal{H})$
    \\[4.8pt]

$\Sem^\dagger$
  & Adjoint QDS $(\Phi_t^\dagger)_{t\in\mathbb{R}^+}$ or adjoint QMC $((\Phi^\dagger)^n)_{n\in\mathbb{N}}$ on
    $\mathcal{B}(\mathcal{H})$ \\[4.8pt]

$\mathcal{L}$ ($\mathcal{L}^\dagger$)
  & (Adjoint) GKLS generator on $\mathcal{B}(\mathcal{H})$,
    see Equation~\eqref{diag_GKLS} [\eqref{GKLS_adjoint}] \\[4.8pt]

$C(\Psi)$
  & Choi operator of a linear map $\Psi$ on
    $\mathcal{B}(\mathcal{H})$
    [see Equation~\eqref{CJ_iso}] \\[4.8pt]

$\mathrm{rank}_{\mathrm{KS}}(\Phi)$
  & KS rank of a quantum channel $\Phi$ \\[4.8pt]

$\spec(\Psi)$
  & Spectrum of a linear map $\Psi$ on $\Bo(\Hi)$,
    cf.~\eqref{spec_def} \\[4.8pt]

$\spec_\mathrm{P}(\Psi)$ ($\spec_\mathrm{B}(\Psi)$)
  & Peripheral (bulk) spectrum of a linear map $\Psi$ on
    $\mathcal{B}(\mathcal{H})$  \\[4.8pt]

$\Attr(\Phi)$ ($\Attr(\Sem)$)
  & Attractor subspace of a quantum channel (QDS)
    $\Phi$ ($\Sem$),
    cf.~\eqref{attr_def_old} [\eqref{fix_attr_QDS}] \\[4.8pt]

$\Fix(\Phi)$ ($\Fix(\Sem)$)
  & Fixed-point subspace of a quantum channel (QDS)
    $\Phi$ ($\Sem$),
    cf.~\eqref{fix_def} [\eqref{fix_attr_QDS}] \\[4.8pt]

    $\Attr(\mathcal{L})$ ($\Ker(\mathcal{L})$)
  & Attractor (kernel) subspace of a GKLS generator $\mathcal{L}$,
    cf.~\eqref{Attr_L_def} [\eqref{Ker_L_def}] \\[4.8pt]

$\mathcal{P}$
  & Steady projection of a quantum channel $\Phi$ (QDS $\Sem$), cf.~\eqref{steady_proj_channel} [\eqref{steady_proj_L}] \\[4.8pt]

 $\mathcal{P}_{\mathrm{P}}$
  & Peripheral projection of a quantum channel $\Phi$ (QDS $\Sem$),
    cf.~\eqref{per_proj_def} [\eqref{per_proj_def_L}] \\[4.8pt]
 $\DoA(\rho_s)$ & Domain of attraction of the steady state $\rho_s$ of a quantum channel $\Phi$ (QDS $\Sem$), cf.~\eqref{dom_attr_def_discrete} [\eqref{dom_attr_def}] 
\\[4.8pt]

$\mathcal{N}(\Phi^\dagger)$ ($\mathcal{N}(\Sem^\dagger)$) & Decoherence-free algebra (DFA) of a UCP map $\Phi^\dagger$ (adjoint QDS $\Sem^\dagger$), cf.~\eqref{eq:decoherence_free_algebra_UCP} [\eqref{eq:decoherence_free_algebra}] \\

\end{xltabular}

	\section*{Conventions}
\begin{itemize}
    \item We always adopted Dirac notation in the text: $\ket{\psi}$ denotes a vector in the Hilbert space $\mathcal{H}$ and $\bra{\psi}$ the continuous linear functional over $\mathcal{H}$ mapping $\ket{\varphi} \in \mathcal{H}$ into the inner product $\braket{\psi}{\varphi}$ between $\ket{\psi}$ and $\ket{\varphi}$ endowing $\mathcal{H}$. In contrast, we used bold italic notation for vectors in $\mathbb{R}^n$ or $\mathbb{C}^n$, such as $\bm{x}$;
    \item We always used in the text the symbol $\oplus$ for the direct sum of \emph{orthogonal} subspaces of $\mathcal{H}$ or $\mathcal{B}(\mathcal{H})$;
    \item A quantum channel on $\mathcal{B}(\mathcal{H})$ 
is denoted by $\Phi$, which is also used for positive (though not necessarily completely positive) maps, see e.g. Proposition~\ref{spec_rad_pos}. The symbol $\Psi$ is instead reserved for \emph{arbitrary} linear maps on $\mathcal{B}(\mathcal{H})$; 

    \item  The involution and the norm endowing any $C^\ast$-algebra are implicitly assumed to be those of $\mathcal{B}(\mathcal{H})$, cf. Appendix~\ref{intro_Cast}.
\end{itemize}

\end{appendices}
	
\addcontentsline{toc}{section}{References}
\newpage

\printbibliography

\end{document}